\documentclass[prl,showpacs,superscriptaddress,floatfix,twocolumn]{revtex4}
\usepackage{graphicx}
\newcommand{\pTwodTwo}[2]{\frac{\partial^2 #1}{\partial^2 #2}}
\newcommand{\pTwodTwoMixed}[3]{\frac{\partial^2 #1}{\partial #2 \partial #3}}
\newcommand{\vv}[1]{{\bf #1}}
\newcommand{\vh}[1]{{\bf \hat{#1}}}
\newcommand{\W}{\Omega}
\newcommand{\w}{\omega}

\newcommand{\Q}{\vv{q}}
\newcommand{\Pp}{{p}_{+}}
\newcommand{\Pm}{{p}_{-}}

\begin{document}

\title{Modulational instability of Rossby and drift waves and generation of zonal jets}
\author{Colm Connaughton}
\email{connaughtonc@gmail.com}
\affiliation {Centre for Complexity Science, University of Warwick, Gibbet Hill
Road, Coventry CV4 7AL, UK}
\affiliation{Mathematics Institute, University of Warwick, Gibbet Hill
Road, Coventry CV4 7AL, UK}
\author{Balasubramanya T. Nadiga}
\email{balu@lanl.gov}
\affiliation{Los Alamos National Laboratory, Los Alamos, New Mexico 87545, USA}
\author{Sergey Nazarenko}
\email{S.V.Nazarenko@.warwick.ac.uk}
\affiliation{Mathematics Institute, University of Warwick, Gibbet Hill
Road, Coventry CV4 7AL, UK}
\author{Brenda Quinn}
\email{B.E.Quinn@warwick.ac.uk}
\affiliation{Mathematics Institute, University of Warwick, Gibbet Hill
Road, Coventry CV4 7AL, UK}

\date{\today}

\begin{abstract}
We study the modulational instability of geophysical Rossby
and plasma drift waves within the Charney-Hasegawa-Mima (CHM) model
both theoretically, using truncated (four-mode and three-mode) models, and
numerically, using direct simulations of CHM equation in the Fourier space.
The linear theory \cite{GIL1974} predicts instability for any amplitude of the
primary wave. For strong primary waves
the most unstable modes are perpendicular to the primary wave, which correspond
to generation of a zonal flow if the primary wave is purely meridional.
For weaker waves, the maximum growth occurs for off-zonal inclined modulations.
For very weak primary waves the unstable waves are close to being in three-wave
resonance with the primary wave.
The nonlinear theory \cite{MN1994} predicts that the zonal flows generated by the linear
instability experience pinching into narrow zonal jets.
Our numerical simulations confirm the theoretical predictions of the linear theory
as well as of the nonlinear pinching. We find that, for strong primary waves,
these narrow zonal
jets further roll up into Karman-like vortex streets. On the other hand, for 
weak primary waves,
the growth of the unstable mode reverses and the system oscillates  between
a dominant jet and a dominate primary wave.  The 2D vortex streets appear
to be more stable than purely 1D zonal jets, and their zonal-averaged speed can reach amplitudes much stronger than is allowed
by the  Rayleigh-Kuo instability criterion for the 1D case. We find that the truncation models
work well for both the linear stage and and often even for the medium-term nonlinear behavior.
In the long term, the system transitions to turbulence helped by the vortex-pairing instability
(for strong waves) and by the resonant wave-wave interactions (for weak waves).
\end{abstract}

\pacs{PACS go here}

\maketitle

\section{\label{sec-intro} Introduction and Motivation}
Zonal flows are prominent features in the atmospheres of giant planets such as Jupiter and Saturn \cite{SIM1999,SLS2000,GNHS2004},
the Earth's atmosphere \cite{LEW1988} and its oceans \cite{GNHS2004,MMNS2008}.
Geophysical jets can regulate small-scale turbulence and transport processes via, for example, the ``Barotropic Governor''
mechanism \cite{JAM1987}. Zonal flows are also important in plasma turbulence of
fusion devices \cite{DIIH2005}. There, they can also regulate the turbulence 
and suppress transport via a drift-wave/zonal-flow
feedback loop  \cite{BNZ1990,BNZ1990B}. The latter process is presently considered the main mechanism
for the Low-to-High (LH) confinement transitions in tokamaks discovered in 
\cite{WAG1982}, - an effect  which is crucial for the success of future fusion 
devices.

Two main zonal flow generation scenarios have been discussed in the literature.
According to the first scenario zonal flows are generated via an inverse energy cascade, which could
be local or nonlocal \cite{BNZ1990,BNZ1990B}. The mechanism for such an inverse cascade is
similar to that of 2D Navier-Stokes turbulence \cite{KRA1967}, but
the presence of the beta-effect makes this cascade anisotropic. This leads to a 
preferential transfer of energy into zonal flows at large scales rather than 
into round vortices as would be the case in Navier-Stokes turbulence.
The beta-effect leads to three-wave resonant interactions which preserve
an additional (to the energy and the potential enstrophy) quadratic invariant, - zonostrophy
\cite{BNZ1991,BAL1991,BAL1997}. Application of the standard Fj{\o}rtoft argument to the three
invariants, the energy,  potential enstrophy and zonostrophy lead to the conclusion
that the energy can only be transferred to large zonal scales \cite{BNZ1991}.
This statistical argument is explained in detail in \cite{NQ2009}.
The second mechanism of zonal flow generation, and the principle topic of this
article, is via modulational instability of a primary meridional Rossby/drift 
wave.  In practice, such  primary waves are themselves the result of an 
instability (typically the baroclinic instability in GFD or the 
ion-temperature-gradient instability in tokamaks)
\cite{LOR1972,GIL1974,ML1980,MN1994,SDS2000,OPSSS2004,SK2008}.

These mechanisms are unlikely to be exclusive in practice and both may coexist
under some conditions. The extent to which one mechanism dominates over the
other is determined by the parameter regime and configurational details.
If the parameter regime were to be such that the baroclinic instability 
resulted in meridional Rossby waves, zonal flows would presumably result from 
the MI mechanism, whereas if the parameter regime
were to be such that the baroclinic instability resulted in more isotropic 
eddies at the Rossby deformation radius, the cascade scenario would likely be 
more relevant.  In our purely barotropic model, these effects are modeled by
the initial condition. A narrow initial spectrum of the waves and large 
initial amplitude promotes the modulational instability mechanism
leading to fast zonal flow generation  bypassing the turbulent cascade stages.
On the other hand, for broad initial spectra, the cascade scenario is
likely to be more relevant.
There is an analogy with the turbulence of surface gravity
waves on water where the inverse cascade and modulational (Benjamin-Fair) 
instability \cite{BF1967} can compete with each other in the generation of 
long waves \cite{OOSB2001}.
A quantitative measure, called the Benjamin-Feir index, was suggested to
estimate probability for triggering the modulational instability \cite{OOSB2001,JAN2003}.
Developing a similar approach for the Rossby/drift wave system would also be
useful. However, we will leave this interesting subject for future
studies, and in the present paper we will only be concerned with the
modulational instability of a monochromatic wave.

We will start by revisiting the linear theory of the modulational
instability which was first analysed by Loretz \cite{LOR1972} and then treated 
in great detail in a beautiful
paper by Gill \cite{GIL1974}. Using numerical and semi-analytical calculations
we highlight the most important properties of Gill's theory.
In particular, we will see how the character of instability changes
with the strength of the carrier wave: from being the classical
hydrodynamic instability of the sinusoidal (Kolmogorov) shear flow for large amplitudes
\cite{AM1960} to becoming a (three-wave) decay instability
of weakly nonlinear waves for small amplitudes \cite{SG1969}.
We will also study the effect of the finite Rossby/Larmor radius
on the instability.

We will then proceed to study the nonlinear stage of the modulational 
instability with direct numerical simulations (DNS), comparing them with
the solutions of the four-mode truncated (4MT) and the three-mode truncated 
(3MT) systems.
We find that at the nonlinear stage, for strong primary waves,
the growth of the zonal mode deviates from the truncated dynamics and
the zonal flow tends to focus into narrow jets, as was theoretically predicted 
in \cite{MN1994}.
These zonal jets subsequently become unstable and acquire
the interesting two-dimensional structure of a double (Karman-like)
vortex street. The vortex street appears to be more stable than
 a plane parallel shear flow with the same zonal profile \cite{MCW2006}
and persists for a relatively long time until (possibly due to dissipation)
a vortex pairing instability sets in and triggers a transition to 
turbulence \cite{MCW2006}.
As the nonlinearity of the primary wave is decreased we find that there is 
a level of nonlinearity below which this sequence of events changes. For 
sufficiently weak primary waves, the jet strength reaches a maximum which is 
still stable. After this maximum is reached, the jet amplitude starts 
decreasing again, continuing to follow the truncated dynamics, and avoids the 
roll-up into vortices. This reversal of 
the jet growth, particularly the maximum jet strength, is well 
predicted by nonlinear oscillatory solutions of the 4MT, and often by the 3MT,
equations. 
The latter are relevant for non-degenerate (in a sense which we shall explain)
 resonant wave triads.
Once the full system deviates from the solutions of the truncated system, as it
inevitably does, it sometimes continues to exhibit oscillatory
behaviour for a while in the weak nonlinearity cases. These  subsequent
oscillations have different periods, however, and are often rather irregular.

Along the way, we will examine the relative performance of the 3MT vs 4MT models
thereby clarifying possible confusions on whether the principal mechanism
of the modulational instability is 3-wave or 4-wave.

%These truncations predict well the linear instability stage, and sometimes

\section{\label{model} The model}

Geophysical and plasma zonal flows are often mentioned together because of the
same simplified nonlinear PDE which was suggested for their description, namely, the Charney-Hasegawa-Mima (CHM) equation \cite{CHA1949,HM1978}:
\begin{equation}
\label{eq-CHM}
\partial_t\left(\Delta \psi - F \psi \right) + \beta \partial_x \psi +  J\left[\psi,\Delta \psi\right]= \nu_n (-\Delta)^n \psi,
\label{chm}
\end{equation}
where $\psi$ is the streamfunction, $F=1/\rho^2$ with $\rho$ being the deformation radius in the GFD context and the ion Larmour radius in the plasma context, $\beta$ is a constant proportional to the gradient of the horizontal rotation frequency or of the plasma
density in the GDF and plasma contexts respectively.
 We introduced
notation for the Jacobean operator,
\begin{equation}
J\left[f,g\right] = (\partial _x f)(\partial_y g) - (\partial _y f)(\partial_x g).
\end{equation}
In the GFD context, the $x$-axis is in the west-east and the $y$-axis
is along the south-north directions respectively.
In plasmas, the $y$-axis is along the plasma density gradient and the $x$-axis is, of course, transverse to this direction.
Also, keeping in mind our numerical simulations which will be described below,
we introduced to the right-hand side (RHS) a hyperviscous dissipation of some degree
$n \ge 2$ (a positive integer) and a small positive coefficient $\nu_n$.

Introducing the Fourier transform of the streamfunction,
$\psi_\vv{k} = \int \psi(\vv{x}) e^{-i (\vv{k} \cdot \vv{x})} \, d \vv{x}$,
the CHM equation, Eq.~(\ref{eq-CHM}), ignoring the hyper-viscosity term for now, is equivalent to the following:
\begin{eqnarray}
& &\partial_t \psi_\vv{k} = + i\, \w_\vv{k}\, \psi_\vv{k} \nonumber \\
&+& \frac{1}{2} \sum_{\vv{k}_1, \vv{k}_2} T(\vv{k},\vv{k}_1,\vv{k}_2)\,
 \psi_{\vv{k}_1}\, \psi_{\vv{k}_2}\, \delta( {\vv{k} - \vv{k}_1 + \vv{k}_2}),
  \label{eq-CHMk}
\end{eqnarray}
where
\begin{equation}
\label{eq-RossbyDispersion}
\w_\vv{k} = -\frac{\beta k_x}{k^2 + F},
\end{equation}
\begin{equation}
\label{eq-CHMInteractionCoefficient}
T(\vv{k},\vv{k}_1,\vv{k}_2) = -\frac{\left(\vv{k}_1 \times \vv{k}_2\right)_z (k_1^2-k_2^2)}{k^2 + F}
\end{equation}
and $\vv{k}=(k_x,k_y)$ and $k = \left| \vv{k} \right|$. The system clearly supports linear waves, known as Rossby or drift waves, in the GFD and plasma contexts respectively. They have the anisotropic dispersion relation given by Eq.~(\ref{eq-RossbyDispersion}). The structure of the nonlinear interaction, Eq.~(\ref{eq-CHMInteractionCoefficient}), is such that the nonlinear term vanishes for a monochromatic wave. Hence Rossby waves are actually exact solutions of the full CHM equation. Much of this article will focus on the stability properties of these solutions.

Originally, the waves arise due to a primary instability, e.g. the baroclinic instability
in GFD \cite{MCW2006}
or the ion-temperature-gradient instability (ITG) in fusion plasmas \cite{RS1961}. The instability
is not included in the CHM equation, and it could be modeled by simulating CHM with
an initial condition or introducing a linear forcing term on the RHS mimicking the
linear instability (this would not take into account the nonlinear mechanisms in the
wave forcing). It is interesting that the GFD-plasma analogy extends to the instabilities
too in that the most unstable mode is "meridional" (i.e. along the $x$-axis)
 and concentrated at the scales of the
order of $\rho$. Thus, in most of our considerations below we will consider the initial (primary) wave which is purely meridional.

%For most of what follows, it will be convenient to deal with Eq.~(\ref{eq-CHMk}) in the %interaction representation. Introduce $\Psi_\vv{k}(t) = \psi_\vv{k}(t) {\rm e}^{-i\, %\w_\vv{k}\, t}$. In terms of $\Psi_\vv{k}$, Eq.~(\ref{eq-CHMk}) is
%\begin{eqnarray}
%\nonumber \partial_t \Psi_\vv{k} &=& \frac{1}{2} \sum_{\vv{k}_1, \vv{k}_2} %T(\vv{k},\vv{k}_1,\vv{k}_2)\,
% \Psi_{\vv{k}_1}\, \Psi_{\vv{k}_2}\, \Delta_{\w_\vv{k},\w_{\vv{k}_1},\w_{\vv{k}_2}} \\
%& &\times \delta( {\vv{k} - \vv{k}_1 + \vv{k}_2}), \label{eq-CHMk2}
%\end{eqnarray}
%where
%\begin{displaymath}
%\Delta_{\w_\vv{k}\ \w_{\vv{k}_1}\ \w_{\vv{k}_2}} = {\rm e}^{i\,(\w_\vv{k} -\w_{\vv{k}_1} %-\w_{\vv{k}_2})\,t }.
%\end{displaymath}

\section{\label{sec-truncatedSytems} Spectral Truncations}

We shall use spectral trunctions of Eq.~(\ref{eq-CHMk}) in our study of the stability properties of Rossby waves. They provide approximations of an intermediate degree of complexity between monochromatic waves and the full PDE. At this
stage, such truncations should be viewed as ad-hoc since, in reality, all 
triads are coupled together in  Eq.~(\ref{eq-CHMk}). Their usefulness will
determined by comparision with DNS solutions of the full system,
Eq.~(\ref{eq-CHMk}). We shall consider two natural truncations: the 3-mode 
truncation and the 4-mode truncation.

\subsection{3--Mode Truncation (3MT) }
The simplest such truncation is to restrict the RHS of Eq.~(\ref{eq-CHMk}) to a single triad containing only 3 modes which we shall denote by $\vv{p}$, $\vv{q}$ and $\vv{p}_-=\vv{p}-\vv{q}$. We construct the truncated equations be taking each wave vector in the triad in turn and assigning it to be $\vv{k}$ in Eq.~(\ref{eq-CHMk}), enumerating all ways of assigning the others and their negatives to $\vv{k}_1$ and $\vv{k}_2$ on the RHS and neglect all terms which involve $\psi_{\vv{k}}$'s within the triad. %Table~\ref{tab-3modeTruncation} illustrates this procedure.
Since $\psi_\vv{k}$ is the Fourier transform of a real field, $\psi_{-\vv{k}} = \bar{\psi_\vv{k}}$. We then arrive at the following equations for the 3-mode truncation:
\begin{eqnarray}
\nonumber \partial_t \psi_\vv{p} + i \w_\vv{p}\,\psi_\vv{p}&=& T(\vv{p},\vv{q},\vv{p}_-)\, \psi_\vv{q} \psi_\vv{p_-}\\
\label{eq-3ModeTruncation}\partial_t \psi_\vv{q} + i \w_\vv{q}\,\psi_\vv{q} &=& T(\vv{q},\vv{p},-\vv{p}_-) \psi_\vv{p} \overline{\psi}_\vv{p_-}\\
\nonumber \partial_t \psi_{\vv{p}_-} + i \w_\vv{p_-}\,\psi_\vv{p_-}&=& T(\vv{p}_-,\vv{p},-\vv{q}) \psi_\vv{p}\, \overline{\psi}_\vv{q}.
\end{eqnarray}
For most of what follows, it will be convenient to deal with Eqs.~(\ref{eq-3ModeTruncation}) in the interaction representation. Introduce $\Psi_\vv{k}(t) = \psi_\vv{k}(t) {\rm e}^{-i\, \w_\vv{k}\, t}$. In terms of $\Psi_\vv{k}$, Eqs.~(\ref{eq-3ModeTruncation}) become
\begin{eqnarray}
\nonumber \partial_t \Psi_\vv{p}&=& T(\vv{p},\vv{q},\vv{p}_-)\, \Psi_\vv{q} \Psi_\vv{p_-} {\rm e}^{i\, \Delta_-\, t}\\
\label{eq-3ModeTruncationMinus}\partial_t \Psi_\vv{q} &=& T(\vv{q},\vv{p},-\vv{p}_-) \Psi_\vv{p} \overline{\Psi}_\vv{p_-} {\rm e}^{-i\, \Delta_-\, t}\\
\nonumber \partial_t \Psi_{\vv{p}_-}&=& T(\vv{p}_-,\vv{p},-\vv{q}) \Psi_\vv{p}\, \overline{\Psi}_\vv{q} {\rm e}^{-i\, \Delta_-\, t},
\end{eqnarray}
where
\begin{displaymath}
\Delta_- = \w_\vv{p} - \w_\vv{q} -\w_\vv{p_-}.
\end{displaymath}
A similar set of equations can be derived for the other natural triad, $(\vv{p},-\vv{q}, \vv{p}_+)$ where $\vv{p}_+=\vv{p}+\vv{q}$:
\begin{eqnarray}
\nonumber \partial_t \Psi_\vv{p}&=& T(\vv{p},-\vv{q},\vv{p}_+)\, \overline{\Psi}_\vv{q} \Psi_\vv{p_-} {\rm e}^{i\, \Delta_+\, t}\\
\label{eq-3ModeTruncationPlus}\partial_t \Psi_\vv{q} &=& T(\vv{q},-\vv{p},\vv{p}_+) \overline{\Psi}_\vv{p} \Psi_\vv{p_+} {\rm e}^{i\, \Delta_+\, t}\\
\nonumber \partial_t \Psi_{\vv{p}_+}&=& T(\vv{p}_+,\vv{p},\vv{q}) \Psi_\vv{p}\, \Psi_\vv{q} {\rm e}^{-i\, \Delta_+\, t},
\end{eqnarray}
where
\begin{displaymath}
\Delta_+ = \w_\vv{p} + \w_\vv{q} -\w_\vv{p_+}.
\end{displaymath}
If $\Delta_+=0$, the triad is exactly resonant. Then
Eqs.~(\ref{eq-3ModeTruncationPlus}) form an exactly integrable set of 
equations which have been extensively studied 
\cite{KL2007,BK2009}.

\subsection{4--Mode Truncation (4MT) }
The 4MT model retains both triads, 
$(\vv{p},\vv{q},\vv{p}_+)$ and $(\vv{p},-\vv{q},\vv{p}_-)$, where 
$\vv{p}_\pm = \vv{p} \pm \vv{q}$ mentioned above. A thorough analysis of
the 4--mode truncation for the Generalised Hasegawa--Mima equation in the
case of weak nonlinearity was presented in \cite{MRD2001}. The truncated 
equations combine Eqs.~(\ref{eq-3ModeTruncationMinus}) and Eqs.~(\ref{eq-3ModeTruncationPlus}):
\begin{eqnarray}
\nonumber \partial_t \Psi_\vv{p}&=& T(\vv{p},\vv{q},\vv{p}_-)\, \Psi_\vv{q} \Psi_\vv{p_-} {\rm e}^{i\, \Delta_-\, t}\\
\nonumber&+& T(\vv{p},-\vv{q},\vv{p}_+)\, \overline{\Psi}_\vv{q} \Psi_\vv{p_-} {\rm e}^{i\, \Delta_+\, t}\\
\nonumber \partial_t \Psi_\vv{q} &=& T(\vv{q},\vv{p},-\vv{p}_-) \Psi_\vv{p} \overline{\Psi}_\vv{p_-} {\rm e}^{-i\, \Delta_-\, t}\\
\label{eq-4ModeTruncation}&+& T(\vv{q},-\vv{p},\vv{p}_+) \overline{\Psi}_\vv{p} \Psi_\vv{p_+} {\rm e}^{i\, \Delta_+\, t}\\
\nonumber \partial_t \Psi_{\vv{p}_-}&=& T(\vv{p}_-,\vv{p},-\vv{q}) \Psi_\vv{p}\, \overline{\Psi}_\vv{q} {\rm e}^{-i\, \Delta_-\, t}\\
\nonumber \partial_t \Psi_{\vv{p}_+}&=& T(\vv{p}_+,\vv{p},\vv{q}) \Psi_\vv{p}\, \Psi_\vv{q} {\rm e}^{-i\, \Delta_+\, t}.
\end{eqnarray}
Strictly speaking, the chosen four modes ($\psi_{0}, \psi_{\vv{q}}, \psi_{+}$
and $ \psi_{-}$) are coupled to further modes and do not form a closed system.
Indeed, even the linear problem closes only if all the satellites $\pm \vv{q} + m \vv{p}$
($m $ is a positive or negative integer) are included \cite{GIL1974}. However, in considering
the linear instability it is traditional to truncate the system to the four modes
only with a justification that the higher order satellites are
less excited in the linear eigenvectors, which turns out to be a very good approximation
if $M \ll 1$ and quite reasonable for $M \sim 1$ and larger \cite{GIL1974}. In this paper we will test predictions of the 4MT system, both linear
and nonlinear, against DNS of the full system.

\subsection{Nonlinearity parameter $M$}

In studying instability of a primary monochromatic wave, we will follow the convention that the wavenumber of this wave is ${\bf p}$ and denoted its 
amplitude by $\Psi_\vv{p}$.
The character of the instability is greatly influenced by the initial amplitude of the
primary wave,  $\Psi_{0}$ =  $\Psi_\vv{p}|_{t=0}$ \cite{GIL1974}.
Following Gill \cite{GIL1974}, we introduce
the nonlinearity parameter
\begin{equation}
M = \frac{\Psi_0 p^3}{\beta}.
\label{Mpar}
\end{equation}
$M$  measures the relative strength of the linear and nonlinear terms at the scale of the carrier wave.
$M\gg 1$ corresponds to the case where the $\beta$-effect plays no role.
For $F=0$ this case reduces to the Euler equation limit and the well-studied instability of the plane parallel sinusoidal shear flow
known as Kolmogorov flow \cite{AM1960}. Note that most papers on the modulational instability within
the plasma context have dealt only with this limit (e.g. \cite{SDS2000,OPSSS2004}).
Case
$M\ll 1$ corresponds to the weak nonlinearity limit dominated by resonant wave triads. In this case
the four constituent modes (carrier wave, modulation and two satellites) can be split
into two coupled triads which produce independent contributions to the instability \cite{GIL1974}.
The instability associated with a single triad is known as the decay instability \cite{SG1969}.
 The condition $M \sim 1$ defines the Rhines scale $k_r$ which
marks a crossover from the hydrodynamic vortex  to the wave behavior \cite{RHI1975}.

\section{Decay Instability of a Rossby Wave}
\label{sec-decayInstability}

\begin{figure*}
\includegraphics[width=14.0cm,angle=-90.0]{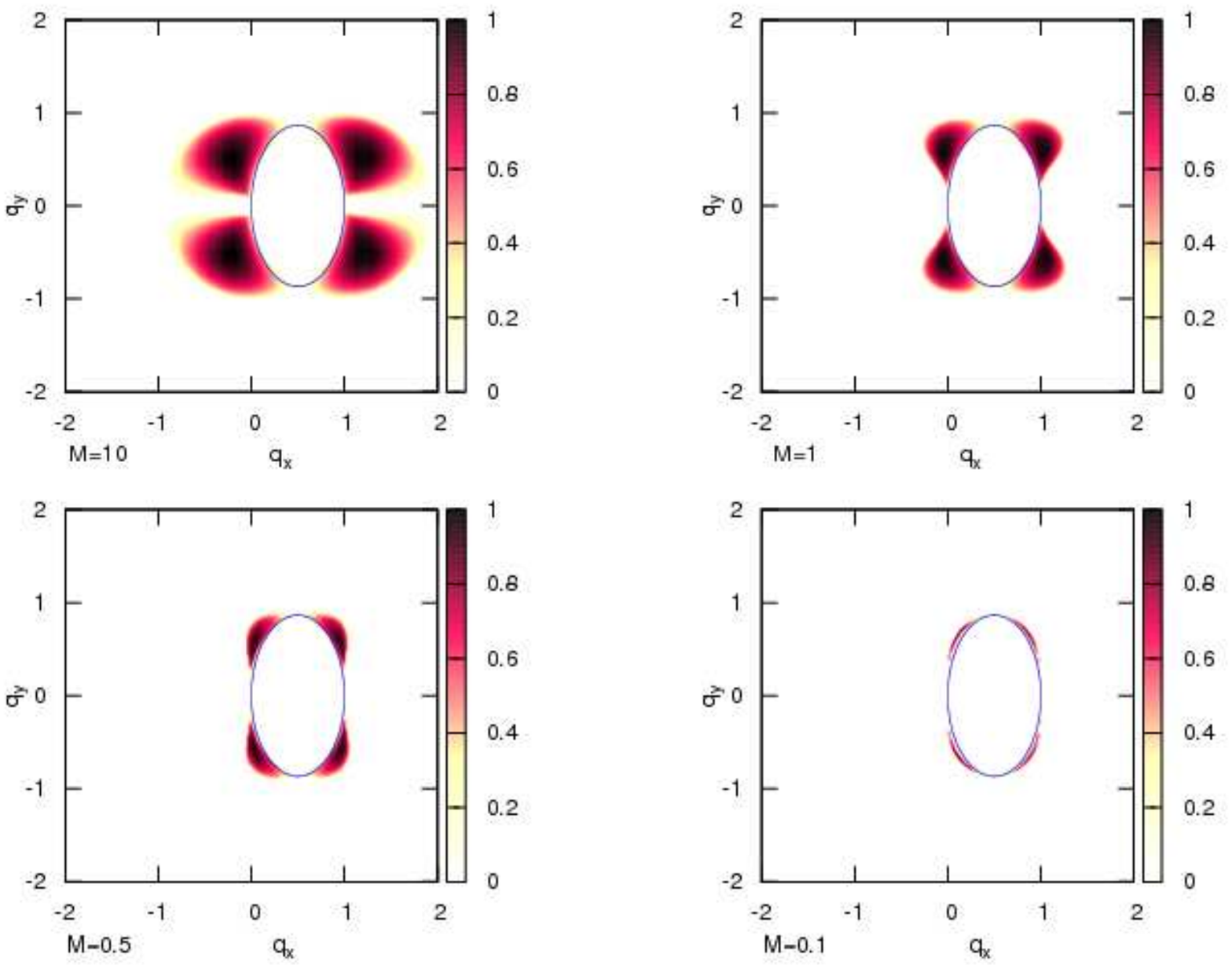}
\caption{\label{fig-DIqMaps-F_0} The growth rate of the decay instability (the negative imaginary part of the roots of Eq.(\ref{eq-decayDispersionReln2})) is plotted as a function of $\vv{q}$ for a fixed meridional carrier wave-vector, $\vh{p}=(1,0)$, for various values of the nonlinearity parameter  $M$. The set of unstable perturbations become concentrated on the resonant manifolds (blue lines) as the nonlinearity of the carrier wave is decreased. }
\end{figure*}

The decay instability is an instability of a carrier wave involving a pair of other modes (i.e. the primary wave decays into two
secondary waves, see e.g. \cite{SG1969}). We shall derive this instability from
the 3MT, Eqs.~(\ref{eq-3ModeTruncationMinus}). Introducing the vector notation 
$\vv{\Psi} = (\Psi_\vv{p},\Psi_\vv{q},\Psi_{\vv{p}_-})$, a monochromatic 
carrier wave is given by $\vv{\Psi}_0 = (\Psi_0,0,0)$ where $\Psi_0$ is a complex constant representing the amplitude of the initial carrier wave. This is an exact solution of Eqs.~(\ref{eq-3ModeTruncationMinus}). We consider the stability of this solution to small perturbations involving the modes $\vv{q}$ and $\vv{p}_-$ by taking $\vv{\Psi} = \vv{\Psi}_0 + \epsilon \vv{\Psi}_1$ with the perturbation given by $\vv{\Psi}_1 = (0, \widetilde{\psi}_\vv{q},\widetilde{\psi}_{\vv{p}_-})$. Linearisation yields the following equations at first order in $\epsilon$:
\begin{eqnarray}
\label{eq-linearised3ModeTrunction}\partial_t \widetilde{\psi}_\vv{q} &=& T(\vv{q},\vv{p},-\vv{p}_-)\, \Psi_0\,\overline{\widetilde{\psi}}_\vv{p_-} {\rm e}^{-i\, \Delta_-\, t}\, \\
\nonumber \partial_t \overline{\widetilde{\psi}}_{\vv{p}_-} &=& T(\vv{p}_-,\vv{p},-\vv{q})\, \overline{\Psi}_0\, \widetilde{\psi}_\vv{q}\, {\rm e}^{i\, \Delta_-\, t}.
\end{eqnarray}
We now seek harmonic solutions:
\begin{eqnarray*}
\widetilde{\psi}_\vv{q}(t) &=& A_\vv{q} {\rm e}^{-i\, \W_\vv{q}\,t}\\
\widetilde{\psi}_{\vv{p}_-}(t) &=& A_{\vv{p}_-} {\rm e}^{-i\, \W_{\vv{p}_-}\,t}.
\end{eqnarray*}
This requires $\overline{\W}_{\vv{p}_-} = -\W_\vv{q}+\Delta_-$. Solving Eqs.~(\ref{eq-linearised3ModeTrunction}) then reduces to finding solutions of the linear system
\begin{displaymath}
A\,\left( \begin{array}{c}A_\vv{q}\\\overline{A}_{\vv{p}_-}\end{array}\right) = 0
\end{displaymath}
 where
\begin{equation}
A = \left(
\begin{array}{cc}
-i \W_\vv{q} & T(\vv{q},\vv{p},-\vv{p}_-)\, \Psi_0\\
T(\vv{p}_-,\vv{p},-\vv{q})\, \overline{\Psi}_0 & i(-\W_\vv{q} + \Delta_-)
\end{array}
\right)
\end{equation}
To obtain non-trivial solutions, we require $\det A = 0$, which yields the dispersion relation:
\begin{equation}
\label{eq-decayDispersionReln1}
\W_\vv{q}(- \W_\vv{q}+\Delta_-) - T(\vv{q},\vv{p},-\vv{p}_-)\, T(\vv{p}_-,\vv{p},-\vv{q})\, \left|\Psi_0\right|^2 = 0.
\end{equation}
This has two roots, $\W_\vv{q}^\pm$ with corresponding eigenvectors:
\begin{equation}
\label{eq-3ModeEigenvector}
\left(
\begin{array}{c}
A_\vv{q}\\
A_{\vv{p}_-}
\end{array}
\right) = \left(
\begin{array}{c}
1\\
\frac{T(\vv{p}_-,\vv{p},-\vv{q})\, \Psi_0}{i\,(\overline{\W}_\vv{q} - \Delta_-)}
\end{array}
\right).
\end{equation}
Instability occurs when $\W_\vv{q}$ has a non-zero imaginary part. 
For an exactly  resonant triad, $\Delta_-=0$. For resonant triads, 
using Eq.~(\ref{eq-CHMInteractionCoefficient}) the roots of 
Eq.~(\ref{eq-decayDispersionReln1}) are
\begin{equation}
\label{eq-resonantDecayInstability}
\W_\vv{q} = \pm i \frac{\left|\Psi_0\right| \left| \vv{p}\times\vv{q}\right|}{\sqrt{(q^2+F)(p_-^2+F)}}\,\sqrt{(p^2-q^2)(p_-^2-p^2)}.
\end{equation}
%a well-known expression.
In this case, instability occurs if $q<p<p_-$.

Before investigating the non-resonant instability further, it is convenient to perform some rescalings.  The dimensionless carrier wave amplitude
will be given by $M$ defined in Eq.~(\ref{Mpar}).
We non--dimensionalise the other terms in Eq.~(\ref{eq-decayDispersionReln1}) as follows:
\begin{eqnarray*}
%\Psi_0 &\to& \frac{\beta M}{k^3},\\
\W&\to& \frac{\beta}{p} \W,\\
F&\to& p^2 F,\\
\vv{p}&\to p \vh{p},\\
\vv{q}&\to& s p \vh{q},
\end{eqnarray*}
where $\vh{p}=(\hat{p}_x, \hat{p}_y)$ and $\vh{q}=(\hat{q}_x,\hat{q}_y)$ are unit vectors pointing in the directions of $\vv{p}$ and $\vv{q}$ respectively. Eq.~(\ref{eq-decayDispersionReln1}) can then be re--arranged to the following form:
\begin{equation}
\label{eq-decayDispersionReln2}
\W(- \W+\hat{\Delta}_-) - M^2 T(s \vh{q},\vh{p},-\vh{p}_-)\, T(\vh{p}_-,\vh{p},-\vh{q}) = 0.
\end{equation}
where $\vh{p}_- = \vh{p} - s \vh{q}$ and $\hat{\Delta}_- = \w_\vh{p} - \w_{s\vh{q}} -\w_\vh{p_-}$. The two roots are
\begin{equation}
\W_\pm = \frac{1}{2}\left( \hat{\Delta}_- \pm\sqrt{(\hat{\Delta}_-)^2 - 4 M^2 T(s \vh{q},\vh{p},-\vh{p}_-)\, T(\vh{p}_-,\vh{p},-\vh{q})}\right)
\end{equation}
To have an instability we require
\begin{displaymath}
\hat{\Delta}_- < 2 M \sqrt{T(s \vh{q},\vh{p},-\vh{p}_-)\, T(\vh{p}_-,\vh{p},-\vh{q})}
\end{displaymath}
which demonstrates that the instability concentrates on the resonant manifold, $\hat{\Delta}_-=0$ as $M\to 0$. This is illustrated in Fig.~\ref{fig-DIqMaps-F_0}. The corresponding analysis for the triad $(\vv{p},-\vv{q},\vv{p}_+)$ produces identical surfaces reflected about the vertical axis reflecting the instability concentrating on the second resonant manifold, $\hat{\Delta}_+=0$. As $M\to 0$, these two surfaces become disjoint from each other except near the origin ${\bf q} =0$.

\section{\label{sec-linearStabilityAnalysis}Modulational Instability: linear  analysis}

\begin{figure*}
\includegraphics[width=17.0cm]{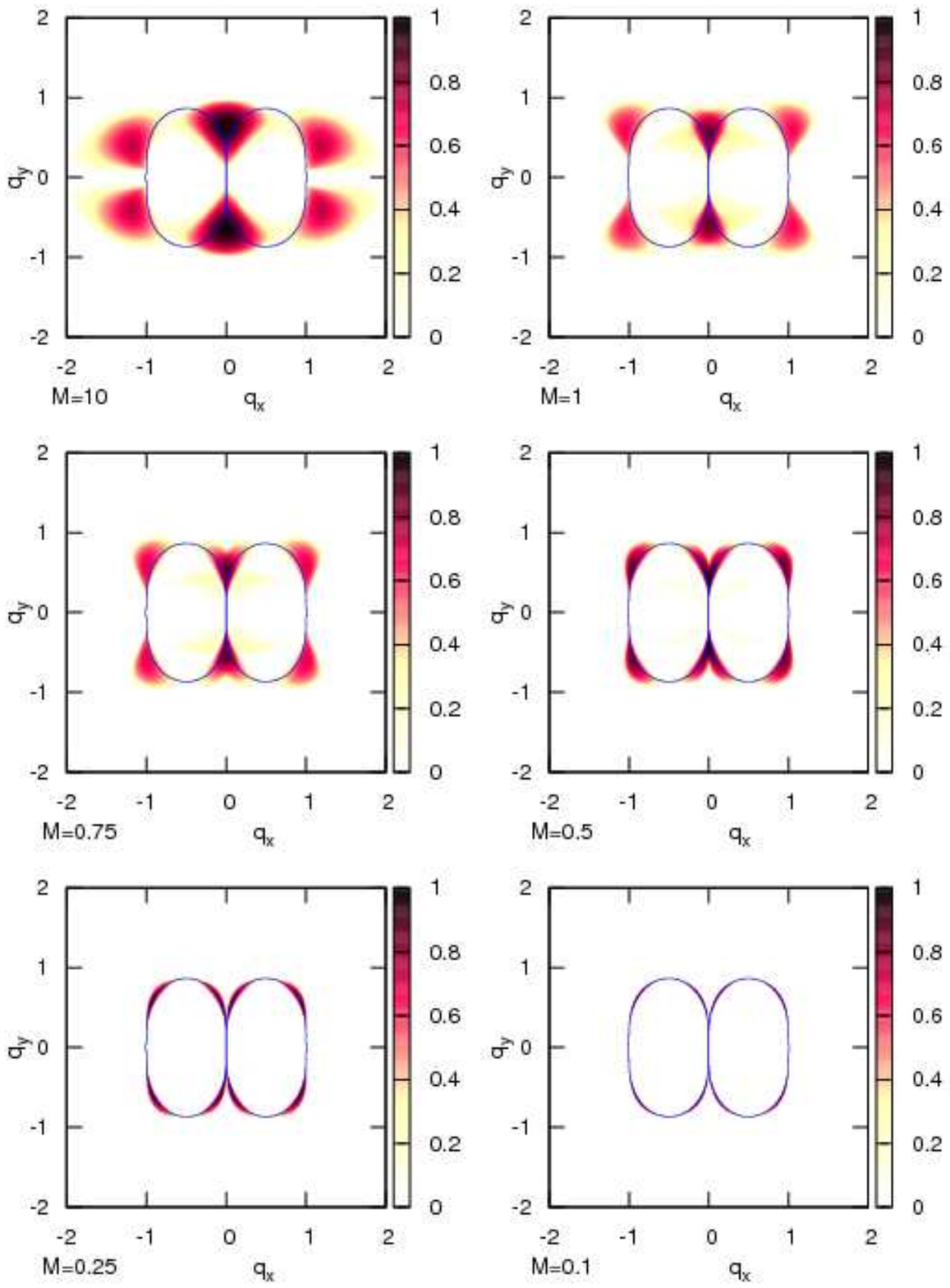}
\caption{\label{fig-collapseToResonantManifolds}  Growth rate of the modulational instability  (the negative imaginary part of the roots of Eq.(\ref{eq-nonDimensionalDispersion})) as a function of $\vv{q}$ for a fixed meridional carrier wave-vector, $\vv{p}=(1,0)$ and $F=0$. The values of the nonlinearity $M$ for the initial carrier wave are $M=10$ (Euler limit), $M=1$, $M=3/4$, $M=1/2$, $M=1/4$ and $M=1/10$ (weakly nonlinear limit). The set of unstable perturbations become concentrated on the resonant manifolds as the nonlinearity of the carrier wave is decreased.}
\end{figure*}

Let us now derive the modulational instability in the same way as we have done for the decay instability. The modulational instability is studied using the 
4MT. We begin by linearising Eqs.~(\ref{eq-4ModeTruncation}) about the pure 
carrier wave solution,
$\vv{\Psi}_0 = (\Psi_0,0,0,0)$ where $\Psi_0$ is a complex constant representing the amplitude of the initial carrier wave. We consider the stability of this solution to small perturbations involving the 3 modes $\vv{q}$, $\vv{p}_-$ and $\vv{p}_+$ by taking $\vv{\Psi} = \vv{\Psi}_0 + \epsilon \vv{\Psi}_1$ with the perturbation given by $\vv{\Psi}_1 = (0, \widetilde{\psi}_\vv{q},\widetilde{\psi}_{\vv{p}_+}, \widetilde{\psi}_{\vv{p}_-})$. Linearisation yields the following equations at first order in $\epsilon$:
\begin{eqnarray}
\nonumber \partial_t \widetilde{\psi}_\vv{q} &=& T(\vv{q},\vv{p},-\vv{p}_-)\, \Psi_0\,\overline{\widetilde{\psi}}_\vv{p_-} {\rm e}^{-i\, \Delta_-\, t} \\
\label{eq-linearised4ModeTrunction} & &+ T(\vv{q},-\vv{p},\vv{p}_+)\, \overline{\Psi_0}\,\widetilde{\psi}_\vv{p_+} {\rm e}^{i\, \Delta_+\, t} \\
\nonumber \partial_t \overline{\widetilde{\psi}}_{\vv{p}_+} &=& T(\vv{p}_+,\vv{p},\vv{q})\, \Psi_0\, \widetilde{\psi}_\vv{q}\, {\rm e}^{-i\, \Delta_+\, t}\\
\nonumber \partial_t \overline{\widetilde{\psi}}_{\vv{p}_-} &=& T(\vv{p}_-,\vv{p},-\vv{q})\, \overline{\Psi}_0\, \widetilde{\psi}_\vv{q}\, {\rm e}^{i\, \Delta_-\, t}.
\end{eqnarray}
We again seek harmonic solutions:
\begin{eqnarray*}
\widetilde{\psi}_\vv{q}(t) &=& A_\vv{q} {\rm e}^{-i\, \W_\vv{q}\,t}\\
\widetilde{\psi}_{\vv{p}_+}(t) &=& A_{\vv{p}_+} {\rm e}^{-i\, \W_{\vv{p}_+}\,t}\\
\widetilde{\psi}_{\vv{p}_-}(t) &=& A_{\vv{p}_-} {\rm e}^{-i\, \W_{\vv{p}_-}\,t}.
\end{eqnarray*}
This requires requires $\W_{\vv{p}_+} = \W_\vv{q}+\Delta_+$ and $\overline{\W}_{\vv{p}_-} = -\W_\vv{q}+\Delta_-$. Solving Eqs.~(\ref{eq-linearised4ModeTrunction}) then reduces to finding solutions of the linear system
\begin{displaymath}
A\,\left( \begin{array}{c}A_\vv{q}\\ A_{\vv{p}_+}\\ \overline{A}_{\vv{p}_-}\end{array}\right) = 0
\end{displaymath}
 where
\begin{equation}
A = \left(
\begin{array}{ccc}
i \W_\vv{q} & T(\vv{q},-\vv{p},\vv{p}_+)\, \overline{\Psi}_0 & T(\vv{q},\vv{p},-\vv{p}_-)\, \Psi_0 \\
T(\vv{p}_+,\vv{p},\vv{q})\, \Psi_0 &i(\W_\vv{q} + \Delta_+) & 0\\
T(\vv{p}_-,\vv{p},-\vv{q})\, \overline{\Psi}_0 & 0 & -i(-\W_\vv{q} + \Delta_-)
\end{array}
\right)
\end{equation}
Setting $\det A = 0$ yields a cubic dispersion relation:
\begin{eqnarray}
\label{eq-MIDispersionReln1}& &\W_\vv{q}(\W_\vv{q}+\Delta_+)( -\W_\vv{q}+\Delta_-)\\
\nonumber& & + T(\vv{q},-\vv{p},\vv{p}_+)\, T(\vv{p}_+,\vv{p},\vv{q})\, \left|\Psi_0\right|^2 ( -\W_\vv{q}+\Delta_-)\\
\nonumber& & - T(\vv{q},\vv{p},-\vv{p}_-)\, T(\vv{p}_-,\vv{p},-\vv{q})\, \left|\Psi_0\right|^2 ( \W_\vv{q}+\Delta_+)  = 0.
\end{eqnarray}
The corresponding eigenvectors are given by
\begin{equation}
\label{eq-4ModeEigenvector}
\left(
\begin{array}{c}
A_\vv{q}\\
A_{\vv{p}_+}\\
A_{\vv{p}_-}
\end{array}
\right) = \left(
\begin{array}{c}
1\\
\frac{T(\vv{p}_+,\vv{p},\vv{q})\, \Psi_0}{-i\,(\W_\vv{q} + \Delta_+)}\\
\frac{T(\vv{p}_-,\vv{p},-\vv{q})\, \Psi_0}{i\,(\overline{\W}_\vv{q} - \Delta_-)}
\end{array}
\right).
\end{equation}
This derivation holds for any system with a quadratic nonlinearity.
Using Eq.~(\ref{eq-CHMInteractionCoefficient}) and performing some algebra we recover the usual form of the dispersion relation specific to the CHM equation \cite{GIL1974} (see also \cite{LOR1972,MN1994,SDS2000,OPSSS2004}):
\begin{widetext}
\begin{equation}
\label{eq-dispersion}
(q^2 + F)\W +\beta q_x + \left|\Psi_0\right|^2 \left|{\bf p} \times\Q\right|^2(p^2-q^2)
\left[ \frac{\Pp^2-p^2}{(\Pp^2+F)(\W+\w) +\beta{\Pp}_x} - \frac{\Pm^2-p^2}{(\Pm^2+F)(\W-\w) +\beta{\Pm}_x}\right] =0
\end{equation}
\end{widetext}
This can be solved numerically, and sometimes 
analytically, for a given set of parameters to determine $\Omega$.
For the purposes of easy comparison of different values of $M$, we nondimensionalise as before.
%in Sec.~\ref{sec-decayInstability}.
 The result is
\begin{equation}
\label{eq-nonDimensionalDispersion}
(s^2 + F)\W + s \hat{q}_x + M^2 s^2(1-s^2)\left|\vh{p}\times\vh{q}\right|^2
\left[ T^+ - T^-\right] =0,
\end{equation}
where
\begin{equation}
T^{\pm}=\frac{\left|\vh{p}\pm s\vh{q}\right|^2 - 1}{(\left|\vh{p}\pm s\vh{q}\right| +F)(-\frac{\hat{p}_x}{1+F} \pm \W) + \hat{p}_x \pm s \hat{q}_x}.
\end{equation}
The roots of this equation are controlled by five parameters, $M$, $F$, $s$, $\theta_p$ and $\theta_q$ where $\theta_p$ and $\theta_q$ are the angles between the x-axis and the carrier wave-vector and perturbation wave-vector respectively.The structure of the instability is strongly dependent on the value of $M$. This is shown in Fig.~\ref{fig-collapseToResonantManifolds}. We see that the modulational instability is, in some sense, the nonlinear sum of the decay instabilities for the two triads,
 and we will clarify this issue in the next section.

\section{Comparison of the 3MT and the 4MT models  with DNS of the Full CHM system}

There is sometimes confusion in the literature, perhaps partially semantic, on 
whether the
modulational instability of the Rossby and drift waves is governed by 3-wave 
or 4-wave
interactions. Here we will clarify this issue.
It was shown by Gill \cite{GIL1974} that as $M\to 0$, the modulational instability
obtained within the 4MT model  localises on the two resonant manifolds for the two triads $\Delta_+=0$ and $\Delta_-=0$. Since these two curves are mostly
disjoint from each other (except for the origin), in the weakly nonlinear limit, the modulational instability is just a simple sum of the two decay instabilities.
Namely, the two unstable eigenvectors of the instability of the 4MT will coincide with the  eigenvectors of the two respective branches
of the decay instability (i.e. the fourth mode in such 4MT eigenvectors will have zero amplitude).
In particular, the maximum growth rates of the 3MT and 4MT instabilities become identical.
For larger values of $M$, the growth rate of the modulational instability is typically larger than that of the corresponding decay instability.

However, for the typical setup where the primary wave is purely meridional and the modulation is purely zonal,
the wavevector ${\bf q}$ is equally close to both branches of the three-wave resonant manifold.
This is because these resonant manifolds cross zero of the ${\bf q}$-space in the direction of the $q_y$-axis, i.e. in the zonal
direction. Thus, the above speculations about the equivalence of the 3MT and the 4MT for weak waves
may not apply to such a setup. Therefore, let us consider the weakly nonlinear case ($M=0.1$) and
examine predictions of the
3MT and the 4MT models and compare them to DNS of the full CHM system in the
following two cases:
\begin{itemize}
\item[(A)]
the primary wave is purely meridional,  ${\bf p} = (10,0)$, and the modulation is purely zonal,
 ${\bf q} = (0,1)$; and
\item[(B)]
the primary wave is purely meridional,  ${\bf p} = (10,0)$, and the modulation is off-zonal. We take
${\bf q} = (9,6)$. This is close to the maximum of the most unstable mode on the resonant curve
We cannot select the exact value of the maximum because the discrete
wave numbers in the periodic box do not typically lie exactly on the resonant
manifolds.  This is a subtlety which can have strong implications for 
numerical simulations of very weakly nonlinear regimes \cite{CNP01} which we 
have been careful to avoid here.
\end{itemize}

\begin{figure}
\includegraphics[width=7.0cm,angle=-90.0]{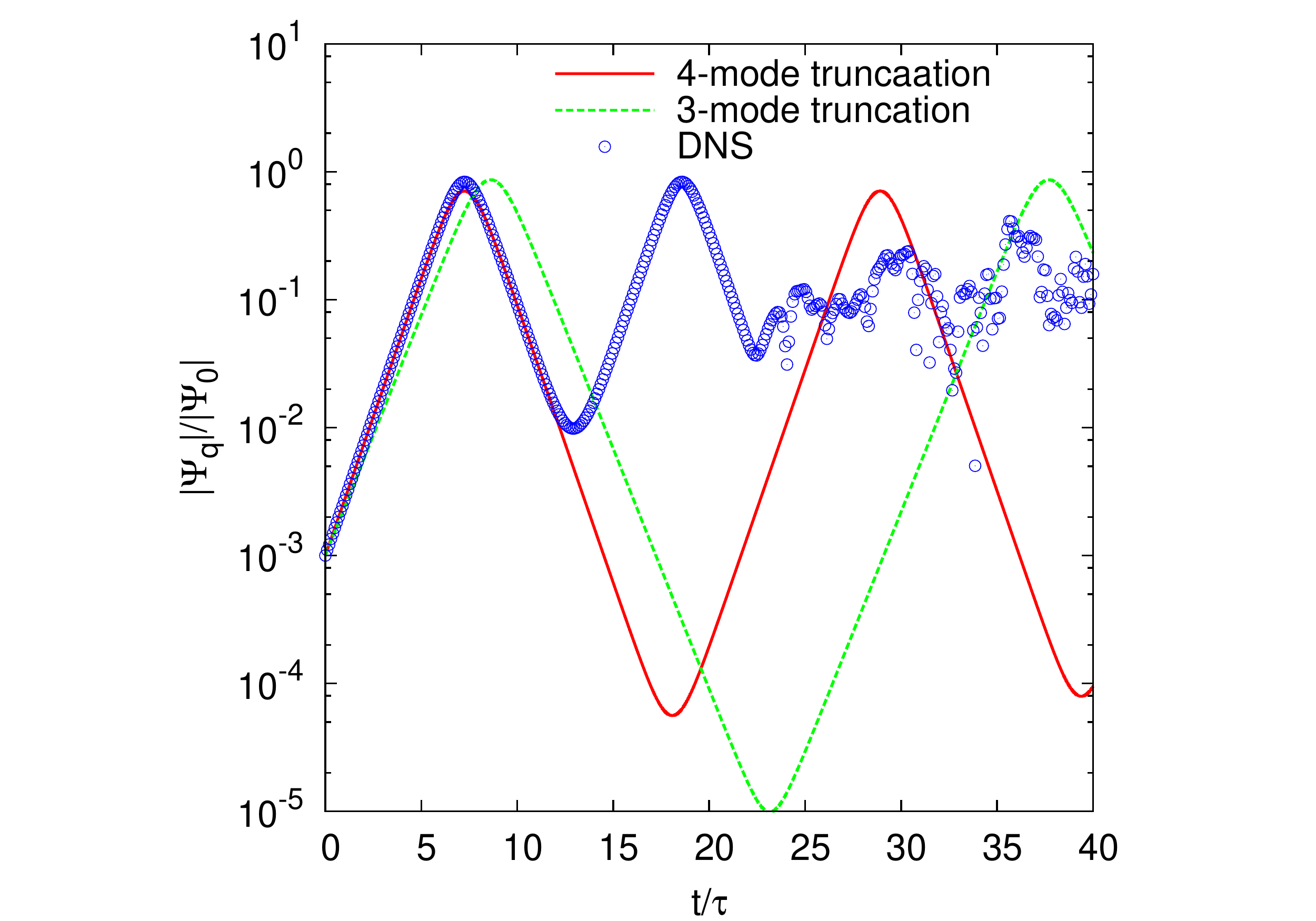}
\caption{\label{fig-compareMIandDILongTimes}
Amplitude of the zonal mode with wavenumber ${\bf q}$ for $M=0.1, \; {\bf p} = (10,0)$
obtained from DNS and from solutions of 3MT and 4MT models. Case (i): purely zonal modulations, ${\bf q}=(0,1)$.}
%Comparison of Modulational and decay instabilies for the case (i):  M=0.1, purely zonal modulations.}
\end{figure}
\begin{figure}
\includegraphics[width=7.0cm,angle=-90.0]{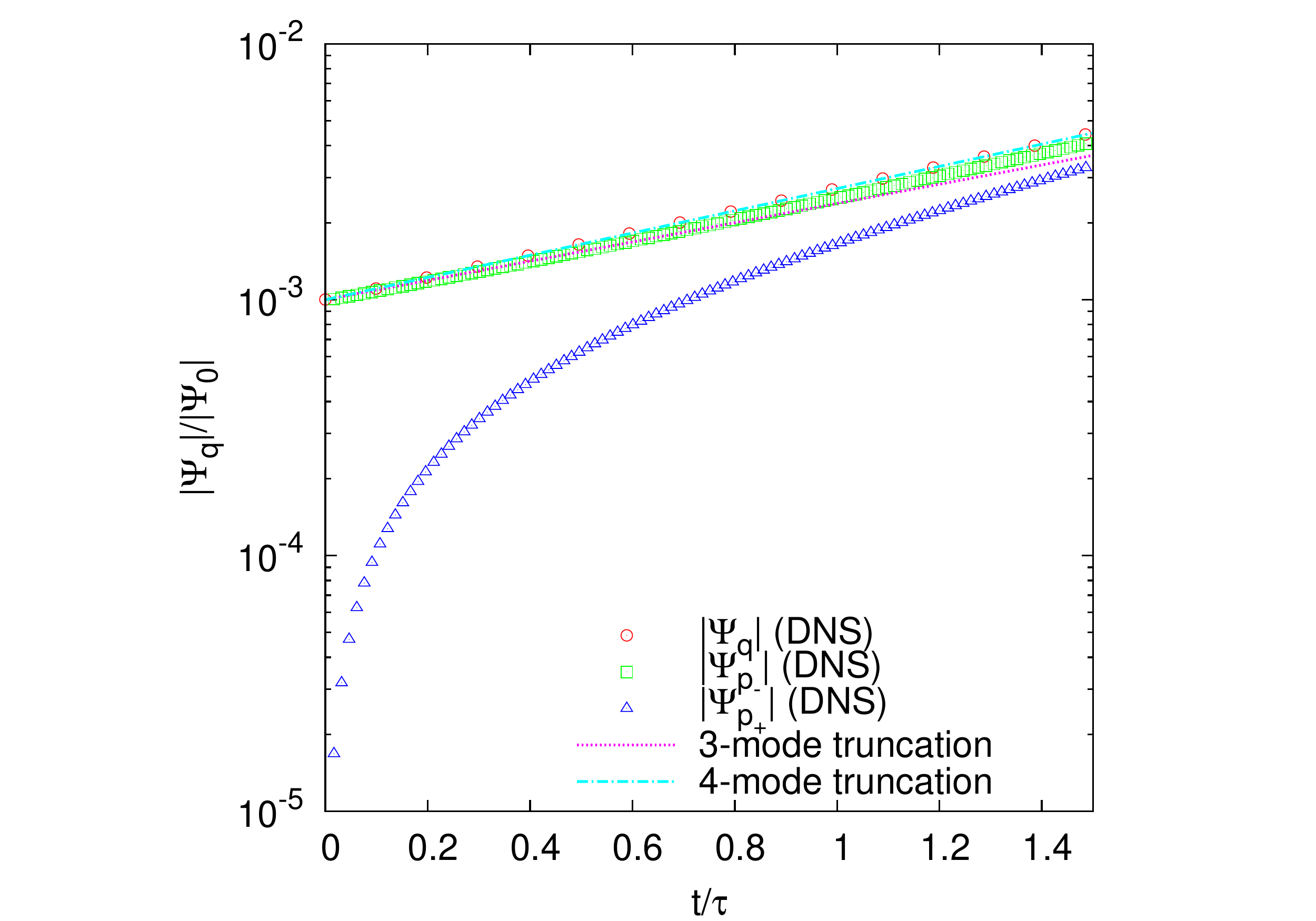}
\caption{\label{fig-compareMIandDIShortTimes} Comparison of modulational and decay instabilities for early times
for the case (i) -purely
zonal modulations (like in Fig.~\ref{fig-compareMIandDILongTimes}).}
\end{figure}

Let us first consider case (A) when the modulation is purely zonal.
Fig.~\ref{fig-compareMIandDILongTimes} compares  $|\Psi_\vv{q}|$
obtained from the solution of Eq.~(\ref{eq-CHMk}) with that obtained from solutions
of 3MT, Eqs.~(\ref{eq-3ModeTruncationMinus}), and 4MT, Eqs.~(\ref{eq-4ModeTruncation}). The initial condition was
constructed from Eq.~(\ref{eq-3ModeEigenvector}), the unstable eigenvector for the decay instability.
We see that the growth rate predicted for the decay instability is not observed.The PDE instead seems
to follow the growth rate for the modulational instability. From the zoomed-in 
plot of the early time evolution shown in 
Fig.~\ref{fig-compareMIandDIShortTimes} we see that the full dynamics very 
quickly
generates the mode $\vv{p}_+$  which is absent from Eqs.~(\ref{eq-3ModeTruncation}). The full system then quickly deviates
from the solution of the 3MT in a time of the order of the inverse of the 
instability growth rate. However,
the set of 4 modes takes much longer to generate any further modes. Thus in this setup, the 4MT Eqs.~(\ref{eq-4ModeTruncation})
provide a much better description of the full dynamics for times up to 10 instability times.

\begin{figure}
\includegraphics[width=7.0cm,angle=-90]{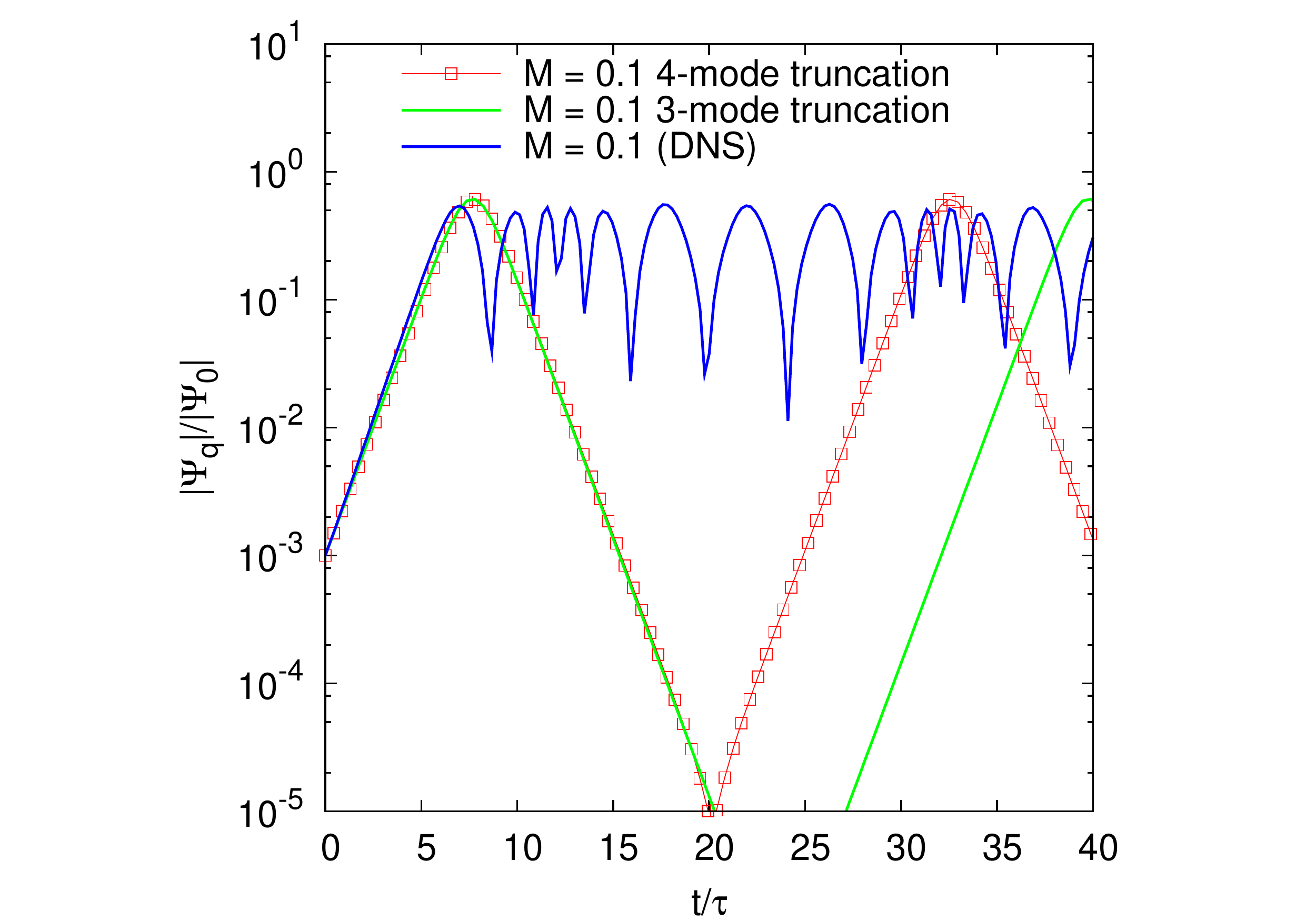}
\caption{\label{fig-growth.q_9_6}
Same as
in Fig.~\ref{fig-compareMIandDILongTimes} but now for
%Comparison of modulational and decay instabilities for
the case (ii):
off-zonal modulations,  ${\bf q} = (9,6)$.}
\end{figure}

Let us now consider the case (B) when the modulation is off-zonal.
Fig.~\ref{fig-growth.q_9_6}  compares  $|\Psi_\vv{q}|$
obtained from the solution of Eq.~(\ref{eq-CHMk}) with that obtained from solutions
of 3MT, Eqs.~(\ref{eq-3ModeTruncationMinus}), and 4MT, Eqs.~(\ref{eq-4ModeTruncation}), for an initial condition
being the unstable eigenvector for the decay instability, Eq.~(\ref{eq-3ModeEigenvector}).
As expected, now the 3MT and the 4MT give practically identical results,
and both of these models agree well with DNS up to the time equal to seven inverse growthrates.
They predict well the maximum of the zonal jet amplitude, although the subsequent stage of decrease
is not described as well as in case (A) by the 4MT model.

From these results we conclude that the 3-wave interaction is indeed the basic 
nonlinear process  when  $M \ll 1$ provided 
the triad is not degenerate, in the sense that it does not contain 
quasi-resonant modes
which are equidistant from two different resonant manifolds as happens 
when the vector ${\bf q}$ is zonal.
In these cases, the 3MT system is just as good as the 4MT and it describes
well the full CHM system for over several characteristic times (i.e.
the inverse instability growthrates).
On the other hand, the most relevant configuration with ${\bf q}$ zonal is,
in fact, degenerate.  In this case, however, the 4MT model works well over many 
characteristic times whereas the 3MT
fails almost immediately. Thus, to have a wider range of applicability, we will study the 4MT model
and abandon the 3MT model in the remainder of the present paper.

Next, let us study the modulational instability arising from the 4MT in greater detail.

\section{\label{sec-linearInstabilityAnalysis} Instability for purely meridional carrier wave and purely  zonal  modulation}

The case of a purely zonal carrier wave ($\vh{p}=(1,0)$) and purely meridional perturbation($\vh{q}=(0,1)$)  is of physical interest and produces considerable
simplification. The dispersion relation then reduces to solving
\begin{widetext}
\begin{equation}
\W^3 + \left(\frac{s^2}{(1+F)(s^2+1+F)}\right)^2\left(\frac{2 M^2(1-s^2)(1+F)^2(s^2+F+1)-(s^2+F) }{s^2 + F} \right) \W =0,
\end{equation}

which has roots
\begin{eqnarray}
\W&=&0\\
\W&=&\pm i \left(\frac{s^2}{(1+F)(s^2+1+F)}\right) \sqrt{\frac{2 M^2(1-s^2)(1+F)^2(s^2+F+1)-(s^2+F) }{s^2 + F}}
\end{eqnarray}
The question of whether the perturbation is unstable reduces to the question of when the quantity under the square root is positive. In this expression, recall
that $s$ is the ratio, $q/p$, of the modulus of the modulation wave-vector to 
the modulus of the primary wave-vector. Letting $s^2=y$, one obtains a quadratic for the quantity under the square root which is positive in the range $s \in (-s_{\rm max}, s_{\rm max})$ where
\begin{equation}
\label{eq-s_max}
s_{\rm max}^2 = \frac{1+2 M^2 F (1+F)^2}{2 M^2(1+F)^2} \left[-1 + \sqrt{1+4\frac{(2 M^2(1+F)^3  -F)(2 M^2(1+F)^2)}{(1+2 M^2 F (1+F)^2)^2}}\right]
\end{equation}
\end{widetext}

\subsection{Case of Infinite Deformation Radius}

\begin{figure}
\includegraphics[width=7.0cm]{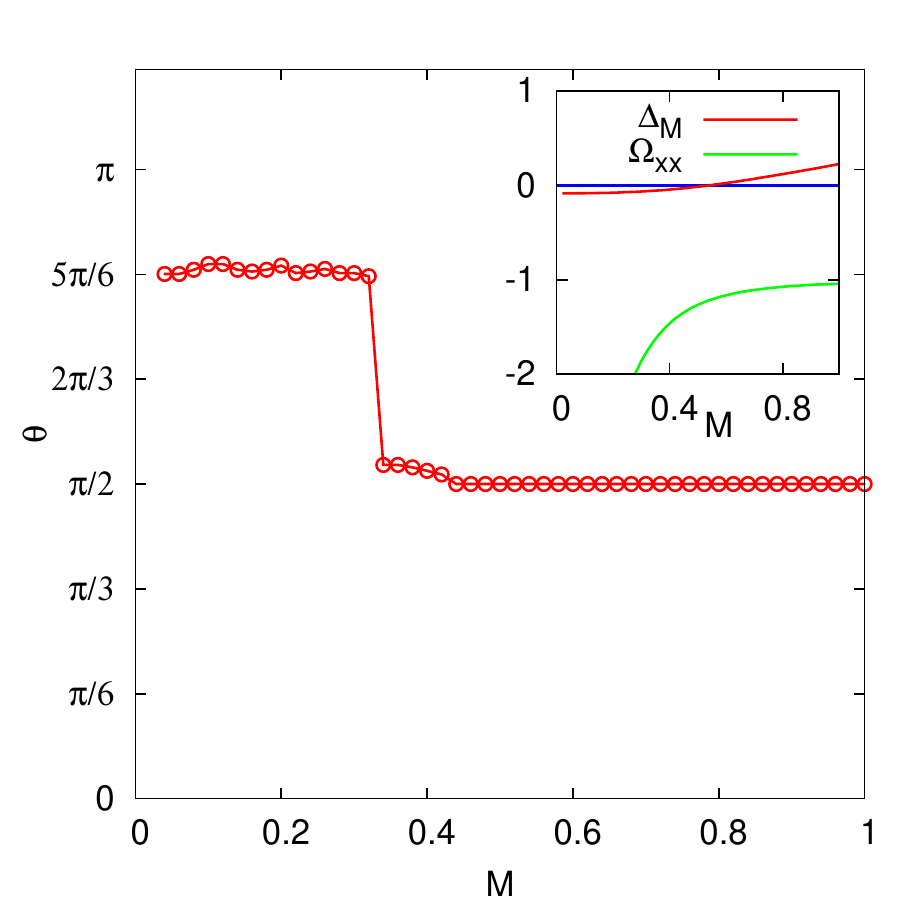}
\caption{\label{fig-theta_max} Angle, $\theta$, between the $\vv{q}$ wave-vector of the maximally unstable perturbation and the $x$-axis as a function of $M$. Inset plots $\Delta_M$ and $\Omega_{xx}$ as a function of $M$ illustrating the transition of the maximum growth rate for on-axis perturbations from a local maximum to a saddle point at $M\approx 0.53$.}
\end{figure}

When $F=0$ the analysis becomes particularly simple. There is always a range of unstable long wavelength perturbations, given by $(0, s_{\rm max})$, for
any value of $M$. $s_{\rm max}$ is given by
\begin{equation}
s_{\rm max} = \sqrt{\frac{-1+\sqrt{1 + 16 M^4}}{4 M^2}}.
\end{equation}
Within this range the growth rate is
\begin{equation}
\W = \left(\frac{s^2}{(s^2+1)}\right) \sqrt{2 M^2 (1-s^4) - s^2 }.
\end{equation}
The growth rate has a single maximum at $s_0=\sqrt{y_0}$ where $y_0$ is the positive root of
\begin{equation}
y^3 + 3 y^2 + (1 + \frac{1}{M^2}) y - 1 =0.
\end{equation}
One can show that $s_0 \to \sqrt{\sqrt{2}-1}$ as $M \to \infty$ and $s_0 = M + O(M^2)$ as $M\to 0$. One would be interested to know when the maximally unstable meridional perturbation is a local maximum with respect to nearby non-meridional perturbations. To ascertain this, one should look at the sign of the determinant
\begin{equation}
\Delta_M(\hat{q}_x, \hat{q}_y) = \left|
\begin{array}{cc}
\pTwodTwo{\W}{\hat{q}_x} &\pTwodTwoMixed{\W}{\hat{q}_x}{\hat{q}_y} \\
\pTwodTwoMixed{\W}{\hat{q}_x}{\hat{q}_y} & \pTwodTwo{\W}{\hat{q}_y}
\end{array}
 \right|
\end{equation}
evaluated at $(\hat{q}_x, \hat{q}_y) = (0,s_0)$. This can be done semi-analytically using {\em Mathematica} and is plotted in the inset of Fig.~\ref{fig-theta_max}. We find that $\Delta_M >0$ with $\pTwodTwo{\W}{\hat{q}_x} <0$ (the
criterion for a local maximum) for $M>M_c$.
$\Delta_M <0$ with $\pTwodTwo{\W}{\hat{q}_x} <0$ (the criterion for a saddle)
for $M>M_c$. The critical value of $M$ is found numerically to be 
$M_c \approx 0.534734$. Numerical explorations show that the local maximum
found for $M>M_c$, is actually global. For $M>M_c$, therefore, the fastest 
growing perturbation is indeed zonal. As $M$ decreases below $M_c$ the most 
unstable pertubation moves to a point with a finite value of $q_x$. The 
maximally unstable perturbation for $M<0.53$ tends to a point on the resonant 
manifold making an angle of $5\pi/6$ with the $x$-axis.
% (this corresponds to the edge of the instability ``cone'' constructed in \cite{OPSSS2004}).
The dependence of this angle on $M$ is shown in Fig.~\ref{fig-theta_max}. A clear transition from an axial maximum to an off-axis maximum is clearly visible.

\subsection{Effects of Finite Deformation Radius}

\begin{figure}
\includegraphics[width=7.0cm]{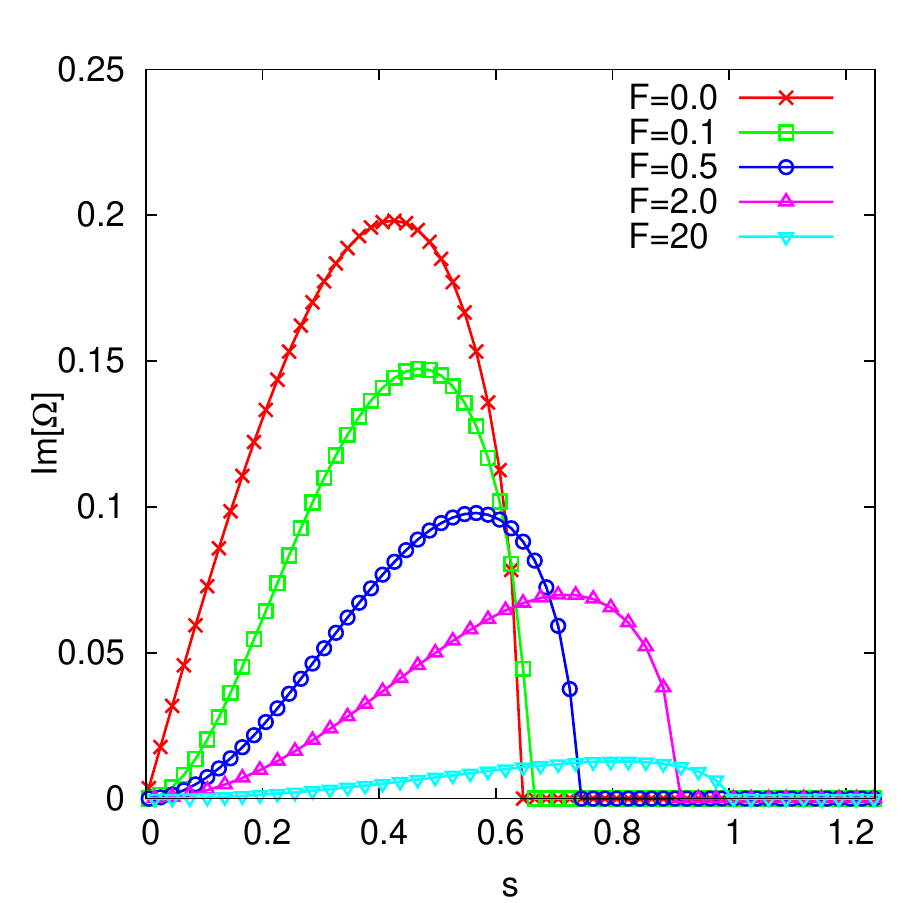}
\caption{\label{fig-effectOfF1} Instability growthrate for purely meridional perturbations with $M=0.5 > \sqrt{2/27}$
for different values of the deformation radius.}
\end{figure}

\begin{figure}
\includegraphics[width=7.0cm]{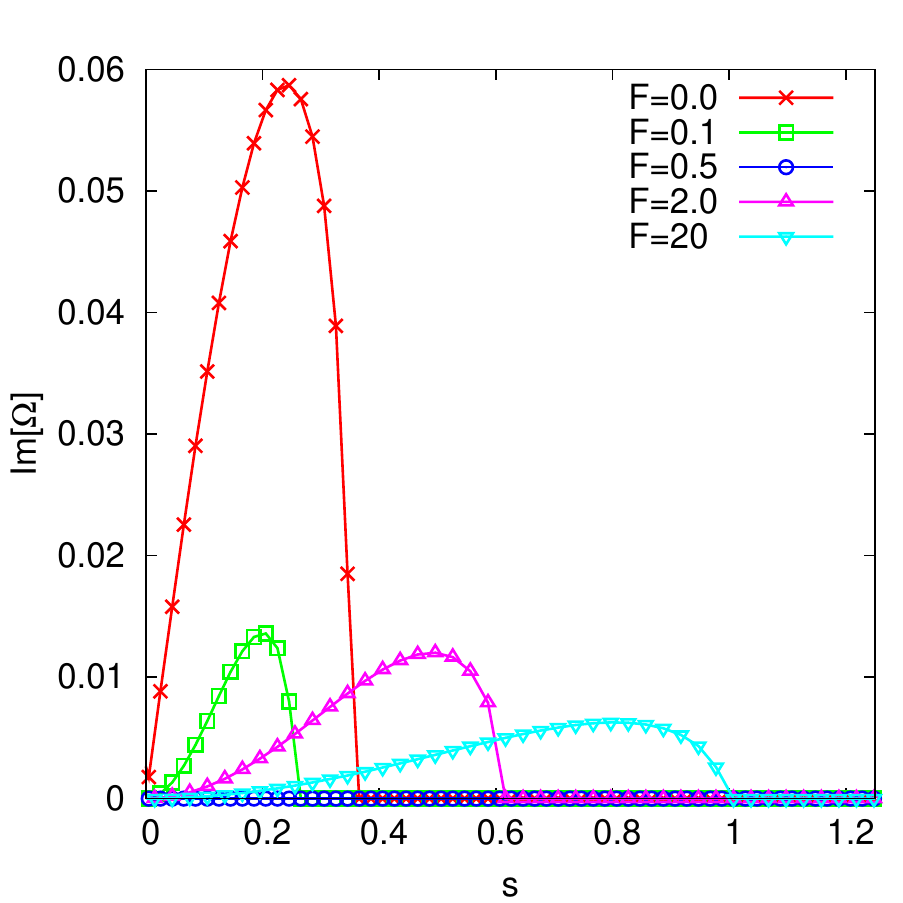}
\caption{\label{fig-effectOfF2} Instability growthrate for purely meridional perturbations with $M=0.25 <\sqrt{2/27}$ for different values of the deformation radius. For this value of $M$, $F_2\approx 0.23$ and $F_3\approx 1.00$. Note that the instability is completely suppressed for intermediate values of $F$ and then emerges again as $F$ increases.}
\end{figure}

We now consider the dependence of MI on the deformation or Larmor radius,  
noting that a finite deformation radius is obtained in the QG system under a 
reduced gravity approximation.
When $F$ is finite, there are 2 regimes, depending on the value of $M$. For an 
interval of instability to exist, we require $s_{\rm max}^2 > 0$. This 
requires that
\begin{equation}
\label{eq-FRoots}
p(F) = 2 M^2 (1+F)^3 -F > 0.
\end{equation}
The discriminant of the corresponding cubic, $p(F)=0$, is 
$-4( -2 M^2 + 27 M^4)$ . Since we are only interested in $F>0$ we can identify 
two regimes.
\begin{itemize}
\item Regime 1: $M> \sqrt{\frac{2}{27}}$\\
Referring to Eq.~(\ref{eq-s_max}), $p(F)=0$ has one real root, $F_1$ 
(which is negative) and $p(F)>0$ when $F>F_1$. Then for any positive value of $F$ there exists a finite range of $s$, $s\in (0, s_{\rm max})$,  for which the perturbation is unstable. $s_{\rm max}$ is given by Eq.~(\ref{eq-s_max}). In this regime, finite deformation radius tends to reduce the growth rate of the instability but cannot suppress it. See Fig.~\ref{fig-effectOfF1}.
\item Regime 2: $M \leq \sqrt{\frac{2}{27}}$\\
Referring to Eq.~(\ref{eq-s_max}), $p(F)=0$ has three real roots, $F_1$, $F_2$ and $F_3$. $F_1$ is negative and $F_2$ and $F_3$ are positive. $p(F)<0$ in the range $(F_2, F_3)$.
In this regime, there are critical values of $F$, $F_1$ and $F_2$ such that the range $s\in (0, s_{\rm max})$ of unstable pertubations only exists if $F<F_1$ or $F>F_2$.  $F_1$ and $F_2$ are obtained by finding the positive roots of Eq.~\ref{eq-FRoots} and $s_{\rm max}$ is again given by Eq.~(\ref{eq-s_max}). In this regime, there is a range of intermediate deformation radii which completely suppresses the instability. See Fig.~\ref{fig-effectOfF2}.
\end{itemize}

\section{\label{sec-roleOfM}Role of the Carrier wave Amplitude.}

%input{roleOfM}

\begin{figure*}
\includegraphics[width=17.0cm]{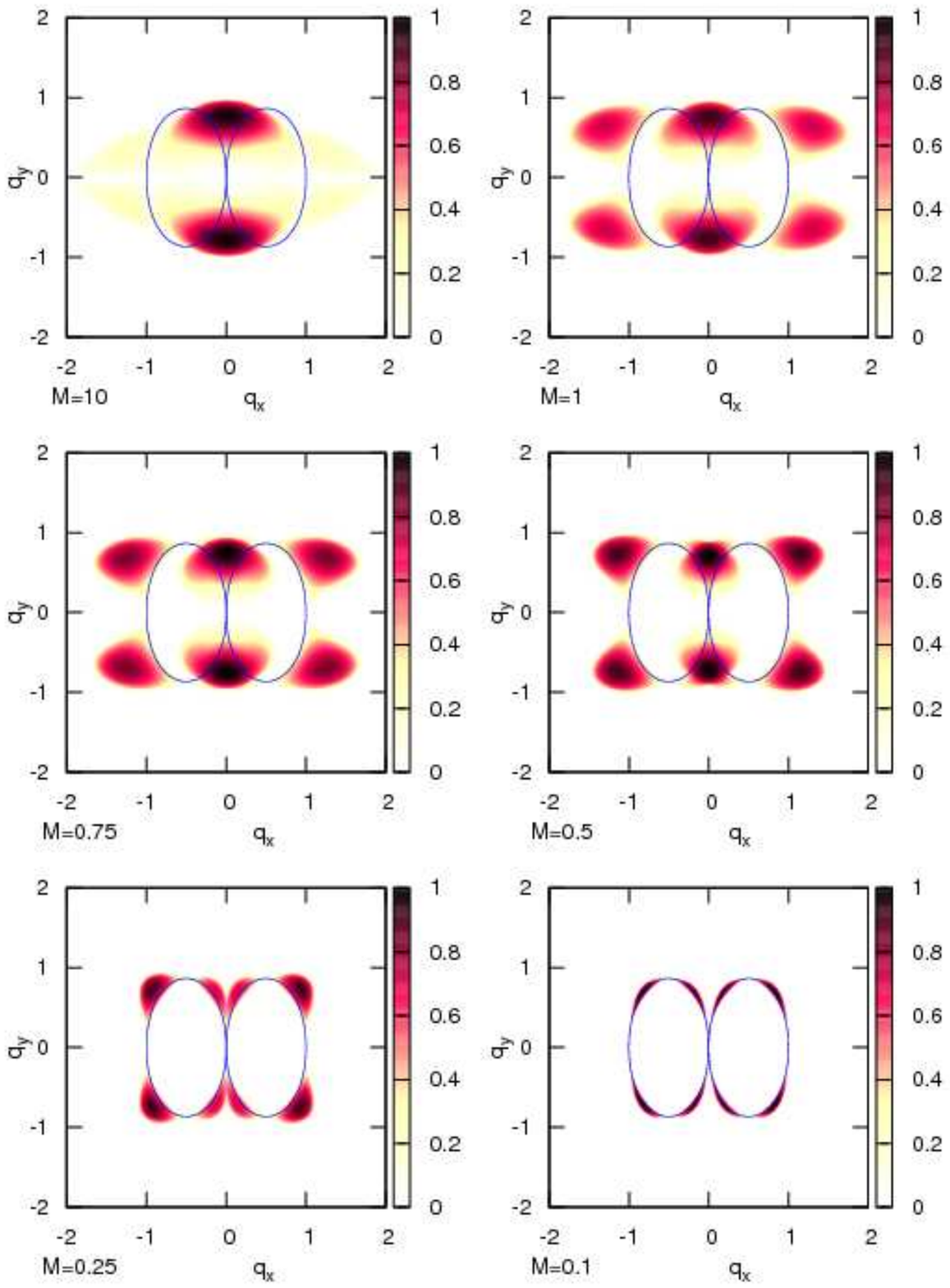}
\caption{\label{fig-qMaps_F_2}
Same as in Fig.~\ref{fig-collapseToResonantManifolds}  but now for a finite deformation radius, $F=2$.}
%Here we plot the level sets of the negative imaginary part of the roots of Eq.(\ref{eq-nonDimensionalDispersion}) as a function of $\vv{q}$ for a fixed meridional carrier wave-vector, $\vv{p}=(1,0)$ and $F=2$. The values of $M$ for the initial carrier wave are $M=10$ (Euler limit), $M=1$, $M=3/4$, $M=1/2$, $M=1/4$ and $M=1/10$ (wave turbulence limit). }
\end{figure*}

\begin{figure*}
\includegraphics[width=17.0cm]{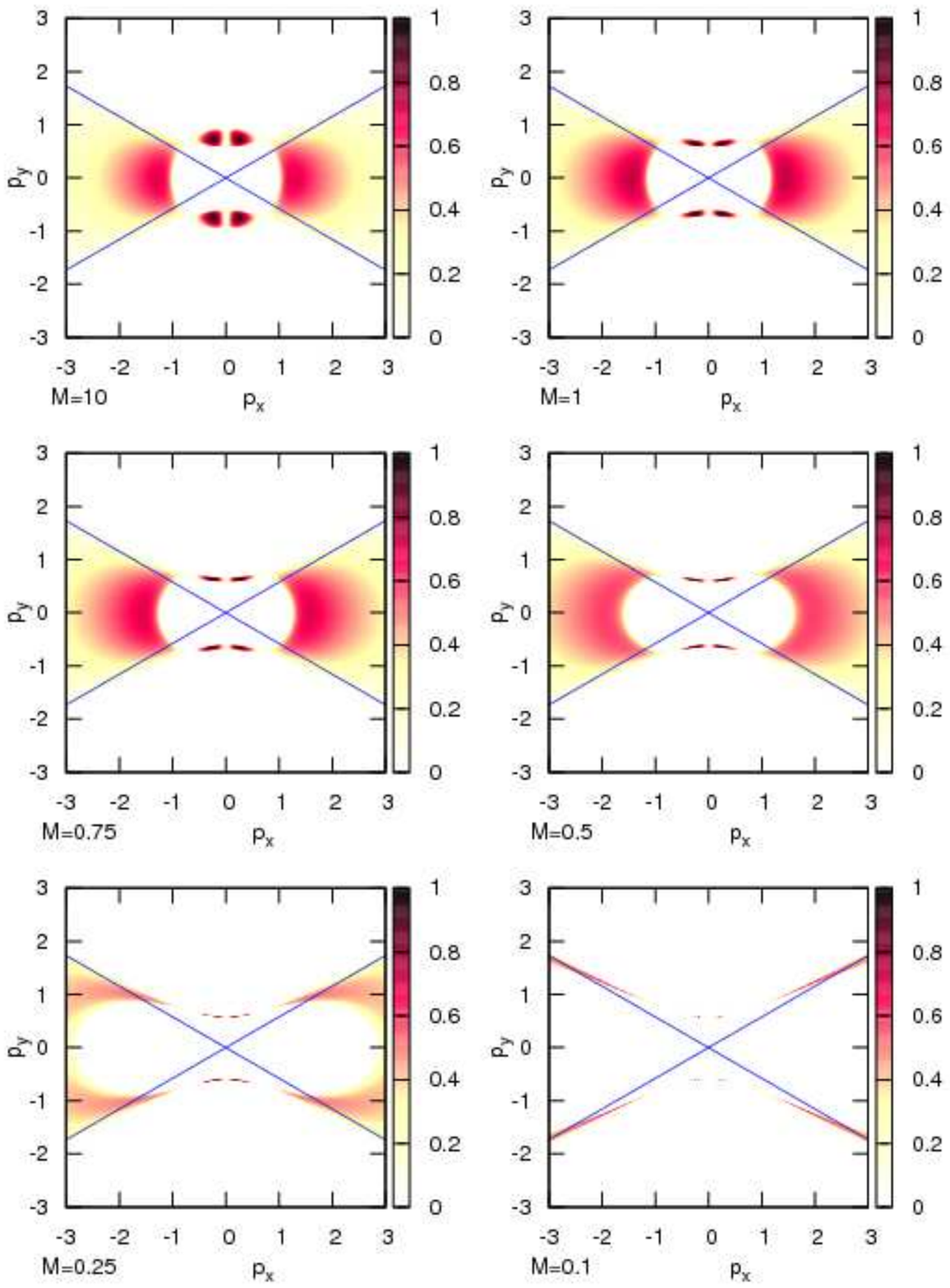}
\caption{\label{fig-kMaps_F_0}
 Growth rate of the modulational instability  (found from  Eq.(\ref{eq-nonDimensionalDispersion}))
%Here we plot the level sets of the negative imaginary part of the roots of Eq.(\ref{eq-nonDimensionalDispersion})
as a function of $\vv{p}$ for a fixed zonal modulation wave-vector, $\vv{q}=(0,1)$ and $F=0$. The values of $M$ for the initial carrier wave are $M=10$ (Euler limit), $M=1$, $M=3/4$, $M=1/2$, $M=1/4$ and $M=1/10$ (weakly nonlinear limit). }
\end{figure*}

\begin{figure*}
\includegraphics[width=17.0cm]{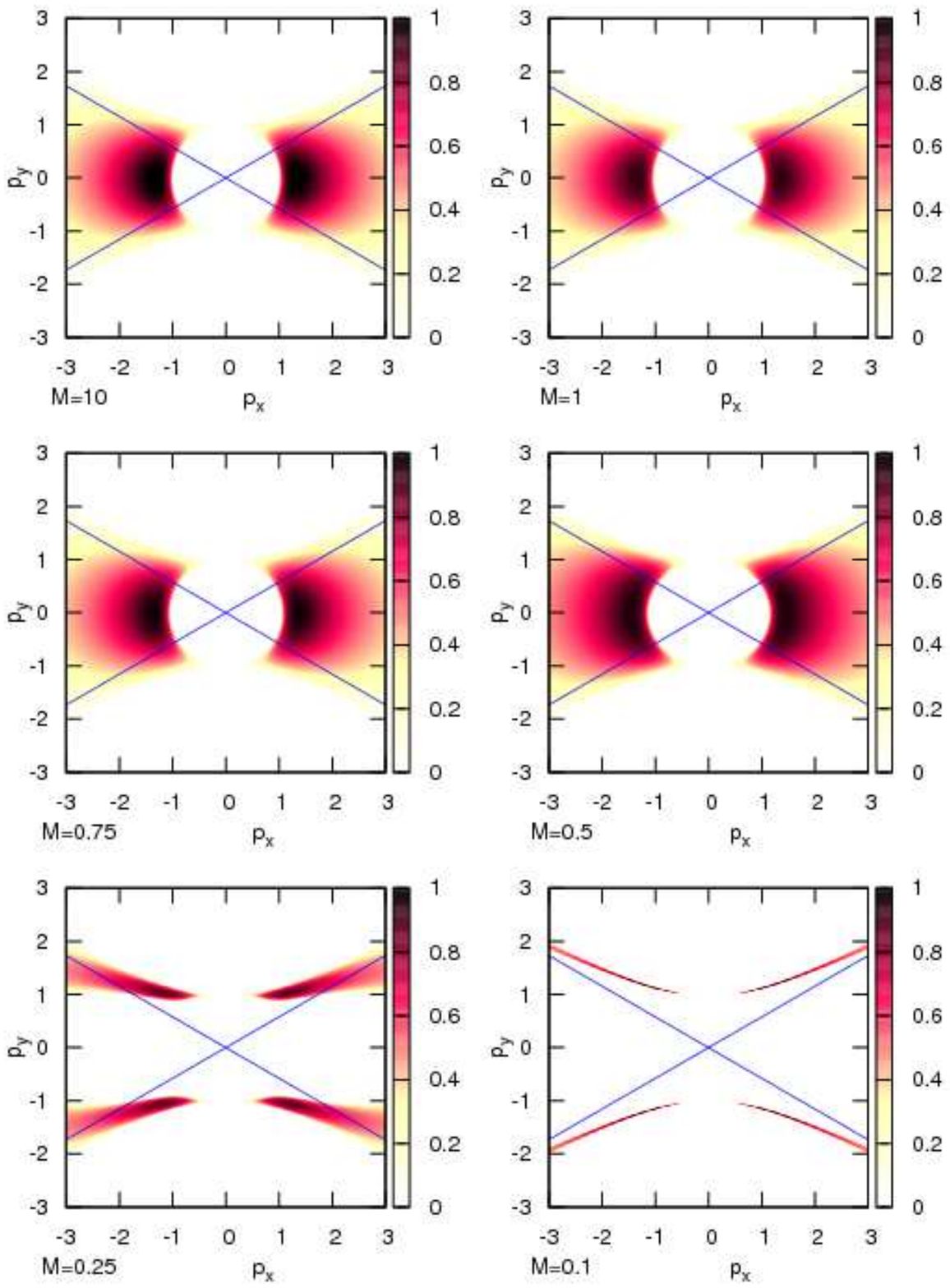}
\caption{\label{fig-kMaps_F_2}
Same as in Fig.~\ref{fig-kMaps_F_0}   but now for a finite deformation radius, $F=2$.}
%Here we plot the level sets of the negative imaginary part of the roots of Eq.(\ref{eq-nonDimensionalDispersion}) as a function of $\vv{k}$ for a fixed zonal modulation wave-vector, $\vv{q}=(0,1)$ and $F=2$. The values of $M$ for the initial carrier wave are $M=10$ (Euler limit), $M=1$, $M=3/4$, $M=1/2$, $M=1/4$ and $M=1/10$ (wave turbulence limit). }
\end{figure*}

We have already mentioned the role of the nonlinearity parameter $M$, for the
most unstable modulation (i.e. zonal for big $M$ and inclined for small $M$), as well as
for the different regimes in the finite $F$ case.
Fig.~\ref{fig-collapseToResonantManifolds} show plots of the instability growthrate
as a function of $\vv{q} =(q_x,q_y)$ for several different values of $M$ for fixed meridional
$\vv{p}$ and $F=0$. In particular we see how the maximum growthrate flips from
zonal to off-zonal $\vv{q}$ when $M$ is reduced (below see also about the collapse of the unstable
region to the resonant curve).
Fig.~\ref{fig-qMaps_F_2} shows plots of the instability growthrate similar to the ones in
Fig.~\ref{fig-collapseToResonantManifolds} but now for finite $F$. 
Qualitatively the finite $F$ plots are similar to those obtained for $F=0$.

Another natural way to visualise the structure of the set of unstable
pertubations is to fix the wavevector of the perturbation  mode, $\vv{q}$, 
and plot the instability growth rate as a function of the primary wavevector,
$\vv{p}$.  Fig.~\ref{fig-kMaps_F_0} does this, plotting the instability 
growth rate as a function of $\vv{p}=(p_x,p_y)$ for several different values 
of $M$ with fixed zonal $\vv{q}$ and $F=0$. We see that nonlinearity 
reduction "eats into" the instability cone, i.e. makes some wavenumbers 
inside the cone stable.  At the same time, the nonlinearity makes some 
wavenumbers outside the cone unstable.  It is important to keep in mind
that, even for large $M$, the maximum growthrate
occurs outside of the cone, for the primary wave orientations closer to zonal 
than to the the meridional direction, see Fig.~\ref{fig-kMaps_F_0}  for $M=10$.
This fact is easy to overlook if one considers only the limit $M \to \infty$ 
(as it is common in the plasma literature) because, in this limit, the growth 
rate maximum is for the meridional primary waves.
On the other hand, the choice of the primary wave direction is often dictated not by
the maximum growthrate of the modulational instability, but by the structure of the
primary instability creating the Rossy and drift waves (ITG instability in plasmas and
the baroclinic instability in GFD).

Finally, Fig.~\ref{fig-kMaps_F_2} shows plots of the instability growth rate
 similar to the ones in
Fig.~\ref{fig-kMaps_F_0} but now for finite $F$. We again see a qualitatively 
similar picture to the $F=0$ case. Note, however, that the global maximum 
growth rate for large $M$ is now obtained for purely meridional primary waves.

Let us now specially consider the
limits $M \gg 1$ and $M \ll 1$.

\subsection{\label{bigM}Limit $M \gg 1$.}

The limit of large nonlinearity $M \gg 1$ is a particularly simple and well studied one
\cite{AM1960,LOR1972,GIL1974,MN1994,SDS2000,OPSSS2004}.
As we mentioned before, the $\beta$-effect becomes unimportant and, for $F=0$,
this case reduces to instability of Kolmogorov flow in Euler equations (i.e. sinusoidal
plane-parallel shear). In this case the most unstable modulation is perpendicular
 to the carrier wave. The instability criterion reduces to \cite{GIL1974}
$$ \cos^2 \phi < \left(1+ \frac{q^2}{p^2} \right)/4, $$
where $\phi$ is the angle between $\vv{p}$ and $\vv{q}$.
For the scale separated case, $q \ll p$, this condition describes an "instability cone"
\cite{MN1994,SDS2000,OPSSS2004}
$$|\phi| < \pi/6.$$
Finite deformation radius modifies this cone to a larger instability area
\cite{MN1994,SDS2000}
$$ F + p_x^2 -3 p_y^2 >0.$$

We repeat that one has to use
the results obtained in the limit $M \to \infty$ with great caution,
because even for large $M$'s the most unstable primary wave
is not predicted correctly in this limit.

\subsection{\label{smallM}Limit $M \ll 1$.}

\begin{figure}
\includegraphics[width=7.0cm]{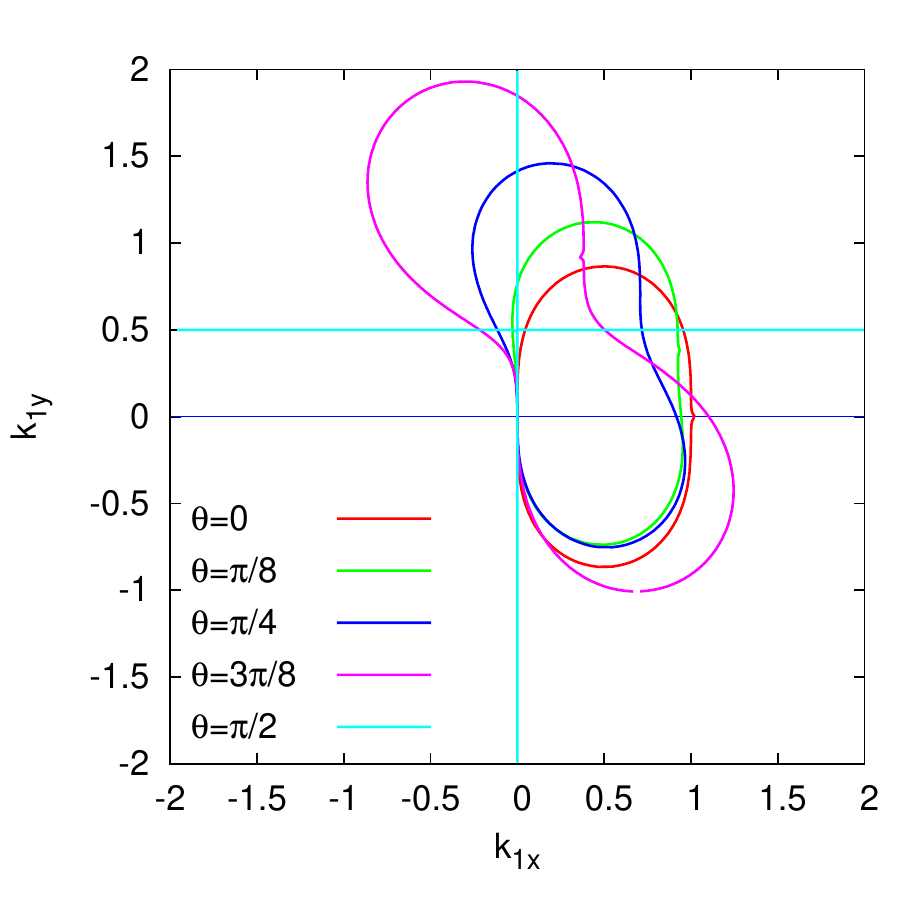}
\caption{\label{fig-resonantManifolds} Shape of the resonant manifold determined by Eq.(\ref{eq-resonantManifolds}) with $\vv{k}_3=(\cos(\theta),\sin(\theta))$ for several values of $\theta$ for the case $F=0$.}
\end{figure}

In the limit of weak nonlinearity, $M\ll 1$, the dynamics is completely wave dominated \cite{GIL1974}.
The nonlinear terms allow waves to interact weakly and exchange energy. Since the nonlinearity is quadratic, wave interactions are triadic (3-wave resonances 
are allowed by the dispersion relation, Eq.~(\ref{eq-RossbyDispersion})). Any 
triad of waves having wave-vectors $\vv{k}_1$, $\vv{k}_2$ and $\vv{k}_3$ 
interact only if they satisfy the resonance conditions:
\begin{eqnarray}
\vv{k}_3 &=& \vv{k}_1 +\vv{k}_2\\
\w(\vv{k}_3) &=& \w(\vv{k}_1) +\w(\vv{k}_2).
\end{eqnarray}
From Eq.(\ref{eq-RossbyDispersion}), this latter relation gives an implicit equation for the resonant manifold of a given $\vv{k}_3=(k_{3x},k_{3y})$:
\begin{eqnarray}
\label{eq-resonantManifolds}
\nonumber & &\frac{k_{1x}}{k_{1x}^2 + k_{1y}^2 + F} + \frac{k_{3x} - k_{1x}}{(k_{3x}-k_{1x})^2 + (k_{3y}-k_{1y})^2 + F}\\
&& -  \frac{k_{3x}}{k_{3x}^2 + k_{3y}^2 + F} = 0.
\end{eqnarray}
Because the system is anisotropic, the shape of resonant manifold depends on the direction of $\vv{k}_3$ as shown in Fig.~\ref{fig-resonantManifolds}.

These resonant manifolds are relevant even for finite nonlinearity since the 
support of the instability concentrates close to the resonant curves as $M$  
is decreased as shown in Fig.~\ref{fig-collapseToResonantManifolds}. Even for 
$M=1$ there is a strong connection between the resonant curves and the shape of
the set of modulationally unstable perturbations.
In fact, Fig.~\ref{fig-collapseToResonantManifolds} shows two resonant curves 
corresponding to two resonant triads,
$$\vv{k}_1 = \vv{p}, \;\;\;\;\; \vv{k}_2 = \vv{q} \;\;\; \hbox{ and} \;\;\; \vv{k}_3 = \vv{p}_+ = \vv{p} + \vv{q}$$
and
$$\vv{k}_1 = \vv{p}_- = \vv{p} - \vv{q}, \;\;\; \vv{k}_2 = \vv{q} \;\;\; \hbox{ and} \;\;\;\;\;
\vv{k}_3 = \vv{p}.$$
Out of four wavenumbers in our truncated system, $\vv{p},\vv{q},\vv{p}_-$ and $\vv{p}_+$, three
are resonant (or nearly resonant) and the remaining one is non-resonant  ($\vv{p}_-$ or $\vv{p}_+$).
As we mentioned before, this picture is correct in non-degenerate situations,
when $\vv{q}$ is not zonal.
Then for $M \to 0$ the amplitude of this non-resonant mode in the instability eigenvector
tends to zero, so effectively there are only three active modes, and one
can use the results obtained above for the 3MT model.
In particular,
Eq.~(\ref{eq-resonantDecayInstability}) gives the instability growthrate:
\begin{equation}
\gamma = \frac{|\psi_0| |\vv{q} \times \vv{p}|
\sqrt{(p^2 -q^2) (p_{+}^2-p^2)}}{\sqrt{(p_{+}^2 + F)(q^2 + F)}}.
\label{decay}
\end{equation}
This expression was previously obtained in the case $F=0$ in \cite{GIL1974}
based on the 4MT model.
One can see that instability of the primary wave occurs if its wavenumber
length lies in between of the wavenumber lengths of the waves it decays into,
$q < p< p_+$.
This condition has a nice dual-cascade interpretation:  to
decay the wave must
be able to transfer its energy to a large scale and its enstrophy to a
smaller scale.
For $F=0$, the typical instability growthrate is $\gamma \sim U_0 p$ where
$U_0=p \psi_0$ is the velocity amplitude of the carrier wave
\cite{GIL1974}.
In the large $F$ case, the instability is slowed by the factor
$F/p^2$ (but not arrested).

Another interesting feature of instability for $M \ll 1$
is seen in Fig~\ref{fig-kMaps_F_0}
where we can see that (for fixed zonal $q$): the unstable region
becomes narrow and collapses onto the sides of the "cone", i.e.
onto the lines $p_y = \pm p_x /\sqrt{3}$. This fact can be explained by
considering the resonant curve for $q \ll p$ where it
behaves as $q_x = -2 (p_x p_y /p^4) q_y^3$.
For instability, this curve has to pass as close as possible to
the vertical (zonal) axis (where we have chosen our $\vv{q}$).
Thus, we need to minimize the above coefficient $(p_x p_y /p^4)$
(e.g. with respect to $p_y$ for fixed $p_x$) which immediately gives
$p_y = \pm p_x /\sqrt{3}$.

For small $M$ the maximally unstable modulation
$\vv{q}$ is off-zonal, which may be important for determining the
final statistical state of the nonlinear evolution.
As we will see later, this state appears to have a predominantly
off-zonal component even if the initial modulation is chosen
to be zonal.

\section{\label{nonlin}Nonlinear Evolution}

From now on we study only systems with infinite deformation radius, $F=0$.
%To test the linear predictions and to study the nonlinear evolution,
%we have performed DNS of CHM system (\ref{eq-CHM}) by pseudo-spectral
%method with $512^2$ resolution and hyperviscosity parameters
%$n=..., \;\; \nu_n =....$.
To test the linear predictions, and to study the nonlinear evolution,
we have performed DNS of the CHM system,  Eq.~(\ref{eq-CHM}), using a
standard  pseudo-spectral method with resolution up to $1024^2$ and
hyperviscosity parameters $\nu_n = 4.5e^{-30}$.
We solve, in tandem, the 4MT system, (\ref{eq-4ModeTruncation}),
and compare it with DNS.  Although the 4MT was used as the departure point
for the linear stability analysis, it is a fully nonlinear set of equations
in its own right. In addition to checking the linear instability predictions 
against DNS, we will also explore the extent to which the nonlinear  dynamics
of the 4MT captures the behaviour of the full PDE. In all cases, we choose 
the initial condition to be along the unstable eigenvector of the 4MT.

%\section{\label{nonlin-unstable} Modulationally Unstable Case}
%\input{nonlinear-unstable}

\subsection{\label{nonlin-unstable-meridional-zonal}
Case of Meridional Carrier Wave and Zonal Modulation.}

\begin{figure*}
\includegraphics[width=17.5cm]{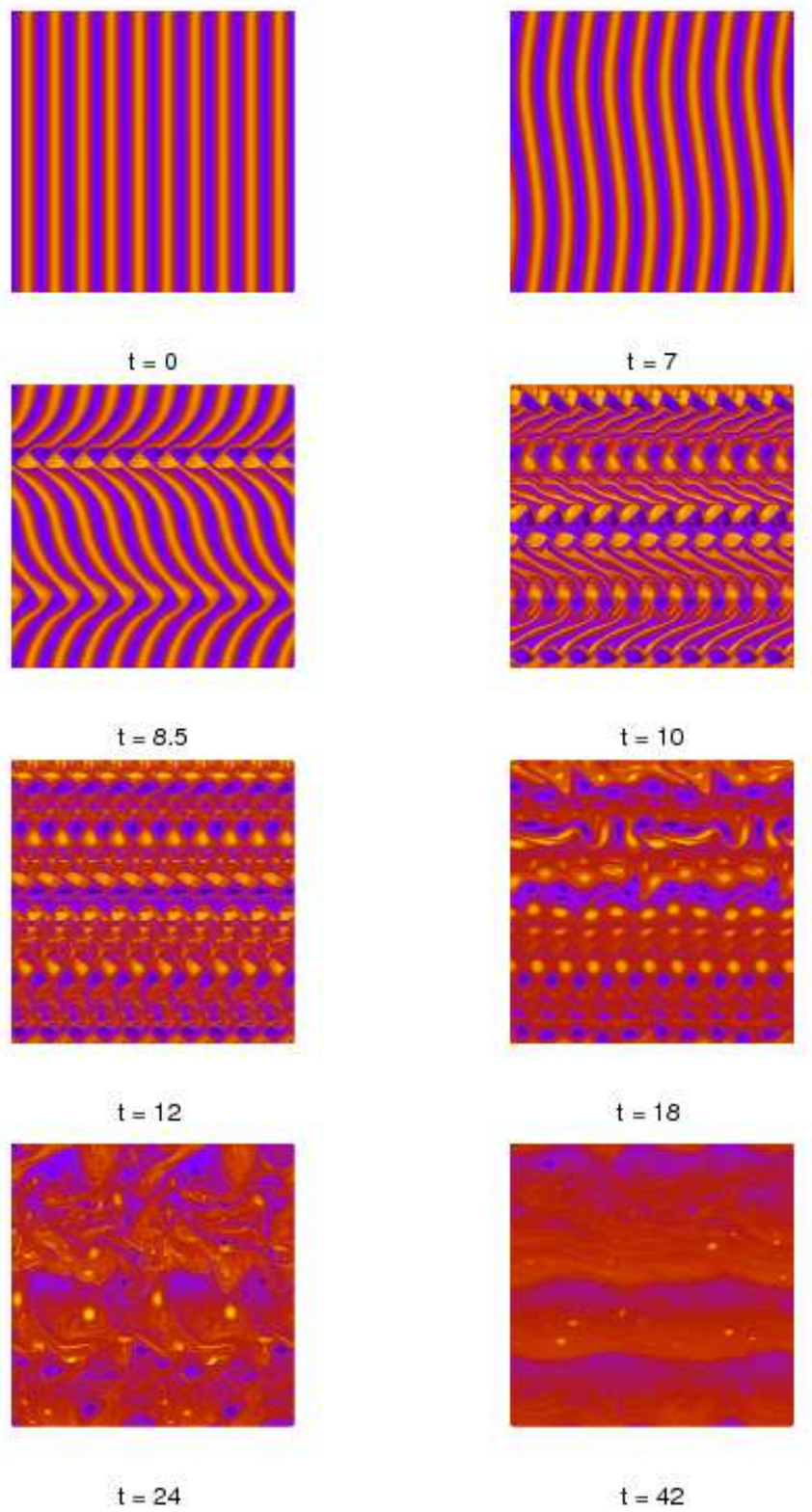}
\caption{\label{fig-snapshots.M10} Vorticity snapshots showing the growth, saturation and transition to turbulence of a zonal perturbation of a meridional carrier wave having $M=10$. }
\end{figure*}

\begin{figure*}
\includegraphics[width=17.5cm]{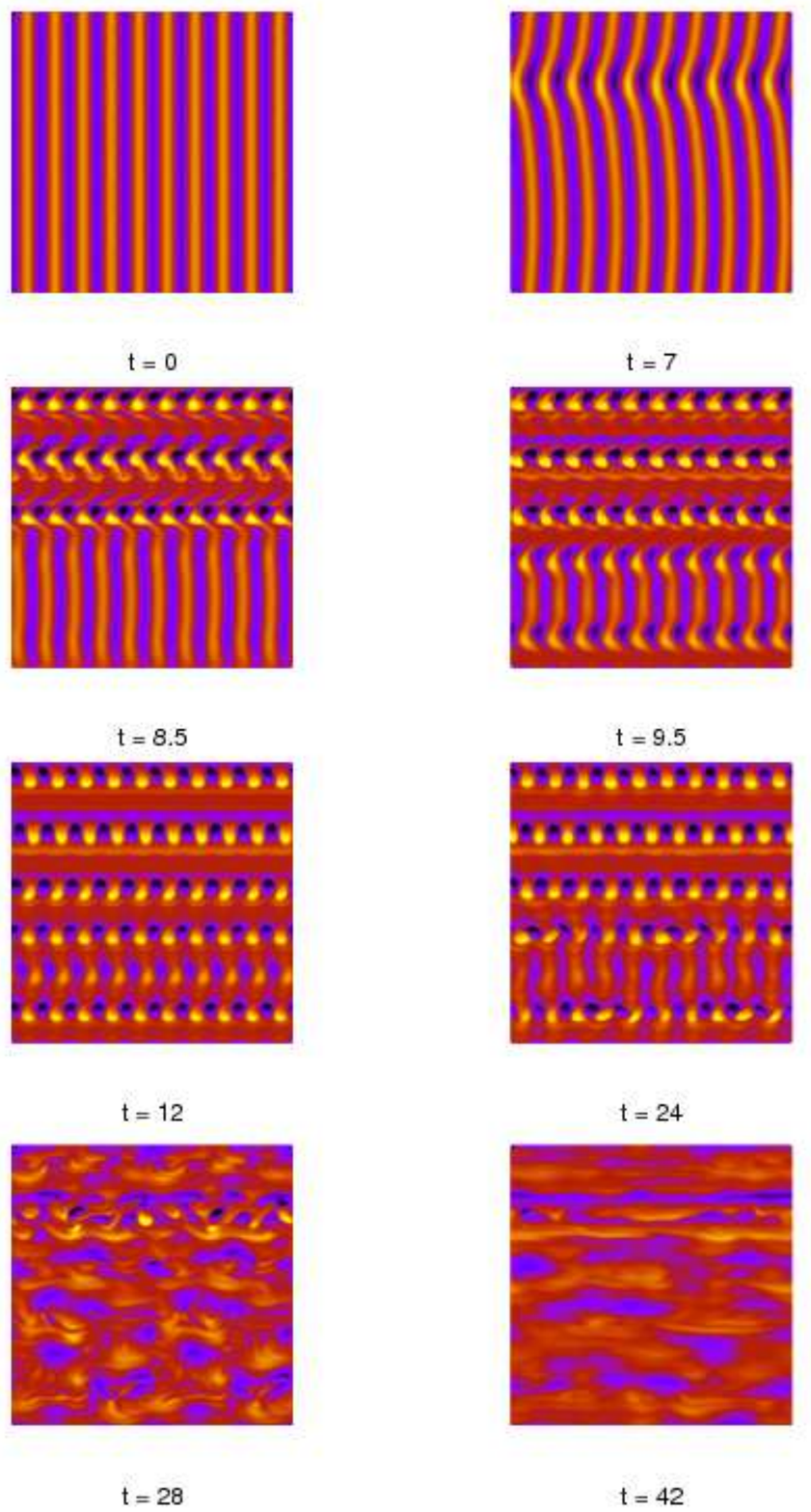}
\caption{\label{fig-snapshots.M1} Vorticity snapshots showing the growth, saturation and transition to turbulence of a zonal perturbation of a meridional carrier wave having $M=1$. }
\end{figure*}

\begin{figure*}
\includegraphics[width=17.5cm]{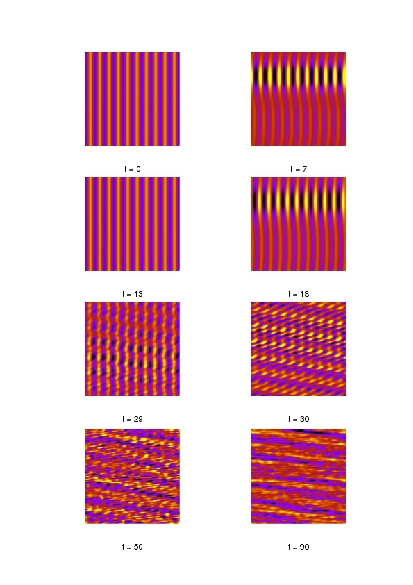}
\caption{\label{fig-snapshots.M0.1} Vorticity snapshots showing the growth, saturation and transition to turbulence of a zonal perturbation of a meridional carrier wave having $M=0.1$. }
\end{figure*}

\begin{figure}
\includegraphics[width=7.0cm,angle=-90]{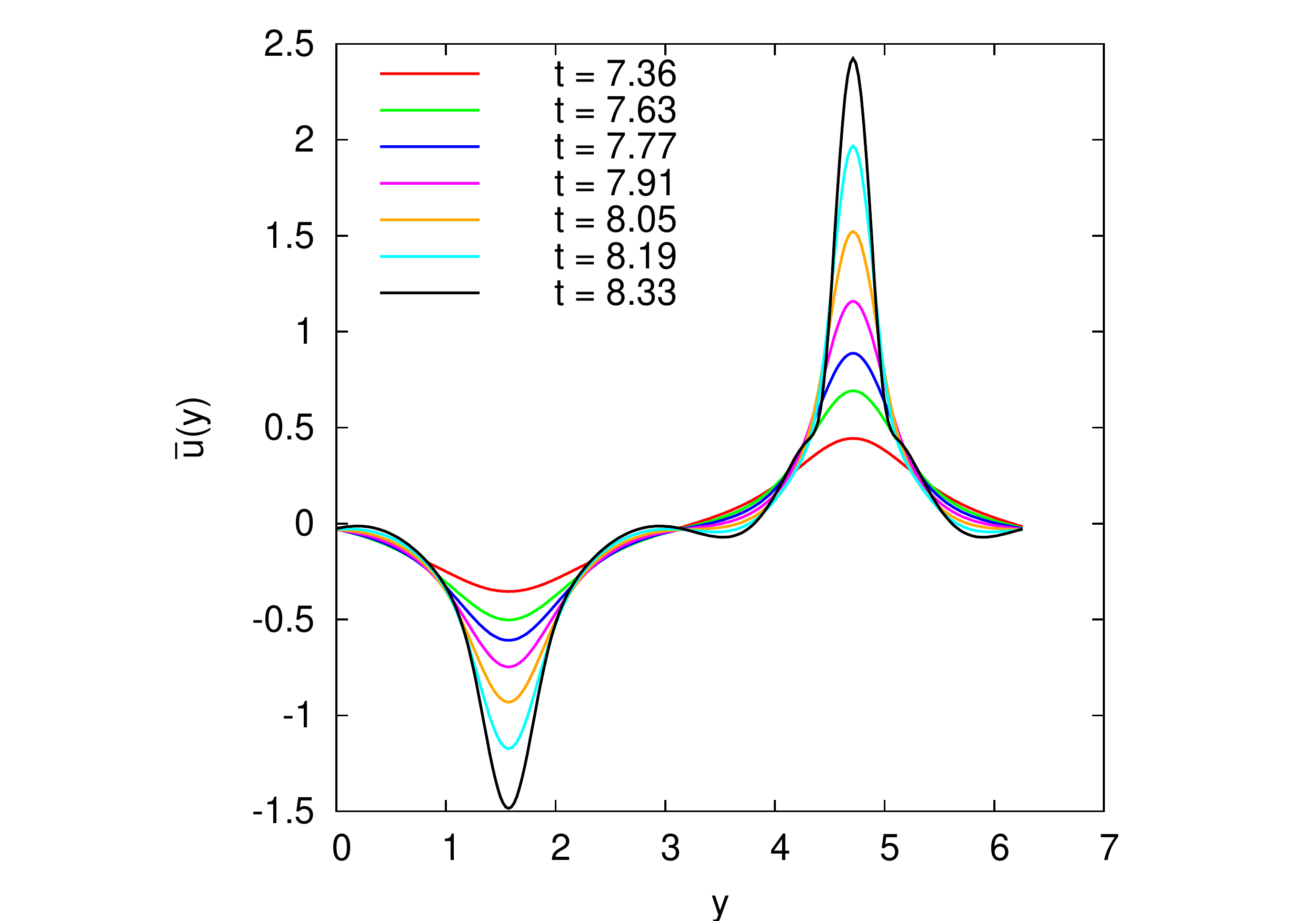}
\caption{\label{fig-Uy.M10} Mean zonal velocity for $M=10$}
\end{figure}

\begin{figure}
\includegraphics[width=7.0cm,angle=-90]{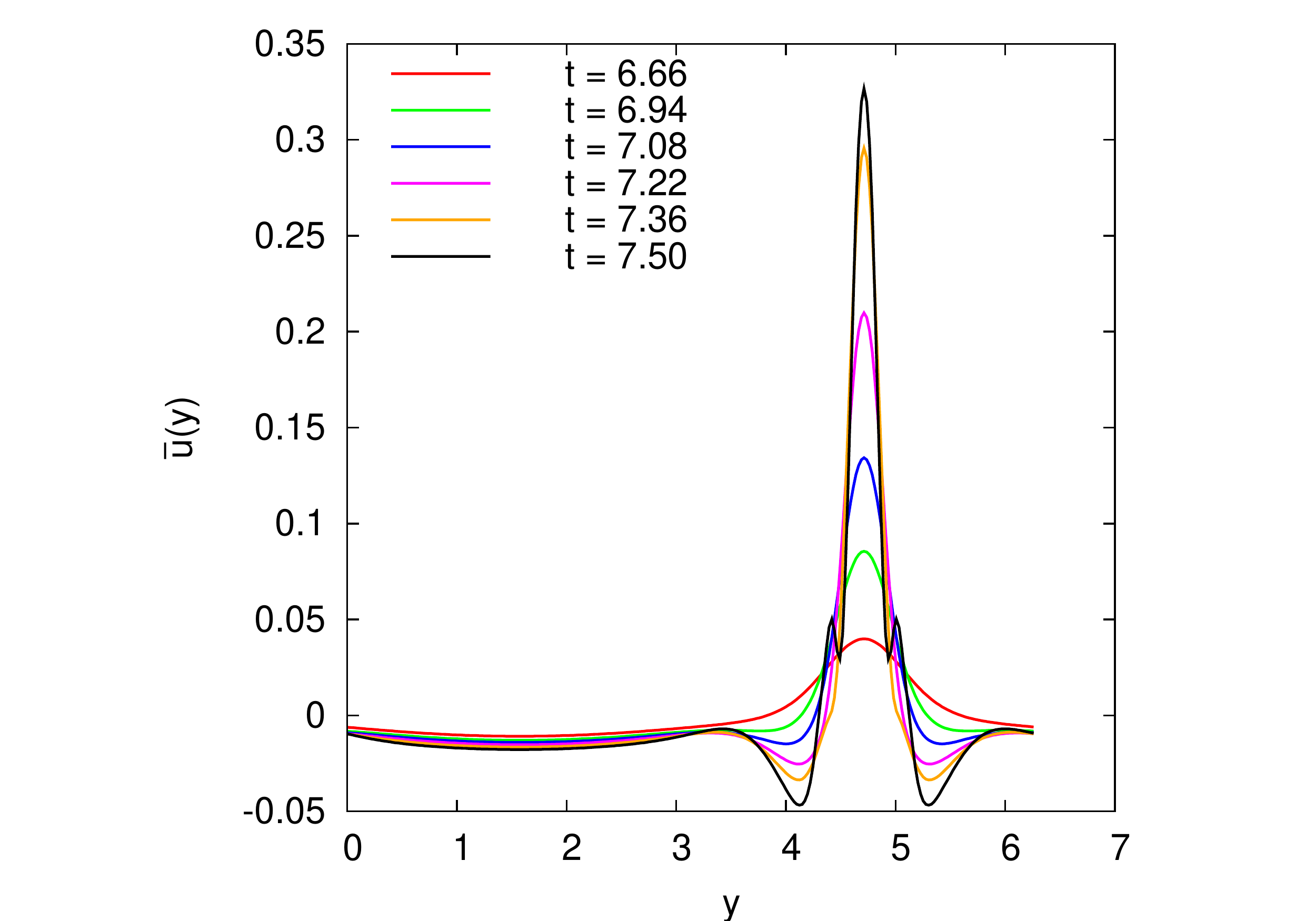}
\caption{\label{fig-Uy.M1} Mean zonal velocity for $M=1$}
\end{figure}

\begin{figure}
\includegraphics[width=7.0cm,angle=-90]{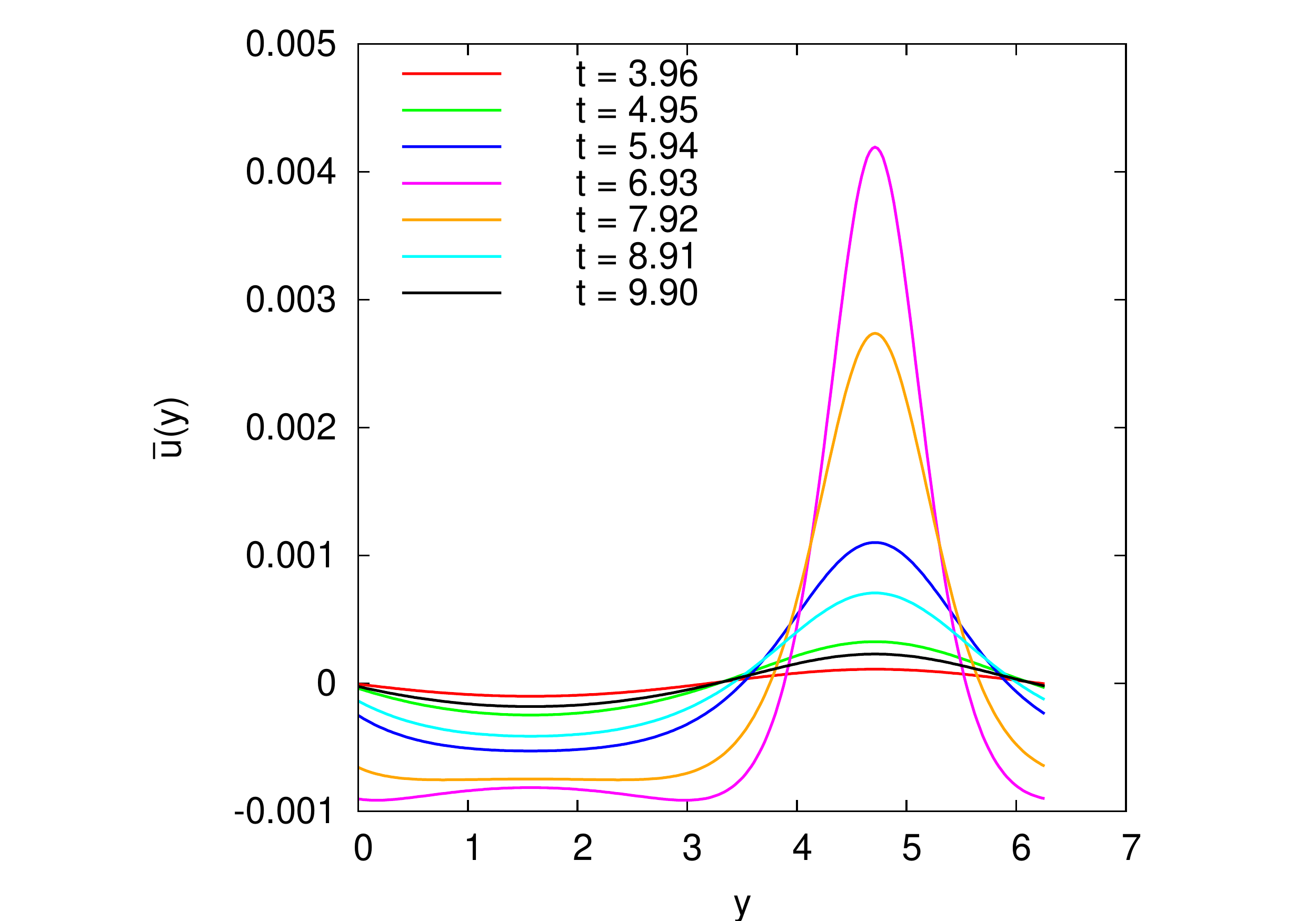}
\caption{\label{fig-Uy.M0.1} Mean zonal velocity for $M=0.1$}
\end{figure}

\begin{figure}
\includegraphics[width=7.0cm,angle=-90]{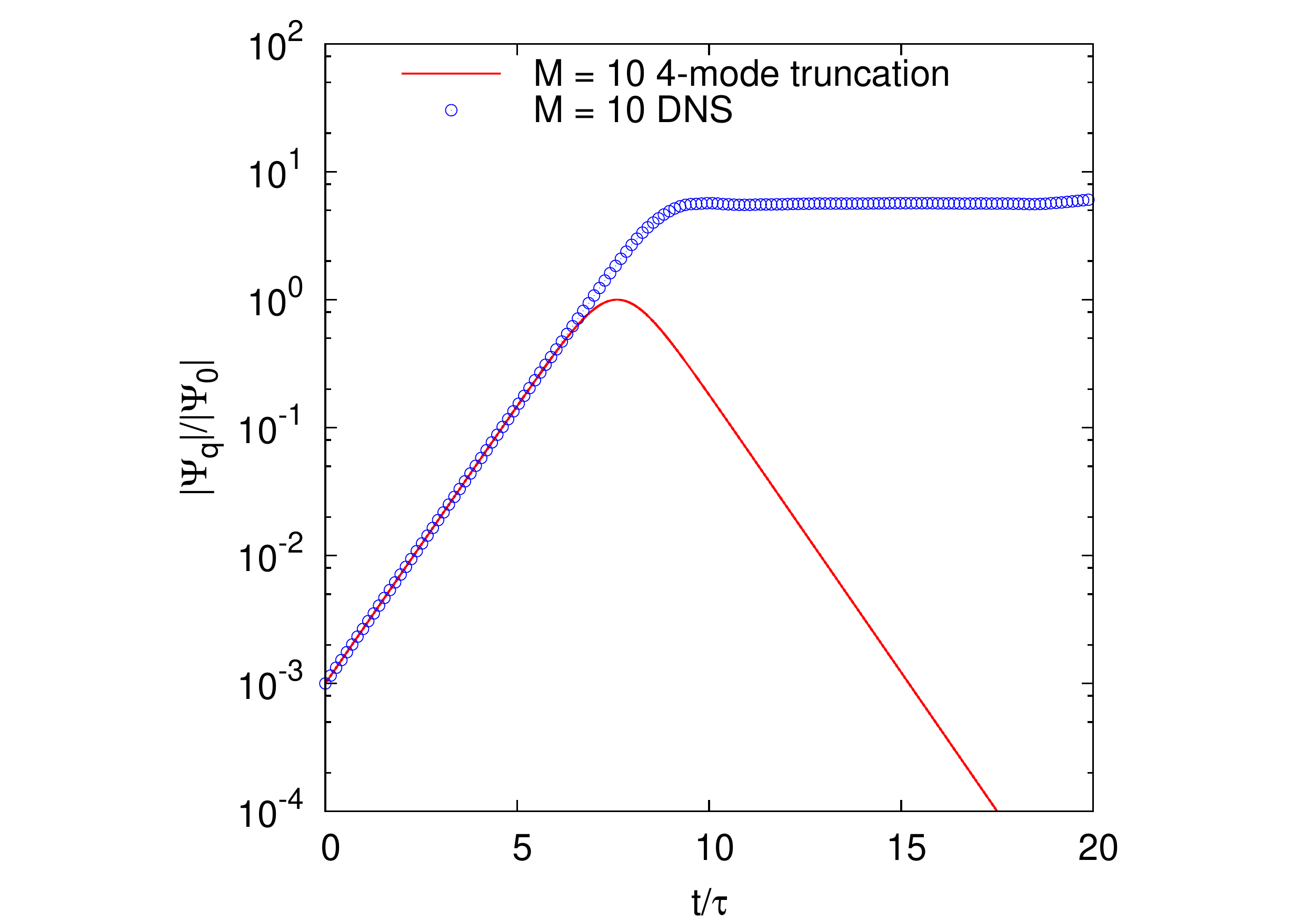}
\caption{\label{fig-growth.M10} Growth of the zonal mode $\vv{q}$ obtained by DNS and by solving 4MT system
for $\vv{p} = (10,0) $ and $\vv{q} = (0,1) $ and for $M=10$.
Time has been scaled by $\tau$ (the inverse of the instability growth rate).}
\end{figure}

\begin{figure}
\includegraphics[width=7.0cm,angle=-90]{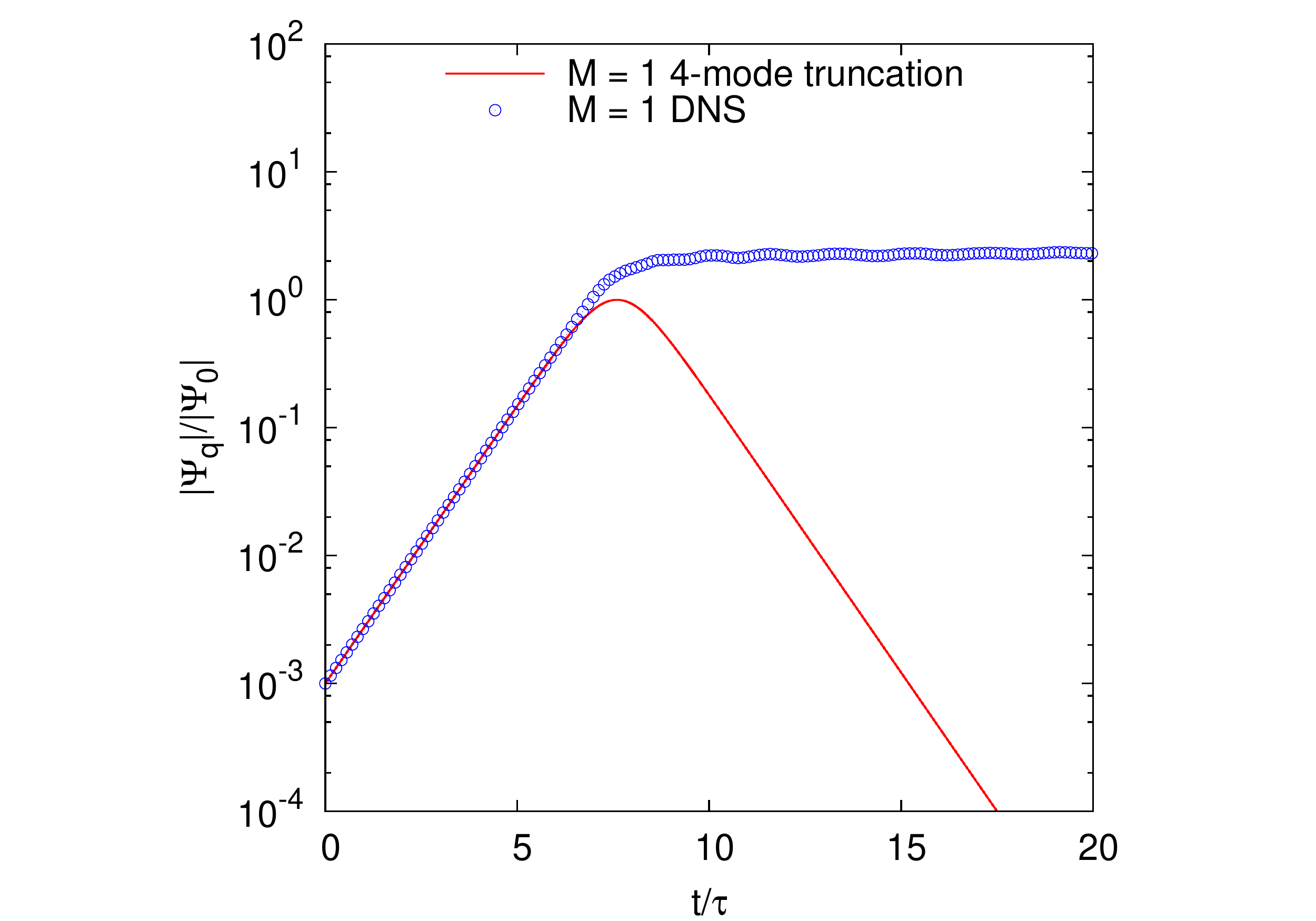}
\caption{\label{fig-growth.M1}
Same as in Fig.~\ref{fig-growth.M10} but
%Growth of the zonal mode $\vv{q}$ obtained by DNS and by solving 4MT system
 for $M=1$}
\end{figure}

\begin{figure}
\includegraphics[width=7.0cm,angle=-90]{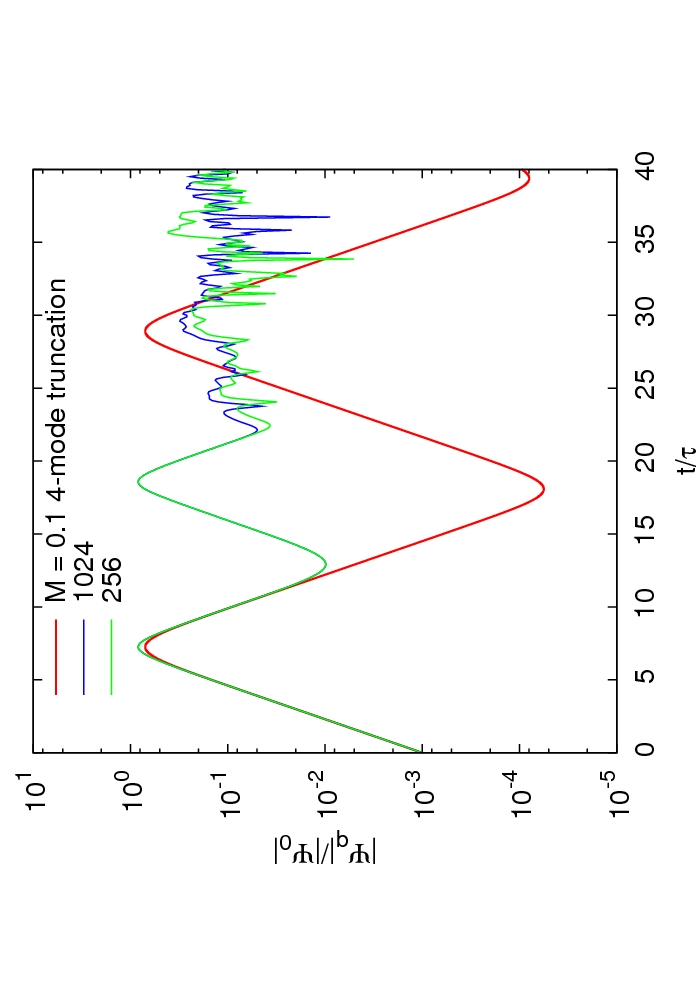}
\caption{\label{fig-growth.M0.1}
Same as in Fig.~\ref{fig-growth.M10} but
%Growth of the zonal mode $\vv{q}$ obtained by DNS and by solving 4MT system
for $M=0.1$. DNS results are presented from calculations at two different resolutions to illustrate that the oscillatory dynamics of the zonal mode are not
influenced by small scale dissipation.}
 \end{figure}

Let us first of all consider the geometry which we dealt with most:
purely meridional carrier wave and purely zonal modulation.
We choose $\vv{p} = (10,0) $ and $\vv{q} = (0,1) $.
A series of frames of the vorticity field for the cases of strong ($M=10$),
medium ($M=1$) and weak ($M=0.1$) nonlinearities  obtained by DNS
are shown in Figs.~(\ref{fig-snapshots.M10}), ~(\ref{fig-snapshots.M1}) and ~(\ref{fig-snapshots.M0.1}) respectively.
The evolution of the mean zonal velocity $\overline u(y)$, averaged over $x$, 
obtained from DNS for the same set of nonlinearities is shown in 
Figs.~(\ref{fig-Uy.M10}),~(\ref{fig-Uy.M1}) and~(\ref{fig-Uy.M0.1}) 
respectively for times close to the formation of the jet.
Finally, evolution of the amplitude, $\left|\psi_{\vv{q}}\right|$, of the
zonal mode for the same runs is shown in Figs~(\ref{fig-growth.M10}), ~(\ref{fig-growth.M1}) and ~(\ref{fig-growth.M0.1}) respectively. For comparison, we
also put the corresponding values of $\left|\psi_{\vv{q}}\right|$ obtained 
from the 4MT.



Immediately, one can see that the initial stage of evolution agrees very well withe
predictions of the linear theory obtained from the 4MT.
Moreover, the 4MT works rather well beyond the linear stage, particularly
in the $M=0.1$ case, where the initial growth reverses in agreement with
the (periodic) behavior of the four-mode system.
For $M=1$, the system's growth does not reverse, but rather experiences a saturation
at the level where the four-mode system reaches maximum and reverses.
The most surprising behavior is seen for $M=10$ where the linear exponential
growth continues well beyond the
point of reversal of the four-wave system, even though the system is clearly nonlinear
at these times and follows a self-similar evolution, see below.

A common feature of the nonlinear saturation stage of the jet growth is 
self-focusing of the zonal jets which become very narrow with respect to the 
initial modulation
wavelength. This self-focusing cannot be described by the
4MT because such anharmonic jet shapes involve strong contributions from higher
harmonics ${\bf p} \pm n {\bf q}$.
For large $M$ and $q \ll k$, such jet "pinching" was predicted theoretically in \cite{MN1994} where
self-similar solutions were obtained describing a collapse of the
jet width. These strong narrow zonal jets are expected to produce transport 
barriers in the
transverse ($y$) direction, which is important in both fusion plasma and the geophysical
contexts.

\begin{figure}
\includegraphics[width=7.0cm,angle=-90]{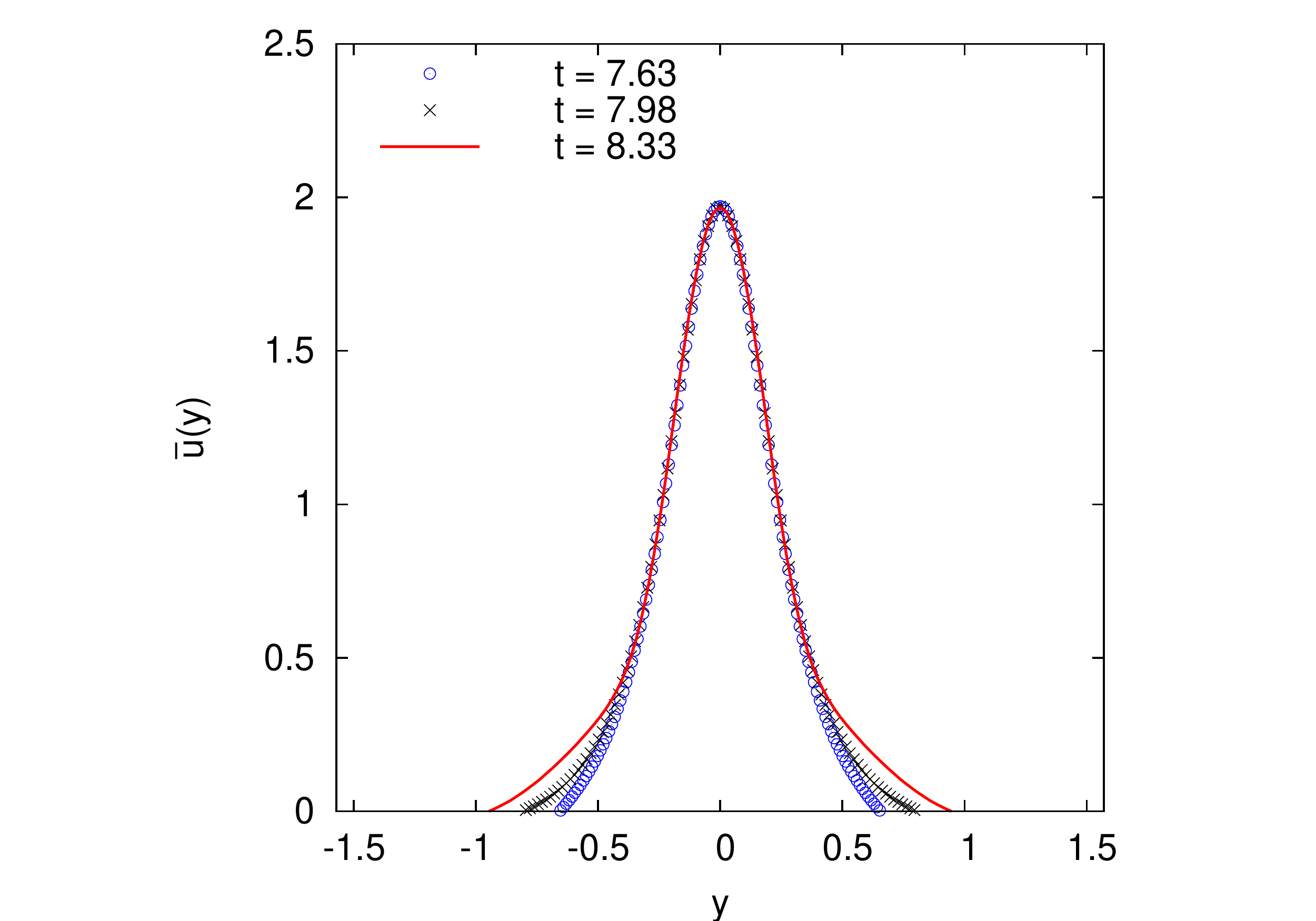}
\caption{\label{fig-selfsim.M10} Zonal velocity $\overline u(y)$  re-scaled with self-similar variables: $\overline u(y,t) = a(t) \, f(b(t) y)$ with
$a(t) = u_0 e^{\gamma t}$ and $b(t) = e^{1.85 t}$ for $M=10$.}
\end{figure}

Figure~(\ref{fig-selfsim.M10}) shows the zonal velocity $\overline u$  re-scaled with
self-similar variables as $\overline u(y,t) = a(t) \, f(b(t) y)$ in the run with $M=10$.
The self-similar stage occurs in the time interval corresponding to the overshoot
in Fig~(\ref{fig-growth.M10}), i.e. after the 4MT has reached its maximum but before
DNS saturated at a plateau.
Empirically, we obtain $a(t) = u_0 \, e^{\gamma t}$ and $b(t) = e^{1.85 t}$.
% where $t_*$ is a jet "collapse".
At least during the decade of amplitude growth observed before pinching occurs,
one can see a good evidence of self-similar
behaviour and, remarkably, the nonlinear growth at the self-similar
stage follows the same exponential law with growthrate $\gamma$
as on the linear instability stage. Note that the  self-similar solutions were obtained in \cite{MN1994}
based on the scale separated
description and, therefore, the self-similar pinching must stop when the
scale separation property breaks down due to the jet narrowing (at which point
a roll-up into vortices occurs, see below).
In the smaller $M$ runs, the overshoot is absent
and the amplitude of the zonal mode decreases after reaching a maximum 
 in correspondence with
the solution of the 4MT. The self-focusing is thereby much reduced and the 
self-similar stage is not clearly observed.

%For lower $M$, the jet pinching does not appear to be self-similar.

One can also see some
qualitative behavior differences for different degrees of nonlinearity $M$.
First of all, we see that the east-west asymmetry is larger for
weaker waves, which is seen as asymmetry of the top and bottom halves
of the vorticity distributions in Figs.~\ref{fig-snapshots.M10},~\ref{fig-snapshots.M1} and~\ref{fig-snapshots.M0.1}.
This is natural considering that for large nonlinearities the beta-effect,
which is the cause of the east-west asymmetry,
is less important.
Further, we see that for large $M$ the nonlinear evolution is vortex dominated
and that the vorticity of the initial carrier wave rolls into vortices organized into Karman-like
vortex streets.  This corresponds to the moment
when the jet velocity reaches a plateau in Figs~(\ref{fig-growth.M10}) and
(\ref{fig-growth.M1}).
On the other hand, in the weak wave case $M \ll 1$
one cannot see vortex roll-ups and the dynamics remains wave dominated.

For large nonlinearities, at the final stages the vortex streets break up due
to a vortex pairing instability, which is followed by a transition to turbulence.
Such turbulence is anisotropic with a pronounced zonal jet component.
On the late frames Fig.~\ref{fig-snapshots.M10} we can see a well formed potential vorticity staircase
structure as described in \cite{DM2008}.

For small nonlinearities, $M \ll 1$, the system's nonlinear evolution starts with self-focusing of the primary wave, but this is followed by a 
quasi-oscillatory behavior where
the system returns close to the initial state. This is very well reproduced by the
four-wave truncation. The same effect was also noted for the Generalised
Hasegawa--Mima model in \cite{MRD2001} and in the atmospheric dynamics context
in \cite{MAH1981}. However, the periodic behavior
% predicted by the four-wave truncation
is not sustained and a transition to an anisotropic turbulent state occurs.
It is interesting that the dominant jet structures observed in such a turbulent state are
off-zonal. This effect may be connected to the off-zonal ``striations'' reported 
for the ocean observations in \cite{MMNS2008}.
We currently regard this connection with some caution since we have not 
performed any time averaging whereas the ocean striations are sufficiently
weak that they only become evident in the averaged data. We hope to 
investigate this further in future work.

\begin{figure}
\includegraphics[width=7.0cm,angle=-90]{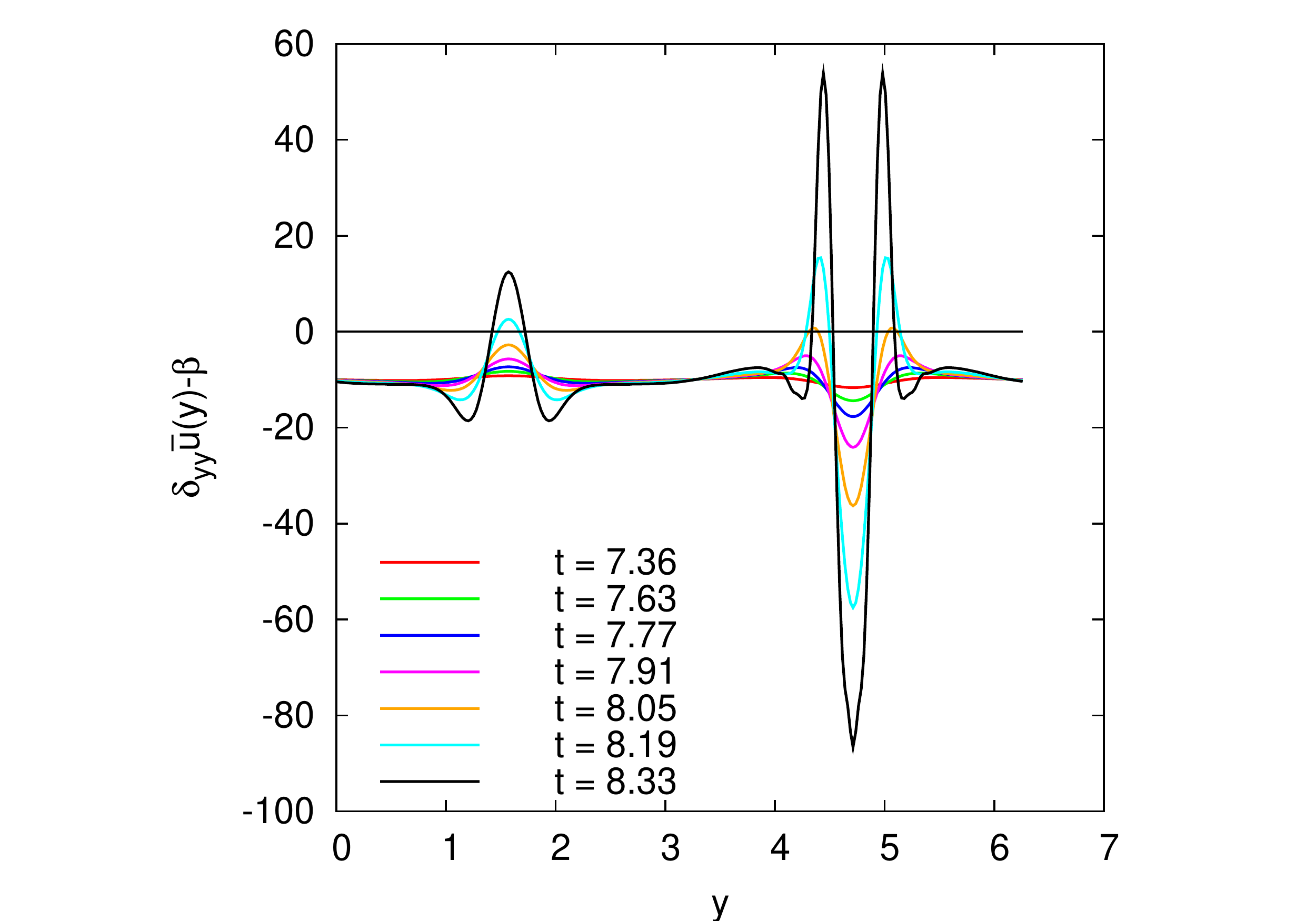}
\caption{\label{fig-rayleigh.M10} The Rayleigh-Kuo profiles for $M=10$}
\end{figure}

\begin{figure}
\includegraphics[width=7.0cm,angle=-90]{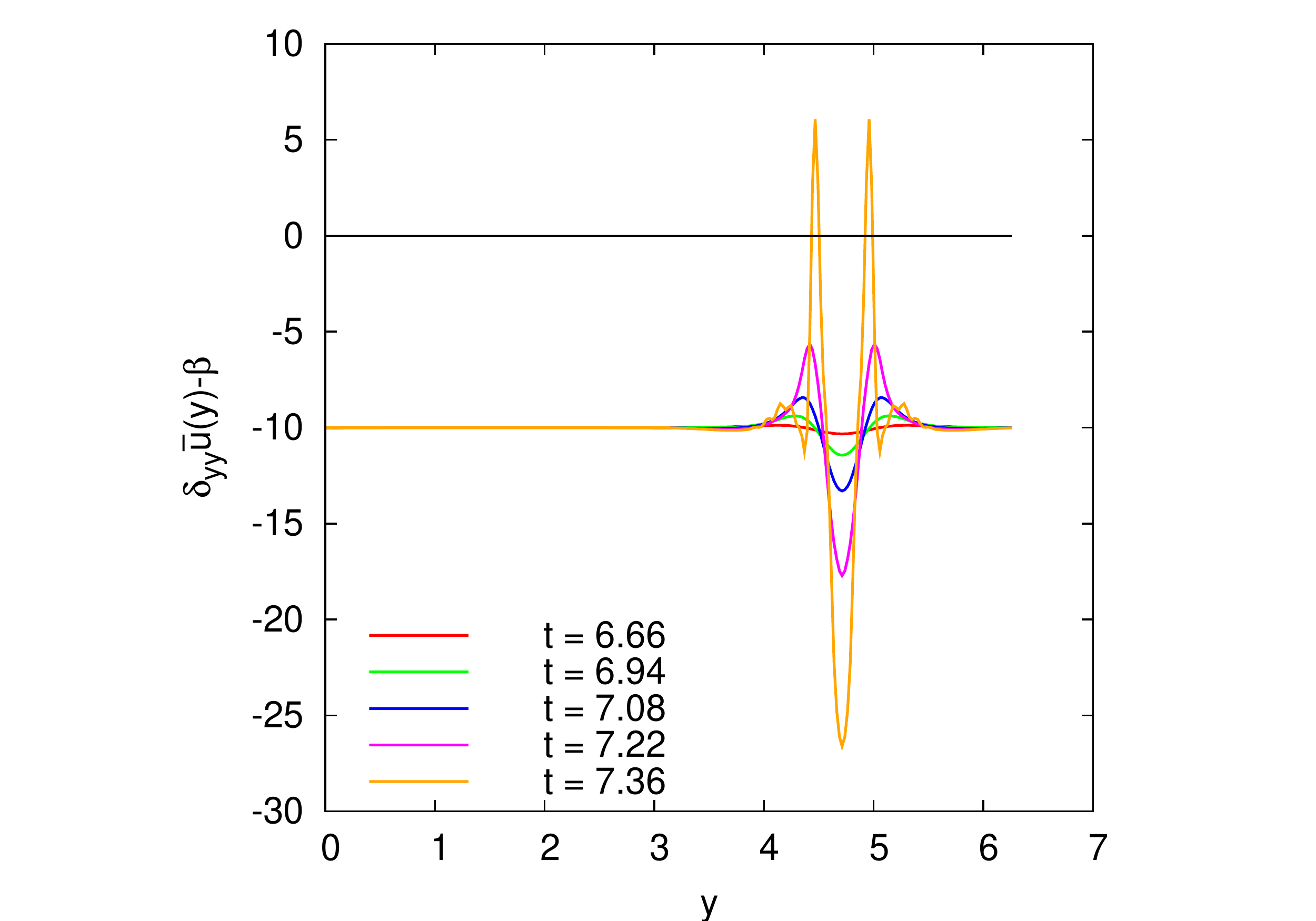}
\caption{\label{fig-rayleigh.M1} The Rayleigh-Kuo profiles for $M=1$}
\end{figure}

\begin{figure}
\includegraphics[width=7.0cm,angle=-90]{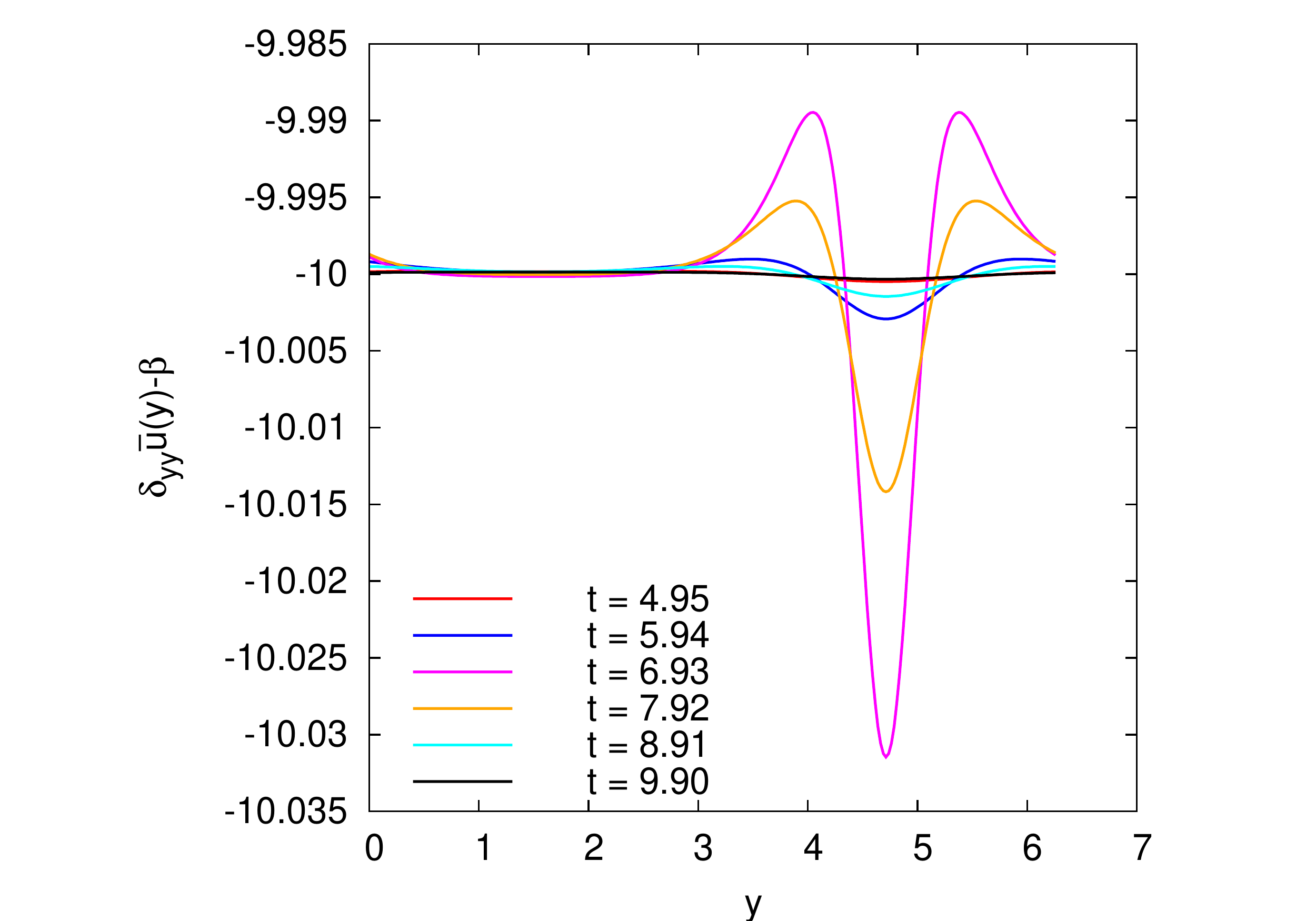}
\caption{\label{fig-rayleigh.M0.1} The Rayleigh-Kuo profiles for $M=0.1$}
\end{figure}

For $M \gtrsim 1$, the vortex streets represent the 2D fine structure of the saturated zonal jets (i.e.
at the plateau part of Figs~(\ref{fig-growth.M10}) and
(\ref{fig-growth.M1})).
Such vortex street configurations are more stable than the plane parallel ($x$-independent)
flows with the same zonal velocity profiles. This can be understood
heuristically (see, for example \cite{MCW2006}, chap. 3) by considering the 
vortices to impart some eddy
viscosity to the mean zonal velocity profile. Recall that stability of the latter is determined
by the Rayleigh-Kuo necessary instability condition \cite{KUO1949},
 \begin{equation}
\partial_{yy} \overline u(y) - \beta >0.
\label{kuo}
\end{equation}
%Figs~(\ref{fig-Uy.M1}) shows an example of the $x$-averaged zonal profile $\overline u(y)$ in the case
%$M=1, \vv{k} = (10,0), \vv{q} =(0,1)$ and
Figs~(\ref{fig-rayleigh.M10}), (\ref{fig-rayleigh.M1}) and (\ref{fig-rayleigh.M0.1})  plot
the profiles of $\partial_{yy} \overline u(y) - \beta$ at different moments in time
corresponding to runs with $M=10$, $M=1$ and $M=0.1$ respectively.
One can see that in Figs~(\ref{fig-rayleigh.M10}) and (\ref{fig-rayleigh.M1})  these
profiles cross the $x$-axis (especially far in the $M=10$ case)
which implies that  the zonal flows get stronger than the limiting values implied
by the Rayleigh-Kuo condition.
We interpret this as a result of a competition between the instability and the
 jet pinching process.  For large $M$ the pinching is self-accelerating (self-similar)
 and it manages to significantly compress/amplify  the unstable jet in the
 finite time needed for the instability to develop (i.e. the inverse growthrate).
On the other hand, in the case $M=0.1$ the jet strength reaches a maximum and then
decreases remaining in the stable range according to the criterion
(\ref{kuo}).

\begin{figure}
\includegraphics[width=7.0cm, angle=-90.0]{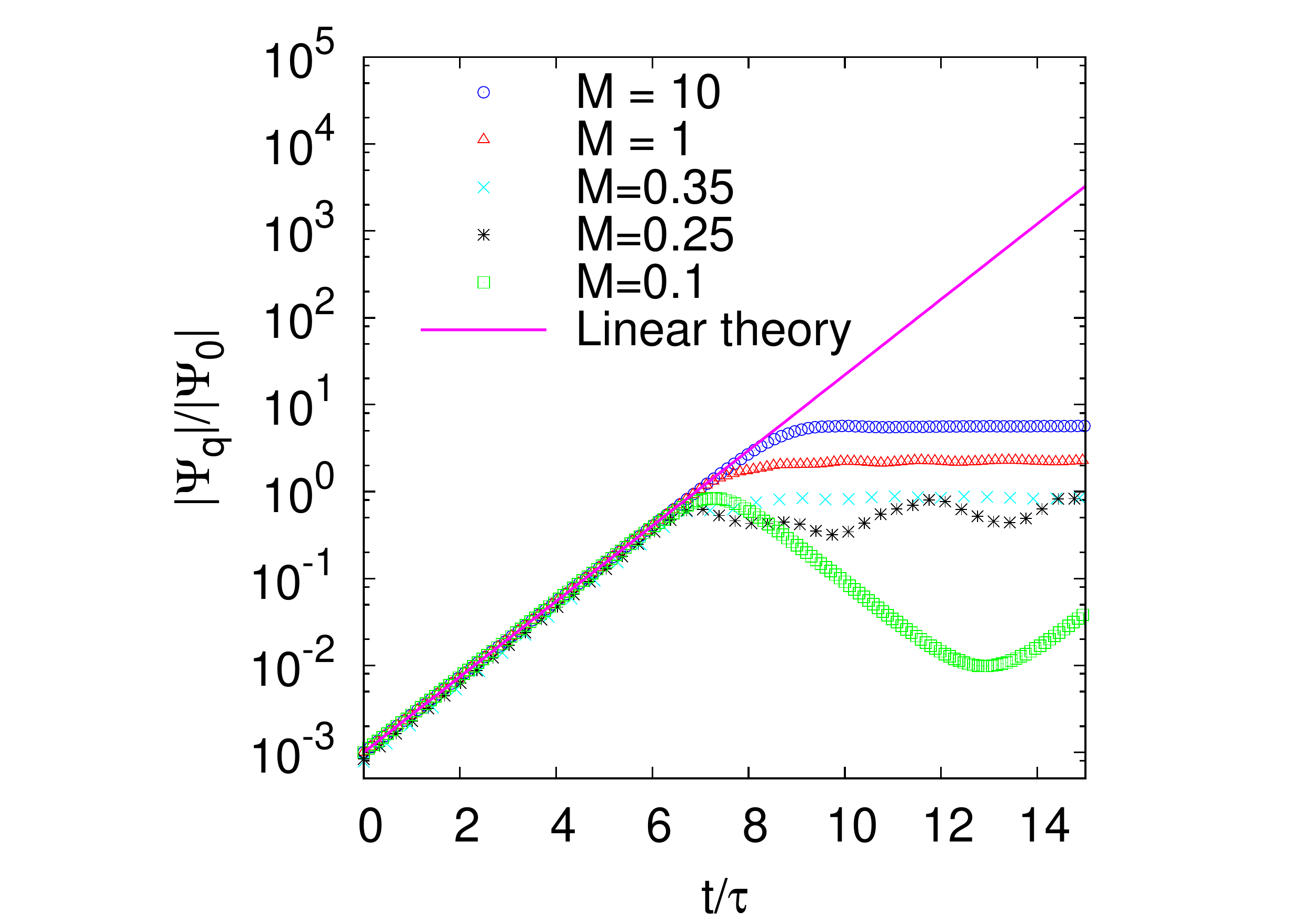}
\caption{\label{fig-growth_earlyTime} Growth of zonal perturbations due to modulational instability of a meridional carrier wave having $\vv{p}=(10,0)$ for several different values of $M$. The amplitude of the zonal mode has been scaled by $\Psi_0$ and time has been scaled by $\tau$ (the inverse of the instability growth rate).}
\end{figure}

These results allow us to draw conclusions about the critical value of 
nonlinearity, $M=M_*$, which separates the two qualitatively different types of behavior: vortex roll-up and
saturation vs the oscillatory dynamics, see Figure~(\ref{fig-growth_earlyTime}).
 If the jet strength maximum, as predicted by the
4MT, exceeds the values of  Rayleigh-Kuo necessary instability condition (\ref{kuo})
then the vortex roll-up occurs and the jet enters into a saturated, relatively
 long-lived plateau stage.
At this moment, the system's behavior starts to depart from  the 4MT model.
On the other hand, if the jet strength maximum,  as predicted by the
4MT, remains below the  Rayleigh-Kuo threshold then the system's growth reverses
and it follows the 4MT dynamics for longer time.


This simple picture leads to a qualitative physical estimate for  $M_* $
and for the saturated velocity of the jet. Let us start with the latter, see
Fig.~(\ref{fig-growth_earlyTime}).

Since the $x$-periodicity is preserved, the step of the vortex street is equal to the
wavelength of the original carrier wave. The vortices in the stable vortex street are approximately
round and the $y$ spacing between the vortices is approximately the same as the $x$-spacing.
Thus, the saturation width of the pinched jet is of order of the wavelength of the initial
carrier wave. Lagrangian conservation of the potential vorticity determines the final
saturated amplitude of the jet. Indeed, the same  vorticity as in the initial carrier wave
rolls into the vortices (the $\beta y$ part of the potential vorticity does not
play much role here since the fluid parcels remain at about the same $y$'s) and the rest of
the vorticity is shed in between of the vortex streets, shredded by shearing and dissipated.
Thus, the jet saturation velocity is of the order of the velocity amplitude of the
initial carrier wave,
 \begin{equation}
\overline u_{max}  \sim \frac{M \beta } {p^2}.
\label{umax}
\end{equation}
This estimate is well confirmed by our numerical results for $M =1$  and $M =10$.
Indeed, taking values of $u_{max} $ from Figs~(\ref{fig-Uy.M10}) and
(\ref{fig-Uy.M1})) we get
\begin{equation}
\overline u_{max}  \approx 3 \frac{M \beta } {p^2}.
\label{umax1}
\end{equation}
 Now, estimating $\partial_{yy} \overline u(y)$ as $p^2 u_{max}$ and using (\ref{umax1}) we can
 rewrite the instability condition (\ref{kuo}) in a very simple form as
 $$M > M_* \sim 1/3.$$
  Our numerics show that $M_* \approx 0.25-0.35$, see Figure~(\ref{fig-growth_earlyTime}). 
  Note that the boundary is not sharp.
    For $M=0.25$ the dynamics is definitely wave-dominated, 
    however some elongated fuzzy vortices are still apparent whereas for $M=0.35$ streets of round vortices are clearly 
    formed with some wave-like oscillations still present.

\subsection{\label{nonlin-unstable-meridional-offzonal}
Case of Meridional Carrier Wave and Off-Zonal Modulation.}

Above we considered the case when the carrier wave is purely meridional and the modulation is purely
zonal. This geometry is important considering that both the baroclinic instability in GFD and the drift-wave
instabilities in plasmas typically have most unstable modes being in the meridional direction.
These modes can be considered as an initial condition for the secondary modulational instabilities
as it is done in the present paper.
At the same time, we have established above that the most unstable modulations
for $M >0.53$ are zonal.

On the other hand, for low $M$ the most unstable modulations are off-zonal.
This, in our opinion, is the reason why the final statistical state in the
system in the $M=0.1$ simulation showed presence of off-zonal anisotropic flows
even though the initial modulation was purely zonal.
Moreover, it is quite likely that in such weakly-nonlinear cases the system will
pick the modulation which is off-zonal already at the initial moments.

\begin{figure*}
\includegraphics[width=17.5cm]{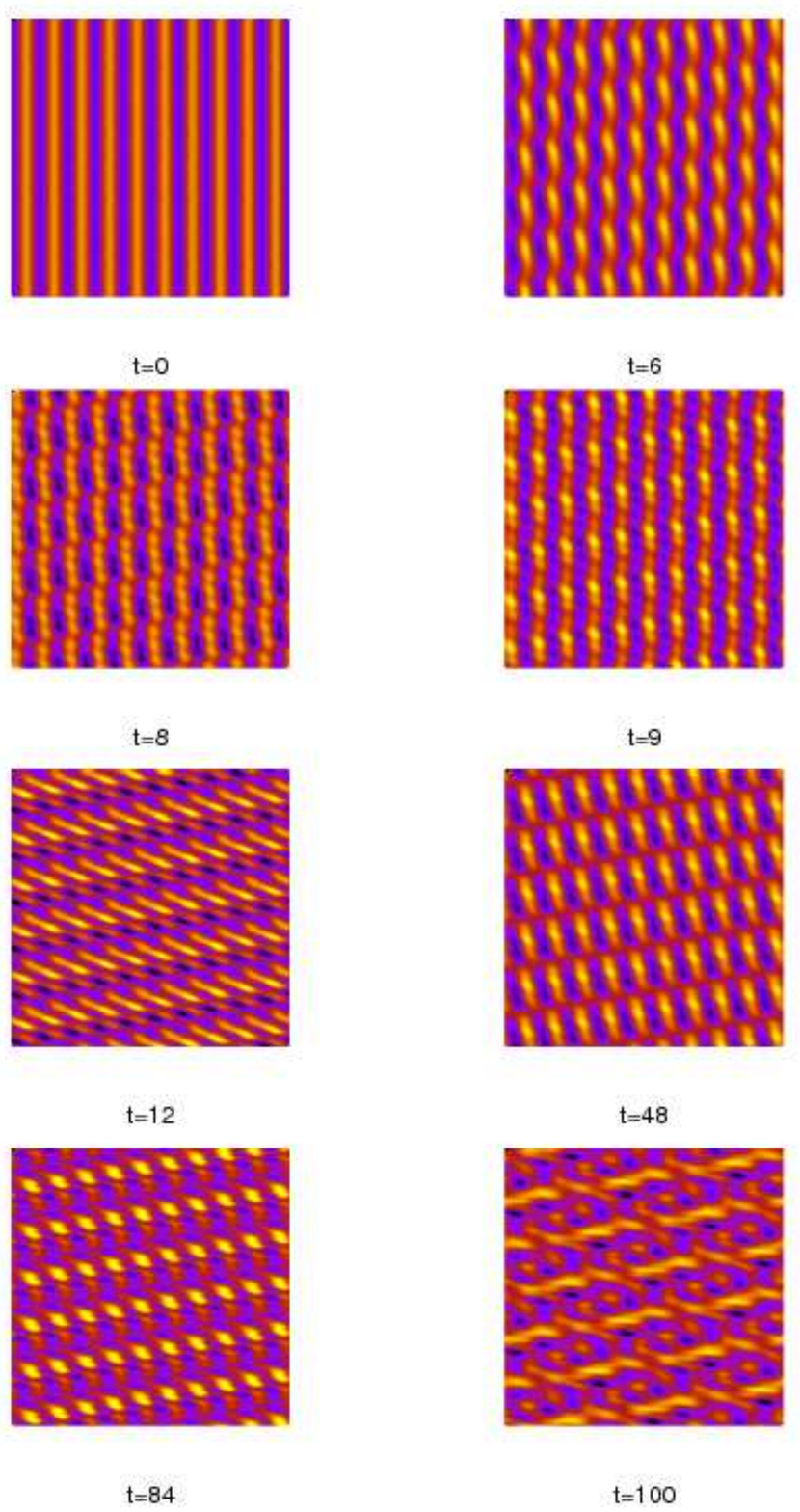}
\caption{\label{fig-snapshots.q_9_6} Vorticity snapshots showing the growth, saturation and transition to turbulence of an off-zonal perturbation of a meridional carrier wave having $M=0.1$. }
\end{figure*}

Thus, here we will consider a case with $M=0.1$
where we start with purely meridional carrier wave, $\vv{p}=(10,0)$ and with the modulation wavevector corresponding to the
 fastest growing mode in this case, namely $\vv{q} = (9,6)$.
Corresponding numerical results for this case are shown in
Fig.~(\ref{fig-snapshots.q_9_6}) (vorticity snapshots) and Fig.~(\ref{fig-growth.q_9_6}) (evolution of the $q$-mode amplitude $|\psi_q|$ and  
respective results obtained from simulating the 4MT and 3MT models).

\begin{figure}
\includegraphics[width=8.0cm,angle=-90]{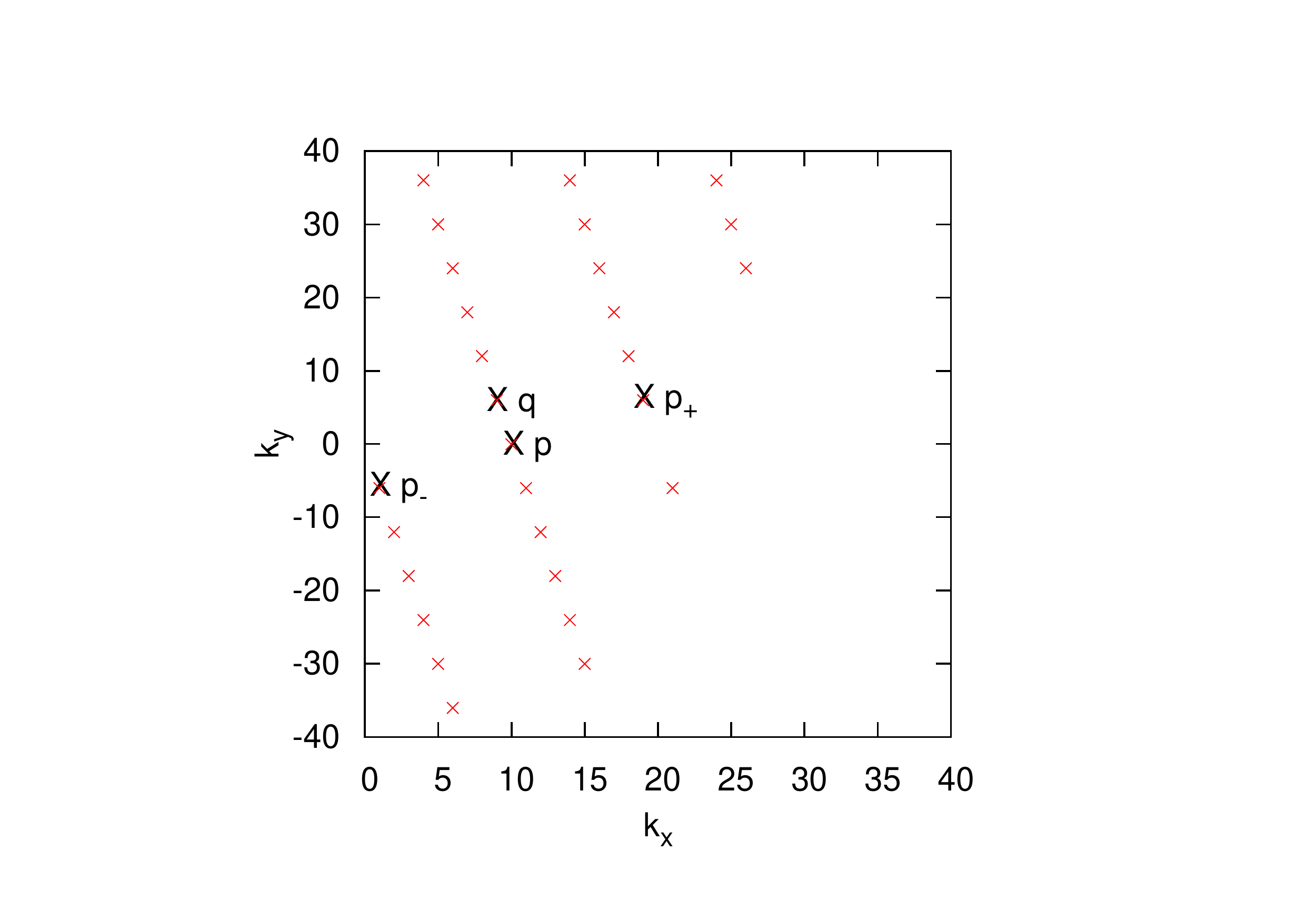}
\caption{\label{fig-crystalline}  Excited modes at $t=60$ with energy within 1\% of the energy of the initial perturbation in run with for $q=(9,6)$, $M=0.1$.}
\end{figure}

First of all, as in all previous cases, we see good agreement of the initial evolution with predictions for the
linear instability obtained based on the 4MT and the 3MT  models.
Moreover, we see that the  4MT and the  3MT in this case qualitatively describe the nonlinear behavior too.
Namely, like in the four-mode system, we see oscillatory behavior, even though the oscillations appear to be irregular.
However, these irregular oscillations are clearly non-turbulent, as one can see from the vorticity
frames in Fig.~(\ref{fig-snapshots.q_9_6}) which shows quite a regular pattern even at $t=100$ (in the units
of the inverse instability growthrate), - by which time the respective $M=0.1$ system with zonal ${\bf q}$ is
completely turbulent, see Fig.~\ref{fig-snapshots.M0.1}. Another way to see that the dynamics are regular
in this case is to look at the 2D $k$-spectra shown in  Fig.~(\ref{fig-crystalline}).  At $t = 0$, the only excited modes  are the carrier wave ${\bf p}$, modulation
${\bf q}$ and two satellites  ${\bf p}\pm {\bf q}$: these modes are marked by bold symbols in 
Fig~(\ref{fig-crystalline}).  At $t=60$ one can see a regular "crystalline"
structure corresponding to a discrete set of nodes $n{\bf p} +m {\bf v}$ (with integer values of $m$ and $n$) with energy within 1\% of the initial carrier wave energy.
Transition to turbulence does eventually occur after a very long time, and the turbulent state
does exhibit off-zonal striations similar to the  respective $M=0.1$ system with zonal initial modulations ${\bf q}$.

%\begin{figure}
%\includegraphics[width=8.0cm,angle=-90]{q_9_6_initial_spectrum}
%\caption{\label{fig-initial_crystalline} 2D $k$-spectrum for $q=(9,6)$, $M=0.1$ at $t=0$ showing excited modes with energy similar to the carrier wave energy.}
%\end{figure}

%%%%%%%%%%%%%%%%%%%%%%%%%%%%%%%%%%%%%%%%%%%%%%%%%
\section{\label{nonlin-stable}Stable Case}

Above we considered in detail various situations where the linear theory based on the 4MT model predicts
instability. We have also investigated the linearly stable case.

\begin{figure}
\includegraphics[width=6.0cm,angle=-90]{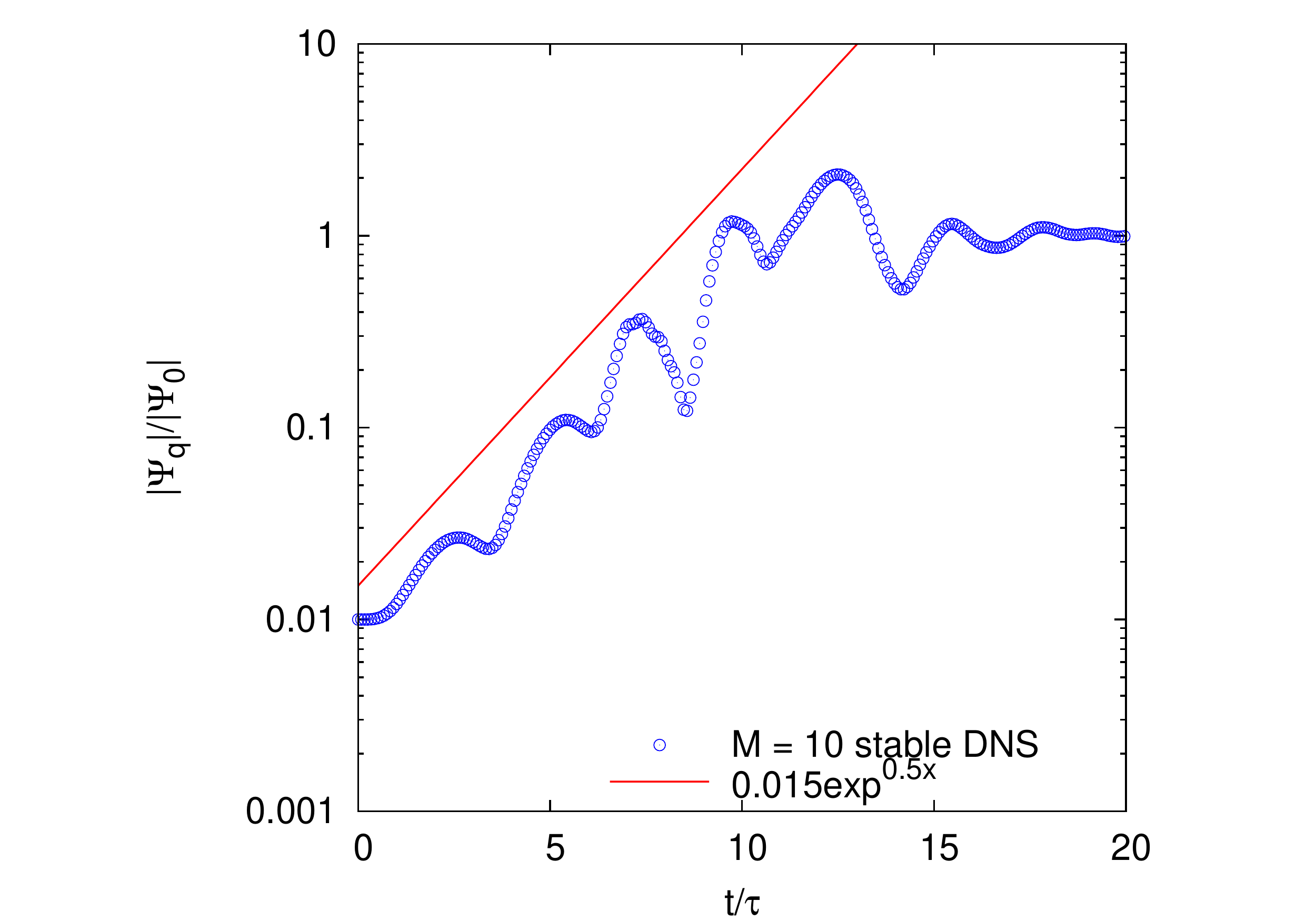}
\caption{\label{fig-stable.M10} Growth of the zonal mode $\vv{q}$ obtained by DNS for $\vv{p} = (8,6) $ and $\vv{q} = (0,1) $ and for $M=10$.}
\end{figure}

\begin{figure}
\includegraphics[width=6.0cm,angle=-90]{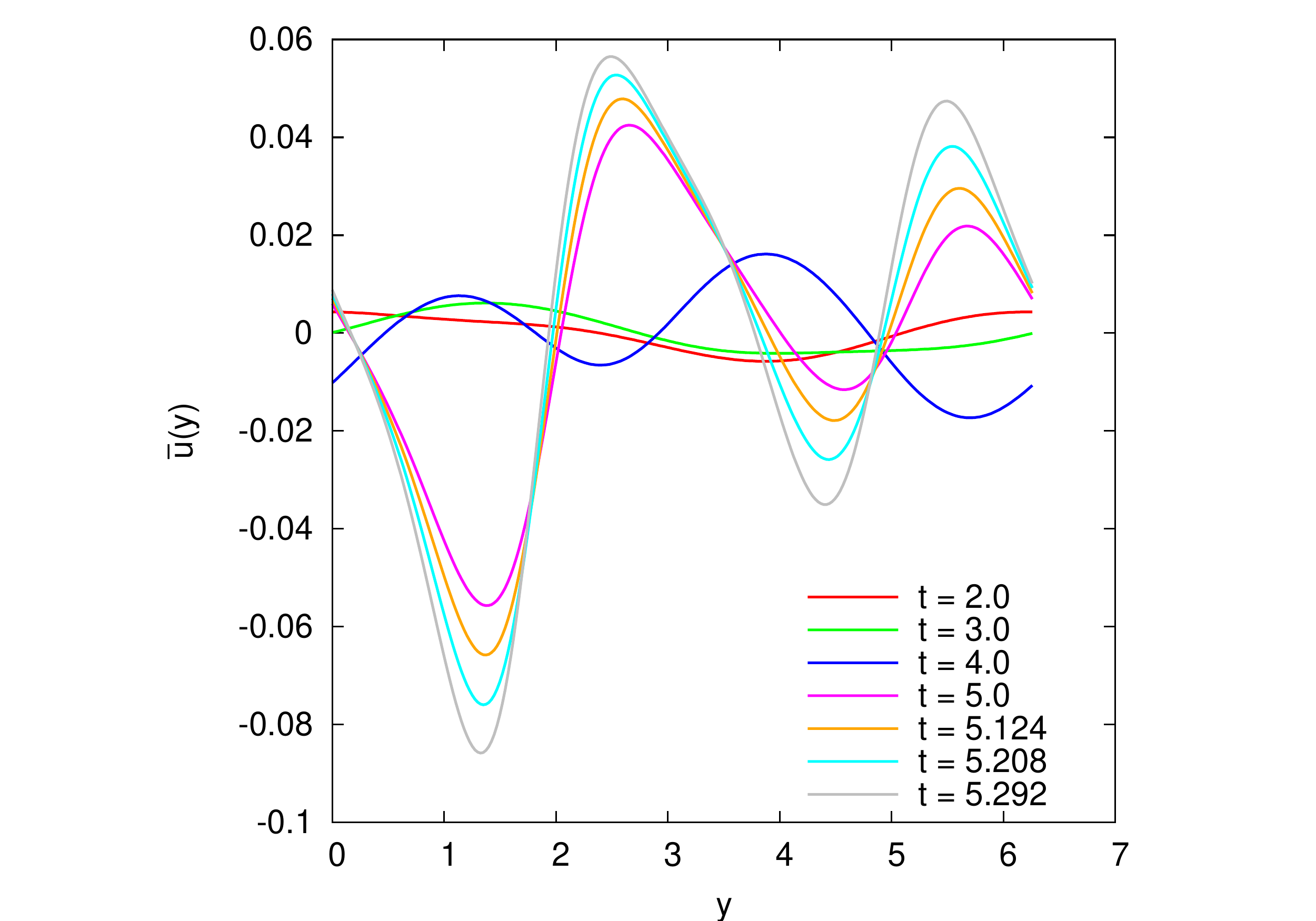}
\caption{\label{fig-Uy.stable.M10} Mean zonal velocity profile for the stable configuration $\vv{p} = (8,6) $, $\vv{q} = (0,1) $ and $M=10$.}
\end{figure}

For small $M$ the zonal mode in the modulationally stable case behaves as expected, following the 4MT theory without growth of the mode.  
In this case, deviations from the 4MT are tiny, hence we omitted  the corresponding graph.
For $M \gg 1$, the situation is more interesting.
Fig.~\ref{fig-stable.M10} shows the evolution of the zonal mode for the run with $\vv{p} = (8,6) $ and $\vv{q} = (0,1) $ and for $M=10$
which corresponds to a linearly stable configuration within the 4MT model.
We see agreement with the 4MT stability prediction at early times, i.e. the zonal mode
 is not growing in Fig.~\ref{fig-stable.M10} for $t \lesssim 1$.
However, after about one timescale  the zonal mode quickly breaks into growth, increasing (more or less exponentially) by two orders of magnitude  Hence, the 
4MT instability criterion must be used with caution if $M \gg 1$.  Further, for $M \gg 1$ stable case
Manin and Nazarenko \cite{MN1994} predicted zonal velocity profile steepening for and this is evident in Fig.~\ref{fig-Uy.stable.M10} where the initial sinusoidal profile develops into a triangular Burger's shock-type profile.

\section{\label{summary} Summary and Conclusions}

In this paper we dealt with the theory and numerical simulations of the modulational
instability of the Rossby/drift waves described by the CHM model.
We have revisited the linear theory of Gill \cite{GIL1974} using the 4-mode truncation and emphasised the role of the carrier wave amplitude/nonlinearity, the
role of the deformation
radius and the role of resonant wave interactions in the case of weakly 
nonlinear carrier
wave. We found a  change of the most unstable modulation from
zonal to off-zonal when the carrier wave nonlinearity parameter $M$ falls below
a critical value, $M >0.53$. This latter effect may be important for understanding
the recent ocean observation of off-zonal jet striations \cite{MMNS2008}.
It is also a likely mechanism for generation of off-zonal random jets in our
numerical simulations at the late development stages of
the modulational instability for $M=0.1$ case.

We established how the modulational instability relates to the decay 
instability obtained within the 3-mode truncation in order to
clarify the question whether the dominant nonlinear mechanism of the modulational
instability is three-wave or four-wave. The 3MT works very well for low nonlinearities
$M$ when the carrier wave and the modulation belong to the same resonant triad which is
non-degenerate, i.e. when it does not include wavevectors too close to $k=0$ point where the
two branches of the resonant curve intersect. This excludes the most popular choice
of purely zonal modulations, for which 3MT appears to be a bad model.
On the other hand, 4MT is more general and it works very well for small nonlinearities $M$
including the case of the purely zonal modulations. Moreover, 4MT also works
well for the initial evolution in the strongly nonlinear cases, $M \gtrsim 1$, including the
linear growth stage and the prediction of the critical nonlinearity $M_*$ for which
transition from the saturated to the oscillatory nonlinear regimes is observed.

We checked the solutions of the truncated system against DNS and concluded that
it is suffient to predict both the linear instability and the early 
nonlinear evolution of
this instability. We showed that DNS agrees very well with the linear predictions in all cases.
In many cases, particularly for small $M$ and in the stable configurations,
the four-wave truncation predicts quite well the early and intermediate
nonlinear evolution phases. The nonlinear evolution for small $M$'s is characterized
by dominant wave dynamics, whereas for large $M$'s the nonlinear evolution leads to
rolling up of the carrier wave vorticity into Karman-like vortex streets.
Such hydrodynamic vortices behave very differently from waves, and it it precisely
at the moment of roll-up that the full system's evolution strongly diverges from the
prediction of the four-wave truncation. After the roll-up the full system enters
into a saturated quasi-stable state which persists for a relatively long time 
but eventually decays due to presence of hyper-viscosity. On other hand, the 
corresponding four-wave system keeps going though an infinite sequence of 
nonlinear oscillations.
If $M$ is small and the roll-ups do not occur (or are delayed)
the full system may follow its four-wave counterpart for much longer:
its initial growth can reverse and may exhibit the nonlinear oscillations
associated with the 4MT.

Finally, we would like to emphasize two physical effects that can be important for both
plasma and GFD systems. For $M \gtrsim 1$ we observe the formation of stable,
narrow zonal jets, in agreement with earlier theoretical predictions of \cite{MN1994}.
As we mentioned, these jets are more stable than one would expect based
on the Rayleigh-Kuo criterion alone because their 2D structure consists of stable
vortex streets. Such narrowjets represent very effective transport barriers
which may be responsible for the LH transitions in tokamaks.
Indeed, narrowness of the jet causes characteristic
pedestal-like radial profiles of the particles and energy typically seen in
tokamaks during H-mode, i.e. the narrow jet provides a thin "insulation" layer (called
internal transport barrier) which keeps the particle and energy densities significantly
higher inside.
The second physical effect we would like to mention occurs at low nonlinearities
$M$. This is the fact that the system tends to select the states with somewhat
off-zonal jet structures. This tendency to favour the off-zonal
structures is seen already on the level of the linear analysis, where as we showed
the most unstable modulation changes from zonal to off-zonal when the nonlinearity is
reduced. Possibly, this mechanism can explain recent ocean observation of off-zonal jet
striations \cite{MMNS2008} (note that the nonlinearity of the ocean Rossby waves
in these situations is likely to be rather low).

In future, it would be interesting to study in more detail how the transport properties
are affected by both zonal and off-zonal jets that arise in the
strongly and weakly nonlinear cases respectively. In particular, it would be interesting
to see how different are the transport barriers provided by coherent vortex streets from
the barriers provided by random jets at the later stages.
It would also be interesting to study situations where a broad spectrum of modulations is
present initially and, in particular, to verify that the system selects the most 
unstable one. If the
carrier wave spectrum is not narrow it would be of interest to study
when the modulational instability wins over the inverse cascade mechanism.

%\begin{figure}
%%\includegraphics[width=7.0cm]{growth.lateTime.M10}
%%\caption{\label{fig-compareODE.M10} Comparison with ODEs for $M=10$..}
%%\end{figure}

%%\begin{figure}
%%\includegraphics[width=7.0cm]{growth.lateTime.M0.5.eps}
%%\caption{\label{fig-compareODE.M0.5} Comparison with ODEs for $M=0.5$..}
%%\end{figure}

%%\begin{figure}
%%\includegraphics[width=7.0cm]{growth.lateTime.M0.1.eps}
%%\caption{\label{fig-compareODE.M0.1} Comparison with ODEs for $M=0.1$..}
%%\end{figure}

%%\begin{figure*}
%%\includegraphics[width=12.5cm]{vorticitySnapshots.M10.eps}
%%\caption{\label{fig-snapshots.M10} Vorticity snapshots showing the growth, saturation and transition to turbulence of a zonal perturbation of a meridional carrier wave having $M=10$. }
%%\end{figure*}

%%\begin{figure*}
%%\includegraphics[width=12.5cm]{vorticitySnapshots.M1.eps}
%%\caption{\label{fig-snapshots.M1} Vorticity snapshots showing the growth, saturation and transition to turbulence of a zonal perturbation of a meridional carrier wave having $M=1$. }
%%\end{figure*}

%%\begin{figure*}
%\includegraphics[width=12.5cm]{vorticitySnapshots.M0.1.eps}
%\caption{\label{fig-snapshots.M0.1} Vorticity snapshots showing the growth, saturation and transition to %turbulence of a zonal perturbation of a meridional carrier wave having $M=0.1$. }
%%\end{figure*}

\bibliography{all}

\begin{thebibliography}{39}
\expandafter\ifx\csname natexlab\endcsname\relax\def\natexlab#1{#1}\fi
\expandafter\ifx\csname bibnamefont\endcsname\relax
  \def\bibnamefont#1{#1}\fi
\expandafter\ifx\csname bibfnamefont\endcsname\relax
  \def\bibfnamefont#1{#1}\fi
\expandafter\ifx\csname citenamefont\endcsname\relax
  \def\citenamefont#1{#1}\fi
\expandafter\ifx\csname url\endcsname\relax
  \def\url#1{\texttt{#1}}\fi
\expandafter\ifx\csname urlprefix\endcsname\relax\def\urlprefix{URL }\fi
\providecommand{\bibinfo}[2]{#2}
\providecommand{\eprint}[2][]{\url{#2}}

\bibitem[{\citenamefont{{Gill}}(1974)}]{GIL1974}
\bibinfo{author}{\bibfnamefont{A.~E.} \bibnamefont{{Gill}}},
  \bibinfo{journal}{Geophys. Fluid Dyn.} \textbf{\bibinfo{volume}{6}},
  \bibinfo{pages}{29} (\bibinfo{year}{1974}).

\bibitem[{\citenamefont{{Manin} and {Nazarenko}}(1994)}]{MN1994}
\bibinfo{author}{\bibfnamefont{D.~Y.} \bibnamefont{{Manin}}} \bibnamefont{and}
  \bibinfo{author}{\bibfnamefont{S.~V.} \bibnamefont{{Nazarenko}}},
  \bibinfo{journal}{Phys. Fluids} \textbf{\bibinfo{volume}{6}},
  \bibinfo{pages}{1158} (\bibinfo{year}{1994}).

\bibitem[{\citenamefont{Simon}(1999)}]{SIM1999}
\bibinfo{author}{\bibfnamefont{A.~A.} \bibnamefont{Simon}},
  \bibinfo{journal}{Icarus} \textbf{\bibinfo{volume}{141}}, \bibinfo{pages}{29}
  (\bibinfo{year}{1999}).

\bibitem[{\citenamefont{{Sanchez-Lavega}
  et~al.}(2000)\citenamefont{{Sanchez-Lavega}, Rojas, and Sada}}]{SLS2000}
\bibinfo{author}{\bibfnamefont{A.}~\bibnamefont{{Sanchez-Lavega}}},
  \bibinfo{author}{\bibfnamefont{J.~F.} \bibnamefont{Rojas}}, \bibnamefont{and}
  \bibinfo{author}{\bibfnamefont{P.~V.} \bibnamefont{Sada}},
  \bibinfo{journal}{Icarus} \textbf{\bibinfo{volume}{147}},
  \bibinfo{pages}{405} (\bibinfo{year}{2000}).

\bibitem[{\citenamefont{Galperin et~al.}(2004)\citenamefont{Galperin, Nakano,
  Huang, and Sukoriansky}}]{GNHS2004}
\bibinfo{author}{\bibfnamefont{B.}~\bibnamefont{Galperin}},
  \bibinfo{author}{\bibfnamefont{H.}~\bibnamefont{Nakano}},
  \bibinfo{author}{\bibfnamefont{H.-P.} \bibnamefont{Huang}}, \bibnamefont{and}
  \bibinfo{author}{\bibfnamefont{S.}~\bibnamefont{Sukoriansky}},
  \bibinfo{journal}{Geophys. Res. Lett.} \textbf{\bibinfo{volume}{31}},
  \bibinfo{pages}{L13303} (\bibinfo{year}{2004}).

\bibitem[{\citenamefont{{Lewis}}(1988)}]{LEW1988}
\bibinfo{author}{\bibfnamefont{J.~M.} \bibnamefont{{Lewis}}},
  \bibinfo{journal}{Bull. Amer. Met. Soc.} \textbf{\bibinfo{volume}{79}}
  (\bibinfo{year}{1988}).

\bibitem[{\citenamefont{Maximenko et~al.}(2008)\citenamefont{Maximenko,
  Melnichenko, Niiler, and Sasaki}}]{MMNS2008}
\bibinfo{author}{\bibfnamefont{N.~A.} \bibnamefont{Maximenko}},
  \bibinfo{author}{\bibfnamefont{O.~V.} \bibnamefont{Melnichenko}},
  \bibinfo{author}{\bibfnamefont{P.~P.} \bibnamefont{Niiler}},
  \bibnamefont{and} \bibinfo{author}{\bibfnamefont{H.}~\bibnamefont{Sasaki}},
  \bibinfo{journal}{Geophys. Res. Lett.} \textbf{\bibinfo{volume}{35}},
  \bibinfo{pages}{L08603} (\bibinfo{year}{2008}).

\bibitem[{\citenamefont{{James}}(1987)}]{JAM1987}
\bibinfo{author}{\bibfnamefont{I.~N.} \bibnamefont{{James}}},
  \bibinfo{journal}{J. Atmo. Sci.} \textbf{\bibinfo{volume}{44}},
  \bibinfo{pages}{3710} (\bibinfo{year}{1987}).

\bibitem[{\citenamefont{{Diamond} et~al.}(2005)\citenamefont{{Diamond}, {Itoh},
  {Itoh}, and {Hahm}}}]{DIIH2005}
\bibinfo{author}{\bibfnamefont{P.~H.} \bibnamefont{{Diamond}}},
  \bibinfo{author}{\bibfnamefont{S.-I.} \bibnamefont{{Itoh}}},
  \bibinfo{author}{\bibfnamefont{K.}~\bibnamefont{{Itoh}}}, \bibnamefont{and}
  \bibinfo{author}{\bibfnamefont{T.~S.} \bibnamefont{{Hahm}}},
  \bibinfo{journal}{Plasma Phys. Control. Fusion}
  \textbf{\bibinfo{volume}{47}}, \bibinfo{pages}{R35} (\bibinfo{year}{2005}).

\bibitem[{\citenamefont{{Balk} et~al.}(1990{\natexlab{a}})\citenamefont{{Balk},
  {Nazarenko}, and {Zakharov}}}]{BNZ1990}
\bibinfo{author}{\bibfnamefont{A.~M.} \bibnamefont{{Balk}}},
  \bibinfo{author}{\bibfnamefont{S.~V.} \bibnamefont{{Nazarenko}}},
  \bibnamefont{and} \bibinfo{author}{\bibfnamefont{V.~E.}
  \bibnamefont{{Zakharov}}}, \bibinfo{journal}{Phys. Lett. A}
  \textbf{\bibinfo{volume}{146}}, \bibinfo{pages}{217}
  (\bibinfo{year}{1990}{\natexlab{a}}).

\bibitem[{\citenamefont{{Balk} et~al.}(1990{\natexlab{b}})\citenamefont{{Balk},
  {Nazarenko}, and {Zakharov}}}]{BNZ1990B}
\bibinfo{author}{\bibfnamefont{A.~M.} \bibnamefont{{Balk}}},
  \bibinfo{author}{\bibfnamefont{S.~V.} \bibnamefont{{Nazarenko}}},
  \bibnamefont{and} \bibinfo{author}{\bibfnamefont{V.~E.}
  \bibnamefont{{Zakharov}}}, \bibinfo{journal}{Sov. Phys. - JETP}
  \textbf{\bibinfo{volume}{71}}, \bibinfo{pages}{249}
  (\bibinfo{year}{1990}{\natexlab{b}}).

\bibitem[{\citenamefont{Wagner et~al.}(1982)\citenamefont{Wagner, Becker,
  Behringer, Campbell, Eberhagen, Engelhardt, Fussmann, Gehre, Gernhardt,
  Gierke et~al.}}]{WAG1982}
\bibinfo{author}{\bibfnamefont{F.}~\bibnamefont{Wagner}},
  \bibinfo{author}{\bibfnamefont{G.}~\bibnamefont{Becker}},
  \bibinfo{author}{\bibfnamefont{K.}~\bibnamefont{Behringer}},
  \bibinfo{author}{\bibfnamefont{D.}~\bibnamefont{Campbell}},
  \bibinfo{author}{\bibfnamefont{A.}~\bibnamefont{Eberhagen}},
  \bibinfo{author}{\bibfnamefont{W.}~\bibnamefont{Engelhardt}},
  \bibinfo{author}{\bibfnamefont{G.}~\bibnamefont{Fussmann}},
  \bibinfo{author}{\bibfnamefont{O.}~\bibnamefont{Gehre}},
  \bibinfo{author}{\bibfnamefont{J.}~\bibnamefont{Gernhardt}},
  \bibinfo{author}{\bibfnamefont{G.~v.} \bibnamefont{Gierke}},
  \bibnamefont{et~al.}, \bibinfo{journal}{Phys. Rev. Lett.}
  \textbf{\bibinfo{volume}{49}}, \bibinfo{pages}{1408} (\bibinfo{year}{1982}).

\bibitem[{\citenamefont{Kraichnan}(1967)}]{KRA1967}
\bibinfo{author}{\bibfnamefont{R.~H.} \bibnamefont{Kraichnan}},
  \bibinfo{journal}{Phys. Fluids} \textbf{\bibinfo{volume}{10}},
  \bibinfo{pages}{1417} (\bibinfo{year}{1967}).

\bibitem[{\citenamefont{{Balk} et~al.}(1991)\citenamefont{{Balk}, {Nazarenko},
  and {Zakharov}}}]{BNZ1991}
\bibinfo{author}{\bibfnamefont{A.~M.} \bibnamefont{{Balk}}},
  \bibinfo{author}{\bibfnamefont{S.~V.} \bibnamefont{{Nazarenko}}},
  \bibnamefont{and} \bibinfo{author}{\bibfnamefont{V.~E.}
  \bibnamefont{{Zakharov}}}, \bibinfo{journal}{Phys. Lett. A}
  \textbf{\bibinfo{volume}{152}}, \bibinfo{pages}{276} (\bibinfo{year}{1991}).

\bibitem[{\citenamefont{{Balk}}(1991)}]{BAL1991}
\bibinfo{author}{\bibfnamefont{A.~M.} \bibnamefont{{Balk}}},
  \bibinfo{journal}{Phys. Lett. A} \textbf{\bibinfo{volume}{155}},
  \bibinfo{pages}{20} (\bibinfo{year}{1991}).

\bibitem[{\citenamefont{{Balk}}(1997)}]{BAL1997}
\bibinfo{author}{\bibfnamefont{A.~M.} \bibnamefont{{Balk}}},
  \bibinfo{journal}{SIAM Review} \textbf{\bibinfo{volume}{39}},
  \bibinfo{pages}{68} (\bibinfo{year}{1997}).

\bibitem[{\citenamefont{{Nazarenko} and {Quinn}}(2009)}]{NQ2009}
\bibinfo{author}{\bibfnamefont{S.~V.} \bibnamefont{{Nazarenko}}}
  \bibnamefont{and} \bibinfo{author}{\bibfnamefont{B.~E.}
  \bibnamefont{{Quinn}}}, \bibinfo{journal}{ArXiv e-prints}
  (\bibinfo{year}{2009}), \eprint{0905.1314}.

\bibitem[{\citenamefont{Lorentz}(1972)}]{LOR1972}
\bibinfo{author}{\bibfnamefont{E.~N.} \bibnamefont{Lorentz}},
  \bibinfo{journal}{J. Atmo. Sci.} \textbf{\bibinfo{volume}{29}},
  \bibinfo{pages}{258} (\bibinfo{year}{1972}).

\bibitem[{\citenamefont{K.~Mima and Lee}(1980)}]{ML1980}
\bibinfo{author}{\bibfnamefont{K.}~\bibnamefont{K.~Mima}} \bibnamefont{and}
  \bibinfo{author}{\bibfnamefont{Y.~C.} \bibnamefont{Lee}},
  \bibinfo{journal}{Phys. Fluids} \textbf{\bibinfo{volume}{23}},
  \bibinfo{pages}{105} (\bibinfo{year}{1980}).

\bibitem[{\citenamefont{Smolyakov et~al.}(2000)\citenamefont{Smolyakov,
  Diamond, and Shevchenko}}]{SDS2000}
\bibinfo{author}{\bibfnamefont{A.~I.} \bibnamefont{Smolyakov}},
  \bibinfo{author}{\bibfnamefont{P.~H.} \bibnamefont{Diamond}},
  \bibnamefont{and} \bibinfo{author}{\bibfnamefont{V.~I.}
  \bibnamefont{Shevchenko}}, \bibinfo{journal}{Phys. Plasmas}
  \textbf{\bibinfo{volume}{7}}, \bibinfo{pages}{1349} (\bibinfo{year}{2000}).

\bibitem[{\citenamefont{Onishchenko et~al.}(2004)\citenamefont{Onishchenko,
  Pokhotelov, Sagdeev, Shukla, and Stenflo}}]{OPSSS2004}
\bibinfo{author}{\bibfnamefont{O.}~\bibnamefont{Onishchenko}},
  \bibinfo{author}{\bibfnamefont{O.}~\bibnamefont{Pokhotelov}},
  \bibinfo{author}{\bibfnamefont{R.}~\bibnamefont{Sagdeev}},
  \bibinfo{author}{\bibfnamefont{P.}~\bibnamefont{Shukla}}, \bibnamefont{and}
  \bibinfo{author}{\bibfnamefont{L.}~\bibnamefont{Stenflo}},
  \bibinfo{journal}{Nonlin. Proc. Geophys.} \textbf{\bibinfo{volume}{11}},
  \bibinfo{pages}{241} (\bibinfo{year}{2004}).

\bibitem[{\citenamefont{Smolyakov and Krasheninnikov}(2008)}]{SK2008}
\bibinfo{author}{\bibfnamefont{A.~I.} \bibnamefont{Smolyakov}}
  \bibnamefont{and} \bibinfo{author}{\bibfnamefont{S.~I.}
  \bibnamefont{Krasheninnikov}}, \bibinfo{journal}{Phys. Plasmas}
  \textbf{\bibinfo{volume}{15}}, \bibinfo{pages}{072302}
  (\bibinfo{year}{2008}).

\bibitem[{\citenamefont{Benjamin and Feir}(1967)}]{BF1967}
\bibinfo{author}{\bibfnamefont{T.}~\bibnamefont{Benjamin}} \bibnamefont{and}
  \bibinfo{author}{\bibfnamefont{J.}~\bibnamefont{Feir}}, \bibinfo{journal}{J.
  Fluid Mech.} \textbf{\bibinfo{volume}{27}}, \bibinfo{pages}{417}
  (\bibinfo{year}{1967}).

\bibitem[{\citenamefont{Onorato et~al.}(2001)\citenamefont{Onorato, Osborne,
  Serio, and Bertone}}]{OOSB2001}
\bibinfo{author}{\bibfnamefont{M.}~\bibnamefont{Onorato}},
  \bibinfo{author}{\bibfnamefont{A.~R.} \bibnamefont{Osborne}},
  \bibinfo{author}{\bibfnamefont{M.}~\bibnamefont{Serio}}, \bibnamefont{and}
  \bibinfo{author}{\bibfnamefont{S.}~\bibnamefont{Bertone}},
  \bibinfo{journal}{Phys. Rev. Lett.} \textbf{\bibinfo{volume}{86}},
  \bibinfo{pages}{5831} (\bibinfo{year}{2001}).

\bibitem[{\citenamefont{Janssen}(2003)}]{JAN2003}
\bibinfo{author}{\bibfnamefont{P.~A.~E.~M.} \bibnamefont{Janssen}},
  \bibinfo{journal}{J. Phys. Ocean.} p. \bibinfo{pages}{863}
  (\bibinfo{year}{2003}).

\bibitem[{\citenamefont{Arnold and Meshalkin}(1960)}]{AM1960}
\bibinfo{author}{\bibfnamefont{V.~I.} \bibnamefont{Arnold}} \bibnamefont{and}
  \bibinfo{author}{\bibfnamefont{L.~D.} \bibnamefont{Meshalkin}},
  \bibinfo{journal}{Usp. Mat. Nauk} \textbf{\bibinfo{volume}{15}},
  \bibinfo{pages}{247} (\bibinfo{year}{1960}).

\bibitem[{\citenamefont{Sagdeev and Galeev}(1969)}]{SG1969}
\bibinfo{author}{\bibfnamefont{Z.}~\bibnamefont{Sagdeev}} \bibnamefont{and}
  \bibinfo{author}{\bibfnamefont{A.~A.} \bibnamefont{Galeev}},
  \emph{\bibinfo{title}{Nonlinear Plasma Theory}}
  (\bibinfo{publisher}{Benjamin}, \bibinfo{address}{New York},
  \bibinfo{year}{1969}).

\bibitem[{\citenamefont{McWilliams}(2006)}]{MCW2006}
\bibinfo{author}{\bibfnamefont{J.~C.} \bibnamefont{McWilliams}},
  \emph{\bibinfo{title}{Fundamentals of Geophysical Fluid Dynamics}}
  (\bibinfo{publisher}{Cambridge University Press}, \bibinfo{year}{2006}).

\bibitem[{\citenamefont{Charney}(1949)}]{CHA1949}
\bibinfo{author}{\bibfnamefont{J.~G.} \bibnamefont{Charney}},
  \bibinfo{journal}{J. Meteor} \textbf{\bibinfo{volume}{6}},
  \bibinfo{pages}{371} (\bibinfo{year}{1949}).

\bibitem[{\citenamefont{Hasegawa and Mima}(1978)}]{HM1978}
\bibinfo{author}{\bibfnamefont{A.}~\bibnamefont{Hasegawa}} \bibnamefont{and}
  \bibinfo{author}{\bibfnamefont{K.}~\bibnamefont{Mima}},
  \bibinfo{journal}{Phys. Fluids} \textbf{\bibinfo{volume}{21}},
  \bibinfo{pages}{87} (\bibinfo{year}{1978}).

\bibitem[{\citenamefont{Rudakov and Sagdeev}(1961)}]{RS1961}
\bibinfo{author}{\bibfnamefont{L.~I.} \bibnamefont{Rudakov}} \bibnamefont{and}
  \bibinfo{author}{\bibfnamefont{R.~Z.} \bibnamefont{Sagdeev}},
  \bibinfo{journal}{Sov. Phys. Dokl.} \textbf{\bibinfo{volume}{6}},
  \bibinfo{pages}{415} (\bibinfo{year}{1961}).

\bibitem[{\citenamefont{Kartashova and L'vov}(2007)}]{KL2007}
\bibinfo{author}{\bibfnamefont{E.}~\bibnamefont{Kartashova}} \bibnamefont{and}
  \bibinfo{author}{\bibfnamefont{V.~S.} \bibnamefont{L'vov}},
  \bibinfo{journal}{Phys. Rev. Lett.} \textbf{\bibinfo{volume}{98}},
  \bibinfo{pages}{198501} (\bibinfo{year}{2007}).

\bibitem[{\citenamefont{Bustamante and Kartashova}(2009)}]{BK2009}
\bibinfo{author}{\bibfnamefont{M.~D.} \bibnamefont{Bustamante}}
  \bibnamefont{and}
  \bibinfo{author}{\bibfnamefont{E.}~\bibnamefont{Kartashova}},
  \bibinfo{journal}{Europhys. Lett.} \textbf{\bibinfo{volume}{85}},
  \bibinfo{pages}{34002} (\bibinfo{year}{2009}).

\bibitem[{\citenamefont{Manfredi et~al.}(2001)\citenamefont{Manfredi, Roach,
  and Denby}}]{MRD2001}
\bibinfo{author}{\bibfnamefont{G.}~\bibnamefont{Manfredi}},
  \bibinfo{author}{\bibfnamefont{C.~M.} \bibnamefont{Roach}}, \bibnamefont{and}
  \bibinfo{author}{\bibfnamefont{R.~O.} \bibnamefont{Denby}},
  \bibinfo{journal}{Plasma Phys. Control. Fusion}
  \textbf{\bibinfo{volume}{43}}, \bibinfo{pages}{825} (\bibinfo{year}{2001}).

\bibitem[{\citenamefont{Rhines}(1975)}]{RHI1975}
\bibinfo{author}{\bibfnamefont{P.}~\bibnamefont{Rhines}}, \bibinfo{journal}{J.
  Fluid Mech.} \textbf{\bibinfo{volume}{69}}, \bibinfo{pages}{417}
  (\bibinfo{year}{1975}).

\bibitem[{\citenamefont{Connaughton et~al.}(2001)\citenamefont{Connaughton,
  Nazarenko, and Pushkarev}}]{CNP01}
\bibinfo{author}{\bibfnamefont{C.}~\bibnamefont{Connaughton}},
  \bibinfo{author}{\bibfnamefont{S.}~\bibnamefont{Nazarenko}},
  \bibnamefont{and}
  \bibinfo{author}{\bibfnamefont{A.}~\bibnamefont{Pushkarev}},
  \bibinfo{journal}{Phys. Rev. E} \textbf{\bibinfo{volume}{63}},
  \bibinfo{pages}{046306} (\bibinfo{year}{2001}).

\bibitem[{\citenamefont{Dritschel and McIntyre}(2008)}]{DM2008}
\bibinfo{author}{\bibfnamefont{D.~G.} \bibnamefont{Dritschel}}
  \bibnamefont{and} \bibinfo{author}{\bibfnamefont{M.~E.}
  \bibnamefont{McIntyre}}, \bibinfo{journal}{J. Atmos. Sci.}
  \textbf{\bibinfo{volume}{65}}, \bibinfo{pages}{855} (\bibinfo{year}{2008}).

\bibitem[{\citenamefont{Mahanti}(1981)}]{MAH1981}
\bibinfo{author}{\bibfnamefont{A.~C.} \bibnamefont{Mahanti}},
  \bibinfo{journal}{Arch. Met. Geoph. Biokl., Ser. A}
  \textbf{\bibinfo{volume}{30}}, \bibinfo{pages}{211} (\bibinfo{year}{1981}).

\bibitem[{\citenamefont{Kuo}(1949)}]{KUO1949}
\bibinfo{author}{\bibfnamefont{H.~L.} \bibnamefont{Kuo}}, \bibinfo{journal}{J.
  Meteor.} \textbf{\bibinfo{volume}{6}}, \bibinfo{pages}{105}
  (\bibinfo{year}{1949}).

\end{thebibliography}
\end{document}